\documentclass[namedreferences,hyperref,optionalrh,solaromanenum]{spr-sola}

\usepackage[title]{appendix}
\usepackage{graphicx}            
\usepackage{color}                       
\usepackage{url}
\usepackage{epstopdf}
\usepackage{hyperref}
\usepackage{lmodern}
\usepackage{float}
\usepackage{amsmath}

\newcommand{\rmd}{{\ \mathrm d}}

\newcommand{\nns}{n_e(s,\varepsilon)}
\newcommand{\nnt}{n_e(\tau)}
\newcommand{\nnr}{n_e(r)}
\newcommand{\SCOM}{s_{COM}}
\newcommand{\TCOM}{\tau_{COM}}
\newcommand{\SPOL}{s_{pol}}

\newcommand{\los}{s}
\newcommand{\robs}{r_{obs}}
\newcommand{\rpos}{r_{pos}}
\newcommand{\rca}{d}
\newcommand{\ints}{\int_s}

\newcommand{\intt}{\int_{\tau}}
\newcommand{\inttdefinf}{\int^{\frac{\pi}{2}}_{-\frac{\pi}{2}}}

\newcommand{\intom}
{\int^{\tau_{+}}_{\tau_{-}}}
\newcommand{\twidth}{(\tau_{+}-\tau_{-})}
\newcommand{\intts}
{\int^{\frac{\pi}{2} - \varepsilon}_{-\frac{\pi}{2}}}

\newcommand{\ds}{\rmd{s}}
\newcommand{\dtau}{\rmd{\tau}}

\chardef\us=`\_

\begin{document}

\begin{frontmatter}

\title{Polarization Diagnostics Applied to Coronal Mass Ejections and the Background Solar Wind}

\author[addressref={aff1},email=sgibson@ucar.edu]{\inits{S.E.}\fnm{Sarah~E.}~\lnm{Gibson}\orcid{0000-0001-9831-2640}}

\author[addressref={aff2},email=deforest@boulder.swri.edu]{\inits{C.E.}\fnm{Craig~E.}~\lnm{DeForest}\orcid{0000-0002-7164-2786}}

\author[addressref={aff3},email=dekoning@colorado.edu]{\inits{C.A.}\fnm{Curt~A.}~\lnm{de Koning}}

\author[addressref={aff4},email=steven.cranmer@colorado.edu]
{\inits{S.R.}\fnm{Steven~R.}~\lnm{Cranmer}%
\orcid{0000-0002-3699-3134}}

\author[addressref={aff1},email=fan@ucar.edu]{\inits{Y.}\fnm{Yuhong}~\lnm{Fan}}

\author[addressref={aff5},email=hum2@aber.ac.uk]{\inits{H.}\fnm{Huw}~\lnm{Morgan}}

\author[addressref={aff6},email=Elena.Provornikova@jhuapl.edu]{\inits{E.}\fnm{Elena}~\lnm{Provornikova}}

\author[addressref={aff1},email=anny@ucar.edu]
{\inits{A.}\fnm{Anna}~\lnm{Malanushenko}}

\author[addressref={aff7},email=david.webb@bc.edu]{\inits{D.}\fnm{David}~\lnm{Webb}}

\address[id=aff1]{National Center for Atmospheric Research, Boulder, CO, 80301, USA}

\address[id=aff2]{Southwest Research Institute, Boulder, CO, 80301, USA}

\address[id=aff3]%
{CIRES, University of Colorado, Boulder, CO, 80301, USA}

\address[id=aff4]%
{Department of Astrophysical and Planetary Sciences,
Laboratory for Atmospheric and Space Physics, University of Colorado, Boulder, CO, USA}

\address[id=aff5]{Aberystwyth University, Aberystwyth, UK}

\address[id=aff6]{Applied Physics Laboratory, Johns Hopkins University, Laurel, MD, USA}

\address[id=aff7]{Institute for Scientific Research, Boston College, Chestnut Hill, MA, USA}

\begin{abstract}

The ratio of radially to tangentially polarized Thomson-scattered white light provides a powerful tool for locating the 3D position of compact structures in the solar corona and inner heliosphere, and the Polarimeter to Unify the Corona and Heliosphere (PUNCH) has been designed to take full advantage of this diagnostic capability. Interestingly, this same observable that establishes the position of transient blob-like structures becomes a {local} measure of the slope of the {global} falloff of density in the background solar wind. {It is thus important to characterize the extent along the line of sight of structures being studied, in order to determine whether they are sufficiently compact for 3D positioning. }
In this paper, we build from analyses of individual lines of sight to 
three-dimensional models of coronal mass ejections (CMEs), allowing us to consider how accurately polarization properties of the transient and quiescent solar wind are diagnosed.
In this way, we demonstrate the challenges and opportunities presented by 
PUNCH polarization data for various quantitative diagnostics.

\end{abstract}
\end{frontmatter}


\section{Introduction}\label{sec1}

The Polarimeter to Unify the Corona and Heliosphere, or PUNCH, is a 
NASA Small Explorer Mission.
PUNCH studies the ``young'' solar wind, by obtaining direct, continuous 3D imaging of the entire corona and inner heliosphere in polarized white light via four synchronous smallsats. It launched in March, 2025, with a two-year prime mission duration. See {\cite{deforest_this_issue}} for an overview of the mission, its instruments, and its scientific objectives.

{As discussed in other papers in this topical issue \citep{malanushenko_this_issue,provornikova_this_issue}, a primary science target for PUNCH is characterizing Coronal Mass Ejections (CMEs).
PUNCH provides a full-circle field of view (FOV) from $1.5-45^\circ$ elongation or $6-180~R_{\odot}$\footnote{{Because elongation angle $\varepsilon$ does not generally map linearly to solar radius, we use $R_\odot$ as a convenience unit representing the approximate apparent solar radius in elongation angle $\varepsilon$ when viewed from $1au$, i.e., $R_\odot=0.25^\circ$. It is not to be confused with $r_\odot$, used elsewhere in this paper, which signifies the physical and fixed solar radius, $r_\odot=695700~km$.}}, mosaicking the fields of view of a central Near Field Imager (NFI) \citep{colaninno_25} with three Wide Field Imagers (WFI) \citep{laurent_25}. In this way, PUNCH fills gaps between existing coronagraphs and heliospheric imagers \citep{brueckner_1995,howard_2008,tomczyk_16,vourlidas_2016,howard_2020,zhukov_25} and obtains continuous wide-field observations of the evolution of CMEs and their substructure from the Sun throughout the inner solar system. This large FOV also yields new insight into the formation, evolution and 3D structure of the stream interaction regions (SIRs) that arise between fast and slow wind streams \citep{neugebauer_66,rouillard_2008} as discussed in this topical issue by {\citet{dekoning_this_issue}}}.

PUNCH obtains both brightness and polarized brightness measurements to image and locate evolving SIR and CME structures in 3D: a measurement of great importance to space weather predictions {\citep{seaton_this_issue}}. Other coronagraphs {including SOHO/LASCO \citep{Moran2004} and STEREO/COR1 \citep{Moran2010} have had both brightness and polarized brightness observations, but no heliospheric imagers until now. PUNCH is the first heliospheric imager to be designed with the sensitivity, spatial FOV, and polarization capability} needed to locate and track bright structures as they propagate, with both its inner coronagraph and its three heliospheric imagers operating as a single ``virtual instrument'' {\citep{deforest_this_issue}}.

In Section~\ref{sec:3dpol} we review how these data may be directly related to a 3D position of a compact mass, or ``superparticle'', along a particular line of sight. We also consider how 
the same image data can provide a completely different diagnostic, a local measure of the global falloff of the background solar wind density. We discuss the transition between these two extremes and how the viewing angle on a compact mass and its width along a line of sight affects the ability to establish its 3D location via polarization, {and discuss} unavoidable uncertainties in polarization analysis due to the size and location of a scattering mass.

In Section~\ref{sec:3dpolmodels}, we  present a variety of forward modeled CMEs,
comparing the {3D position} of the CME front to ground truth simulation knowledge of the position of mass along the line of sight.
We also discuss how front-back ambiguity associated with polarization analysis can be resolved using the wide field of view and sensitivity of PUNCH images, and how establishing the 3D position of CME substructures might be used to diagnose magnetic flux rope chirality.

In Section~\ref{sec:conclusions}, we present our conclusions.

\begin{figure}[ht]%
\centering
\includegraphics[width=0.95\textwidth]{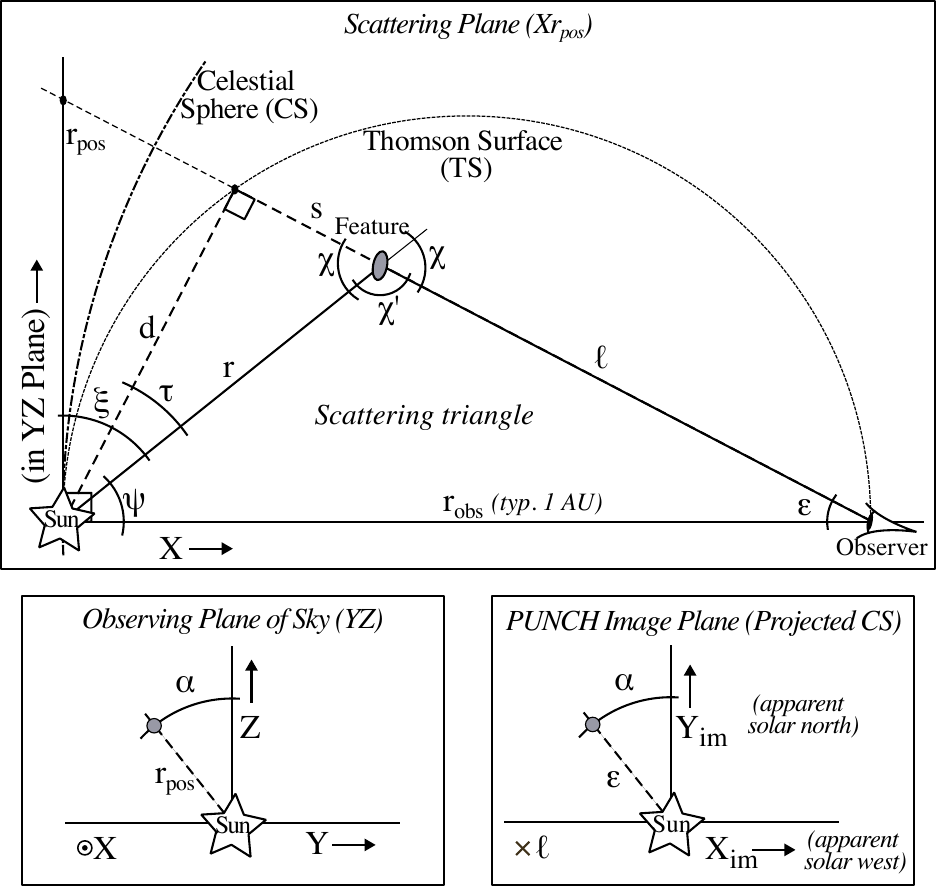}
\caption{Geometry of the system. The superparticle is the grey oval, and the observer is shown as the eyeball on the bottom right, looking along a line of sight with elongation angle $\varepsilon$ (see Appendix~\ref{secAdef} for definitions of other variables). Also indicated is the cross section of the Thomson Surface (TS), which is significant because 
it is the location of closest approach to the Sun for each line of sight, and thus strongest illumination.
Near the Sun, this occurs in or near to the plane of the sky, but for wider elongation angles, the radial closest approaches of the (non-parallel) lines of sight forms the spherical surface shown here. Note that in the mapping of the bottom-left image, the sky plane is the $Y-Z$ plane as is commonly assumed for close-in coronagraph images (since PUNCH is in Earth orbit, the $X-Y-Z$ coordinate system used here are assumed to be Radial-Tangential-Normal (RTN) (see \citet{burlaga_84,franz_02}). For lines of sight farther out, it is more common (and appropriate for the PUNCH mission) to map the structure onto the celestial sphere in an azimuthal-equidistant mapping with grid spacing constant in elongation angle ($\varepsilon$) as seen projected on the $X_{im}-Y_{im}$ image plane (bottom-right inset). See \citet{deforest_this_issue} for further discussion of this figure.\label{fig-geom}}
\end{figure}

\section{Polarization Diagnostics }\label{sec:3dpol}

\subsection{Background}

{The brightness and polarization of the white light corona was first considered theoretically by \citet{schuster_1879}, and then in comparison to observations of solar eclipses 
\citep{young_1911,Minnaert1930,Baumbach1938,ohman1947,vandeHulst1950}.
These analyses demonstrated how} radiation from the solar photosphere is Thomson-scattered in the
optically-thin corona, and how the linearly polarized component of that
radiation (polarized brightness, or $pB$)
 becomes an excellent diagnostic of the line-of-sight (LOS) integral
of $n_e$ \citep[see also][]{Altschuler1972,Munro1977,Inhester2015}. The falloff with distance of $n_e(r)$ has been studied using $pB$ by multiple authors, including
\citet{Saito1970,Munro1977,Fisher1995,Guhathakurta1996,Gibson1999,Edwards2023}.
Others have combined coronagraph data with radio and in-situ measurements
to find density parameterizations that cover distances from the low
corona to interplanetary space
\citep[e.g.,][]{Bird1990,Leblanc1998,Mann2023}.

Many of the density diagnostic studies cited above used observations of $pB$ from ground-based coronagraphs. Total brightness ($tB$)  -- i.e., the sum of radial and tangential brightnesses -- is difficult to measure from the ground because of the largely unpolarized sky brightness background (except during a solar eclipse). With the launch of space-based coronagraphs, it became possible to routinely observe both $pB$ and $tB$,
making it possible to establish the position along the line of sight of a ``superparticle'', as described in the next section.

We note that at least two definitions of $pB$ have entered the literature, as more fully detailed in \citet{deforest_2022}. In this paper, we consider $pB$ to be the {difference} in transmitted radiance through two ideal linear polarizers oriented tangentially and radially to the Sun, respectively, on an instrument image plane. In other words, $pB$ is a modified Stokes parameter analogous to Stokes $Q$ but with reference angle rotating according to the image plane geometry rather than fixed in the image plane {(see \citet{deforest_this_issue} Appendix 2 for further discussion)}. 

Finally, we note that measurements of the white-light corona involve the
combination of the electron-scattered K-corona and the dust-scattered
F-corona (the latter being the cause of ``zodiacal light'').
The F-corona is essentially unpolarized in the inner corona
(elongation angles $\varepsilon < 5-10^{\circ}$), and so
it does not contribute significantly to $pB$ there \citep{Lamy2020}.
However, the F-corona does become polarized to approximately 7\% at larger distances \citep{morgancook_2020}. Further, total brightness ($tB$) includes unpolarized light and so has a contribution from the F-corona at all heights. Fortunately, the F-corona is very smooth compared to the K-corona, and robust methods have been established to model and remove it from both $pB$ and $tB$ \citep{morgancook_2020,lamy2021,lamy2022}.

\subsection{Locating Position of Superparticles}\label{sec:local}

Studies of the corona
close to the Sun are able to use the approximation that the set of LOS observing rays
are all parallel to one another.
In that limit, one can also define the ``plane of the sky'' (POS)
as the plane perpendicular to those rays that also contains the center
of the Sun.
However, for large elongation angles (as observed by PUNCH's Wide Field Imagers) this limit no longer applies and
it makes more sense to refer to the Thomson Surface (TS; Figure~\ref{fig-geom}), sometimes called
the Thomson Sphere, as the locus of points that generally maximize the
optically thin scattering signal \citep[see, e.g.,][]{Vourlidas2006,
Howard2012,Howard2013,DeForest2013}. 

Bearing this in mind, we now consider a compact mass, or superparticle, in 3D Cartesian coordinates, as shown in Figure~\ref{fig-geom}. {(We will explicitly define conditions where a structure's central position can be found, such that it may be considered ``localized'', in Section~\ref{sec:define_local}.)} The LOS elongation angle $\varepsilon$ gives the radial distance $r_{pos}$ projected on the $Y-Z$ plane, and this along with the object's projected position angle $\alpha$ allows us to calculate a
projected (or, for small $\varepsilon$, actual) Y and Z position, as shown in the bottom left insert of Figure \ref{fig-geom}. To find the 3D position we then need to determine the object's position $x$ along the $X$ axis. 
As we show below, the {\it degree of polarization,} $p\equiv{pB}/{tB}$,
can be directly related to the scattering angle $\chi$ of light off of the object.
$\chi$ is the complement of $\tau$, the angular position of the object along the line of sight from the TS, and $x$ can be determined from either $\chi$ or $\tau$, in combination with $\varepsilon$ (see Section~\ref{sec:poltocart}). The full 3D Cartesian position, $x, y, z$ is thus established.

This property of polarization has been exploited in previous works, \citep[e.g.,][]{hildner_75,poland_munro_76,crifo_83,Moran2004,Dere2005,Moran2010,Mierla2009,Mierla2010,Mierla2011,Dai_2014,chifu_thesis,pagano_2015,bemporad_2015,Lu_2017,Floyd2019,Dai_2020,Mierla2022}.
One example that we will return to in Section~\ref{sec:forward_chi} is \citet{DeForest2017}, who used polarization data in STEREO-A COR2 coronagraph data to map the position of the front and core of a CME in three dimensions. When they reprojected it onto the plane of sky observed by the SOHO LASCO coronagraph, they found consistent cotemporal structures between the two longitudinally displaced satellites.

The concept of a superparticle is of course an idealized one, and any solar wind structure will have some extent along the line of sight. It is therefore useful to consider exactly how the position extracted from polarization relates to the position of the center of mass of compact structures along the line of sight, in order to then investigate what might be diagnosed when mass is not so compact.

\subsubsection{Center of Mass}\label{sec:com_local}

In the case of the center of mass (COM) calculation for a distribution of density ($n_e$) along a line of sight,  we are calculating,
\begin{eqnarray}  \label{Eq-COM}
\SCOM = \frac{\ints{\nns \cdot \los \cdot \ds}}{\ints \nns \cdot \ds}
\end{eqnarray}
where $s$ is the distance along the line of sight from the TS to the scattering point. 
From Figure~\ref{fig-geom} we see that,
\begin{eqnarray}
s &=& \rca \cdot \tan(\tau)  \label{eq-distang}\\
ds &=& \ \frac{d}{cos^2(\tau)}\dtau \label{eq-distangdelta}
\end{eqnarray}
where $\rca$ is the {distance} of closest approach of the line of sight, i.e., the intersection with the TS, and $\tau$ is the angle along the line of sight from this to each scattering point.
Note that $\rca$ can be directly related to the LOS elongation angle $\varepsilon$ and the observer's distance from the Sun. Taking observations from Earth, we have:
\begin{eqnarray} \label{Eq-ElongRca}
\rca &=& r_{obs} \cdot \sin(\varepsilon) \\
&=& 1 \textrm{au} \cdot \sin(\varepsilon) \nonumber \\
&=& 215 R_{\odot} \cdot \sin(\varepsilon)\nonumber\,.
\end{eqnarray}

So for a concentration of density at $\tau = \tau_o$,
\begin{eqnarray}  \label{Eq-COMc}
\SCOM 
&=& 
\rca
\cdot \tan(\tau_o) \\
\TCOM &=& \tau_o
\end{eqnarray}

\subsubsection{Polarization Ratio}\label{sec:poltocart}

As shown in e.g., \citet{likathesis}, 

\begin{eqnarray}
pB &\equiv& B_T - B_R \nonumber \\
&=& \ints \mathcal{G}_1 \cdot \nns  \cdot \ds \nonumber \\
tB &\equiv& B_T + B_R \nonumber \\
&=& \ints(2 \cdot \mathcal{G}_2  - \mathcal{G}_1) \cdot \nns \cdot \ds
\end{eqnarray}
where $B_R$, $B_T$ are radially and tangentially polarized light respectively, and 
\begin{eqnarray}
\mathcal{G}_1(r,\chi) &=& \mathcal{F}_o \cdot [(1-u) \cdot A(r) + u \cdot B(r)]  \cdot \sin^2(\chi) \nonumber\\
&=& \mathcal{G}_1^{'}(r) \cdot \sin^2(\chi) \nonumber\\
\mathcal{G}_2(r) &=& \mathcal{F}_o \cdot [(1-u) \cdot C(r)+ u \cdot D(r)]
\label{G1G2}
\end{eqnarray}
where
\begin{equation}
 \mathcal{F}_o =
\frac{\mathcal{C}_1}{(1 - \frac{u}{3})} 
\label{Fo}
\end{equation}
$\mathcal{C}_1$ is a constant, $A, B, C, D$ are the standard radially-dependent van der Hulst coefficients of Thomson scattering,
and
$u$ is the limb darkening coefficient. 

So, the {\it degree of polarization} due to Thomson scattering of the K-corona,
$pB$/$tB$, can be written as 
\begin{eqnarray} 
p &=& 
\frac{\ints \mathcal{G}_1^{'}(r)\cdot \sin^2(\chi) \cdot \nns \cdot \ds}{\ints(2 \cdot \mathcal{G}_2(r) - \mathcal{G}_1^{'}(r)\cdot \sin^2(\chi)) \cdot  \nns \cdot \ds}  \\
\textrm{and recalling that {(see Appendix~\ref{secAdef}):}
}\nonumber \\
\sin^2(\chi) &=& \cos^2(\tau) \nonumber \\
s &=& d \cdot \tan(\tau) \nonumber \\
ds &=& \frac{d}{\cos^2(\tau)}\dtau \nonumber
\end{eqnarray}
we can express this in terms of the angle $\tau$,
\begin{eqnarray}
p &=&
\frac{\intt \mathcal{G}_1^{'}(r) \cdot \nnt \cdot\dtau}{\intt{(\frac{2\mathcal{G}_2(r)}{\cos^2(\tau)}} - \mathcal{G}_1^{'}(r)) \cdot \nnt \cdot\dtau}\label{Eq-pfull-tau}
\end{eqnarray}

For a superparticle positioned at $\tau = \tau_o$ with no other emission along a line of sight $\varepsilon$, the integrals may be ignored and the degree of polarization simplifies to:
\begin{eqnarray} \label{Eq-ploc}
p &=& \frac{\mathcal{G}_1^{'}(r(\varepsilon,\tau_o)) \cdot n_e(\tau_o)} {(\frac{2\mathcal{G}_2(r(\varepsilon,\tau_o))}{\cos^2(\tau_o)} - \mathcal{G}_1^{'}(r(\varepsilon,\tau_o))) \cdot n_e(\tau_o)} \nonumber \\
&=& \frac{\cos^2(\tau_o)}{(2 \cdot \frac{\mathcal{G}_2(r(\varepsilon,\tau_o))}{\mathcal{G}_1^{'}(r(\varepsilon,\tau_o))} - \cos^2(\tau_o))}
\end{eqnarray}

We define the {\it polarization ratio} of radially to tangentially polarized light 
as,
\begin{eqnarray}\label{Eq-polratloc}
PR 
&\equiv& \frac{B_R}{B_T}
\\
&=& \frac{tB - pB}{tB + pB}
\\
&=&\frac{1-p}{1+p} \label{eq-PR}\\
&=& 1 - \frac{\mathcal{G}_2(r(\varepsilon,\tau_o))}{\mathcal{G}_1^{'}(r(\varepsilon,\tau_o))}\cdot \cos^2(\tau_{o})\nonumber
\end{eqnarray}

For everywhere except very close to the Sun, it can be shown that (see e.g. \citet{howard_tappin_2009,dekoning_this_issue}) the van de Hulst coefficients become,

\begin{equation}
    A = C = \frac{3}{2} B = \frac{3}{2}D = \mathcal{C}_2 r^{-2}
    \label{vdh_far}
\end{equation}

so $ \mathcal{G}_2 = \mathcal{G}_1^{'}$,
and we can simplify further:
\begin{eqnarray} \label{Eq-plocsimp}
p = \frac{\cos^2(\tau_{o})}{2 - \cos^2(\tau_{o})} \\
PR = 1 - \cos^2(\tau_{o}) = \sin^2(\tau_{o}) 
\end{eqnarray}
which matches the far-field formula derived by \citet{deforest_13b}.

If we define an angular distance from the TS in terms of polarization ratio, 
\begin{equation}\label{Eq-tauPR}
\tau_{pol}=\operatorname{asin}(\sqrt{PR})
\end{equation}
For the $\delta$-function mass distribution centered at $\tau = \tau_o$ (i.e., the superparticle that we have been discussing), this is simply
\begin{equation}
\tau_{pol}=|\tau_o|
\label{Eq-tpol_sp}
\end{equation}
$\tau_{pol}$ is unsigned and may arise from either of the two $\pm \tau_o$ solutions symmetric about the TS. The sign ambiguity is due to the $\operatorname{asin}$ function (the ambiguity in the square root is resolved by the geometry of the scattering plane -- see Appendix B of \citet{deforest_this_issue} for discussion). 

Similarly, 
\begin{eqnarray}\label{Eq-COMsimp}
\SPOL &=&  \rca \cdot \tan(\tau_{pol})  = \rca \cdot \frac{\sqrt{PR}}{\sqrt{1-PR}} \nonumber \\ 
 &=&    \robs \cdot\sin(\varepsilon) \cdot \frac{\sqrt{PR}}{\sqrt{1-PR}}
\end{eqnarray}

If $\tau_o= \tau_{pol} = 0$, meaning the superparticle is located at the TS, we see that $s_{pol} = PR = 0$, just as would be expected.

From Figure~\ref{fig-geom} we see that in general, $\pm \SPOL$ can be related to two  positions along the $X$-axis,
\begin{eqnarray}
x_{pol} &=& d \cdot \sin(\varepsilon) \pm s_{pol} \cdot \cos(\varepsilon)\nonumber \\
&=& d \cdot \sin(\varepsilon) \pm d \cdot \cos(\varepsilon)\tan(|\tau_{pol}|)  \nonumber \\
&=& \robs \cdot \sin(\varepsilon) \cdot [\sin(\varepsilon) \pm \cos(\varepsilon) \tan(|\tau_{pol}|)] \nonumber \\
&=& \robs \cdot \sin(\varepsilon) \cdot [\sin(\varepsilon) \pm \cos(\varepsilon) \frac{\sqrt{PR}}{\sqrt{1-PR}}] 
\label{Eq-Xpol}
\end{eqnarray}

Inspection of Figure~\ref{fig-geom} shows us that the complete three-dimensional location of the superparticle 
can be specified in Cartesian coordinates from Equations~\ref{Eq-Xpol} and
\begin{eqnarray}
y_{pol} &=&  (r_{pos} - x_{pol} \cdot \tan(\varepsilon)) \cdot sin(\alpha) 
\label{Eq-Ypol}
\end{eqnarray}
\begin{eqnarray}
z_{pol} &=& (r_{pos} - x_{pol} \cdot \tan(\varepsilon)) \cdot cos(\alpha)
\label{Eq-Zpol}
\end{eqnarray}
In summary: if sufficiently compact {(more on this in Section~\ref{sec:define_local})}, we can directly relate PUNCH observations of $pB$, $tB$ to the 3D COM position of a superparticle --- subject to a sign ambiguity as to whether it is in front or behind the TS. Once this ambiguity is resolved {by polarization data (see Section \ref{sec:front-back})}, the 3D position of the superparticle is obtained.

\subsection{Diagnosing Radial Density Falloff}\label{sec:spherwind}

The density falloff in the background corona and solar wind is generally radial, i.e., $n_e \propto r^{-c}$ \citep{Saito1970}, with density exponent $c$ decreasing from values of {order
10 in the first few tenths of a solar radius }{\citep{gibstreamwsm,guhawsmhole} where wind flow is subsonic and scale height effects dominate, to the asymptotic value of 2 expected for
a constant-velocity solar wind in interplanetary space \citep{parker57}. We note that the fast solar wind is expected to reach its final speed closer to the Sun than the slow solar wind, which continues to accelerate to large distances so that $c > 2$ \citep{bunting_2024}.
For the range of distances relevant to PUNCH that we primarily consider
in this paper, the expected range of variation in $c$ is between $2$ and $4$, e.g., 
as shown in the observationally-constrained model of \citet{Leblanc1998},
$c \approx 3.35$ at $r = 5 \, R_{\odot}$, and it decreases to
2.30 at $r = 10 \, R_{\odot}$, then to 2.002 at $r = 100 \, R_{\odot}$.
Using data from {\em Parker Solar Probe,} \citet{Maruca2023} found
evidence for solar wind acceleration between 30 and 150 $R_{\odot}$,
with an associated density exponent $c \approx 2.26$ in that region.
{One-dimensional turbulence-driven models provide similar ranges, with the
full set of models from \citet{Cranmer2007} and \citet{Cranmer2013}
having median values of $c$ of 2.97, 2.47, and 2.07 at $r = 5, 10$, and
100~$R_{\odot}$, respectively.}

Fitting the exponent $c$ to observations has generally required measurements at multiple radial heights in order to measure the slope of the density falloff. As we will now show, the polarization ratio $PR$ provides a direct, local measurement of this global property. As a parallel with the superparticle analysis, we again consider the relation between COM and the diagnostic provided from white light polarization data.

\subsubsection{Center of Mass}\label{sec:com_global}

Consider a spherically symmetric density falloff {with distance},
\begin{equation} \label{Eq-spherdens}
n_e(r) = n_o \cdot r^{-c}
\end{equation}

Combining this with Equations~\ref{Eq-COM} - \ref{eq-distangdelta} and using
\begin{equation}
    r = \frac{\rca}{\cos(\tau)}
\end{equation}

we find for each line of sight of a given elongation angle, $\varepsilon$, {and limits of integration along the line of sight $\tau_-$ to $\tau_+$,}

\begin{equation}
    \SCOM(\varepsilon) = \frac{\rca(\varepsilon) \cdot \intom{\sin(\tau) \cdot \cos^{c-3}(\tau)} \cdot\dtau}{\intom{\cos^{c-2}(\tau)} \cdot\dtau}.
\end{equation}

Taking as an example the spherically expanding constant speed solar wind case, $c=2$, it follows that
\begin{equation}
    \SCOM(\varepsilon) = \frac{\rca(\varepsilon) \cdot [ln(\cos(\tau_-)) - ln(\cos(\tau_+))]}{\tau_+-\tau_-}
    \label{tcom_2}
\end{equation}
For such a distribution of spherically symmetric density falling off with distance, $\nnt$ peaks at the TS ($s = \tau = 0$) for each line of sight. Thus, Equation~\ref{tcom_2} returns a value of zero for lines of sight 
that are symmetric about the TS, i.e., $\tau_+=-\tau_-$.

However, in solving the integrals for
such extended distributions, it is essential to address the inherent asymmetry of lines of sight about the TS.
From Figure~\ref{fig-geom} we see that 
in general, lines of sight coming from the observer will have limits of integration $\tau_{+}=\frac{\pi}{2} -\varepsilon,
    \tau_{-}=-\frac{\pi}{2}$,
so for the $c=2$ example,
\begin{eqnarray}
    {\SCOM}(\varepsilon) &=& \rca(\varepsilon) \cdot \frac{1}{\pi-\varepsilon}  \cdot (ln[{\cos(\frac{\pi}{2})] - ln[\cos(\frac{\pi}{2} -\varepsilon})])\\
    &=& \frac{r_{obs}}{\pi-\varepsilon} \cdot \sin(\varepsilon) \cdot (-\infty - sin(\varepsilon))
\end{eqnarray} 
The general solution for the asymmetric case, for lines of sight  $0 < \varepsilon < \frac{\pi}{2}$, 
is:
\begin{eqnarray}
   {\SCOM} = -\infty \\
   {\TCOM} = -\frac{\pi}{2}
   \label{Eq-tcomasymm}
\end{eqnarray}

\subsubsection{Polarization Ratio}\label{sec:pol_global}

The $\tau_{pol}$ calculated from the polarization ratio using Equation~\ref{Eq-tauPR} no longer diagnoses the COM for such extended distributions. Instead, it diagnoses the radial falloff of density, as we now describe.

As already noted, we can simplify by assuming the scattering occurs far enough from the Sun for it to be considered a point source (true for the PUNCH FOV $> 5 R_{\odot}$). In that limit of large $r$ {where the limb darkening term drops out}, it follows from Equations~\ref{G1G2}-\ref{Fo} and \ref{vdh_far} that
\begin{eqnarray}
\mathcal{G}_2 &=& \mathcal{G}_1^{'} 
 = \mathcal{C}_1 \cdot 
 (\frac{1}{6}) \cdot \mathcal{C}_2 \cdot r^{-2}
\end{eqnarray}
Equation~\ref{Eq-pfull-tau} becomes,
\begin{eqnarray} \label{eqn_pfullfar}
p 
&=&  \frac{\int r^{-2} \cdot \nnt \cdot\dtau}{\int r^{-2} \cdot \nnt \cdot (2 sec^2(\tau) - 1) \cdot\dtau}\label{eq-pointsource}
\end{eqnarray}

For the spherically symmetric falloff $\nnr = n_o \cdot r^{-c}$, again using $ r = \frac{\rca}{\cos(\tau)}$ and adding limits of integration along an observer's lines of sight $\varepsilon$, it follows that,
\begin{eqnarray} \label{Eq-poldensspher}
p&=& \frac{\intom \cos^q(\tau) \cdot\dtau}{\intom 2 \cos^{(q-2)}(\tau) - \cos^q(\tau) \cdot \dtau}
\end{eqnarray}
where $q=c+2$.

From 
\citet{gradshteyn2007} we see that, for $q>0$,
\begin{eqnarray}
    \intom  \cos^q(\tau) \cdot\dtau = \frac{\cos^{q-1}(\tau_+) \cdot \sin(\tau_+)}{q} \\ \nonumber
    -\frac{\cos^{q-1}(\tau_-)\cdot \sin(\tau_-)}{q} \\ \nonumber
    + \frac{q-1}{q} \intom \cos^{q-2}(\tau) \cdot\dtau
    \label{cosqeqn}
\end{eqnarray}

Integrating from $\tau_- = -\frac{\pi}{2}$ to $\tau_+ =\frac{\pi}{2}$, this reduces to
\begin{eqnarray}
    \inttdefinf  \cos^q(\tau) \cdot\dtau = \frac{q-1}{q} \inttdefinf \cos^{q-2}(\tau) \cdot\dtau
\end{eqnarray}
so Equation~\ref{Eq-poldensspher} becomes
\begin{eqnarray} 
p_{symm} &=& \frac{\frac{q-1}{q} \inttdefinf \cos^{q-2}(\tau)\cdot\dtau}{\inttdefinf 2 \cos^{q-2}(\tau) - \frac{q-1}{q} \inttdefinf \cos^{q-2}(\tau) \cdot\dtau} \nonumber \\ 
&=& \frac{1}{2\frac{q}{q-1} - 1} \nonumber \\
&=& \frac{q-1}{2q - q + 1} \nonumber \\
&=& \frac{q-1}{q+1} \nonumber \\
&=& \frac{c+1}{c+3}\label{Eq-polspher}
\end{eqnarray}

{We note that Equation~\ref{Eq-polspher} was first presented in this form by \citet{Saito1970}, with related expressions found in work referenced therein, i.e., \citet{schuster_1879} and \citet{Baumbach1938}.}

From this we see that
\begin{eqnarray}\label{Eq-polratspher}
PR_{symm} &=& \frac{1-p_{symm}}{1+p_{symm}} \nonumber \\
&=& \frac{1}{c+2}
\end{eqnarray}

\begin{figure}[hb!]
\centering
\includegraphics[width=0.75\textwidth]{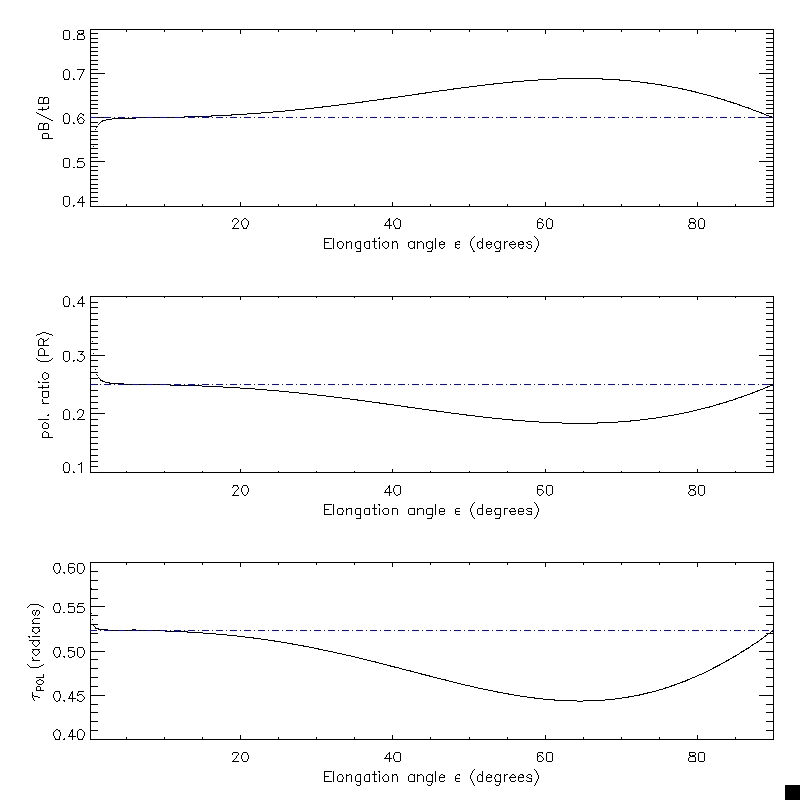}
\caption{Diagnostics obtained for a density distribution $n_e(r)=n_o \cdot r^{-2}$. Top: Ratio of white-light polarized brightness to total brightness ($p$), middle: polarization ratio ($PR$), and bottom: $\tau_{pol}$, all plotted vs. elongation angle $\varepsilon$. The blue dash-dotted line shows the asymptotically regained symmetric solutions, $p_{symm}=3/5, PR_{symm}=1/4$, and $\tau_{pol_{symm}}=\operatorname{asin}(\sqrt (1/4))$. 
Differences are seen between this simplified point-source solution and the solid-black-line complete-form solution, which takes into account the extended disk of the Sun and truncates lines of sight at the observer, as discussed in the text. }\label{Fig-PolPRVol2}
\end{figure}

If we calculate $\tau_{pol}$ for such a symmetric distribution, we find that
\begin{equation}
\tau_{pol_{symm}} = \operatorname{asin}(\sqrt{PR_{symm}}) 
= \operatorname{asin}(\sqrt{\frac{1}{c+2}})\label{Eq-tpol_spher}
\end{equation}

Equation~\ref{Eq-tpol_spher} is analogous to Equation~\ref{Eq-tpol_sp}, but conceptually represents a radically different diagnostic. The latter is a diagnostic of position along the line of sight for a superparticle, while the former is a diagnostic of density slope for a radially-symmetric expanding solar wind.

\begin{figure}[hb!]
\centering
\includegraphics[width=0.99\textwidth]{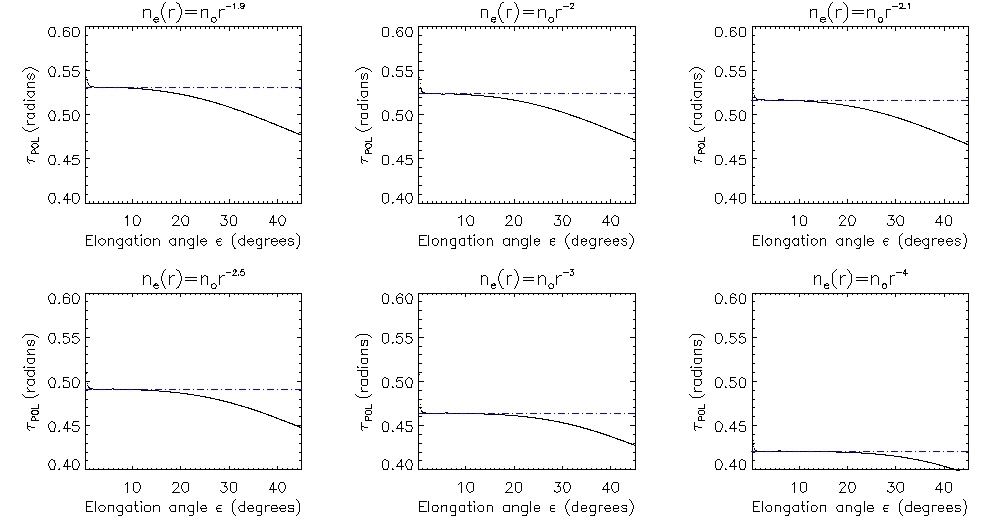}
\caption{Forward modeled $\tau_{pol}$ for $n_e=r^{-c}$ spherically symmetric density distributions, for choices of (top) $c=1.9,2.0,2.1$ and (bottom) $c=2.5,3,4$, for elongation angles $\varepsilon < 45^\circ$. }\label{Fig-TauVolAll45}
\end{figure}

To establish how generally this slope diagnostic might be used, we need to consider the effects of
 asymmetric integration limits for lines of sight $\varepsilon > 0$. In these cases, we see from Equation~\ref{cosqeqn} that an additional correction term dependent LOS elongation angle $\varepsilon$ needs to be added:
 \begin{eqnarray}
   F_{corr}(\varepsilon) = \frac{\sin^{q-1}(\varepsilon) \cdot \cos(\varepsilon)}{q} 
\end{eqnarray}
This leads us to the general (solar point-source) asymmetric solution, 
\begin{equation}
p = \frac{\frac{q-1}{q} \intts \cos^{q-2}(\tau) \cdot\dtau + F_{corr}(\varepsilon)}{\intts 2 \cos^{(q-2)}(\tau) \cdot\dtau - \frac{q-1}{q} \intts \cos^{q-2}(\tau) \cdot\dtau - F_{corr} (\varepsilon)} \label{Eq-polasym}
\end{equation}

It is clear that as $\varepsilon \rightarrow 0$, $F_{corr}(\varepsilon) \rightarrow 0$, and similarly as $\varepsilon \rightarrow \frac{\pi}{2}$, $F_{corr}(\varepsilon) \rightarrow 0$. For both of these limits, the symmetric solutions {shown in} Equations~\ref{Eq-polspher}-\ref{Eq-tpol_spher} are regained.

To demonstrate this, we again consider the case of $q = 4, c = 2, n_e = n_o \cdot r^{-2}$. In this case, Equation~\ref{Eq-polasym} becomes
\begin{eqnarray}
p &=& \frac{\frac{1}{4} [\intts 3\cos^2(\tau) \cdot\dtau + \sin^3(\varepsilon) \cdot \cos(\varepsilon)]}{\frac{1}{4}[\intts 5\cos^2(\tau) \cdot\dtau - \sin^3(\varepsilon) \cdot \cos(\varepsilon)]}\nonumber \\ \nonumber
&=& \frac{3(\frac{\pi - \varepsilon}{2} + \frac{1}{4} \sin(\pi - 2\varepsilon)) + \sin^3(\varepsilon) \cdot \cos(\varepsilon)}{5(\frac{\pi - \varepsilon}{2} + \frac{1}{4} \sin(\pi - 2\varepsilon))- \sin^3(\varepsilon) \cdot \cos(\varepsilon)} \\ \nonumber
&=& \frac{3}{5} (\varepsilon \rightarrow 0) \nonumber
\end{eqnarray}

Figure~\ref{Fig-PolPRVol2} shows the quantities $p, PR, \tau_{pol}$ for the $n(r) = n_o \cdot r^{-2}$ falloff, using the complete form Equation~\ref{Eq-pfull-tau} (that is, not assuming a solar point source). As expected, the symmetric solutions are regained as
$\varepsilon \rightarrow \frac{\pi}{2} (90^\circ)$.
As $\varepsilon \rightarrow 0$, the complete solution (black line) diverges from the point-source solution due to the proximity to the solar disk (see Equation~\ref{G1G2} and \citet{vandeHulst1950}). 
Figure~\ref{Fig-TauVolAll45} shows how polarization would change for various choices of slope $c$ in the PUNCH field of view, $\varepsilon = 0 - 45^\circ$.
In general, the departure from the symmetric, simplified solution is minor for ($\varepsilon < 15^\circ$), indicating it should be possible to distinguish between these slopes. 

\begin{figure}[ht!]
\centering
\includegraphics[width=0.6 \textwidth,trim={40 30 40 30},clip]{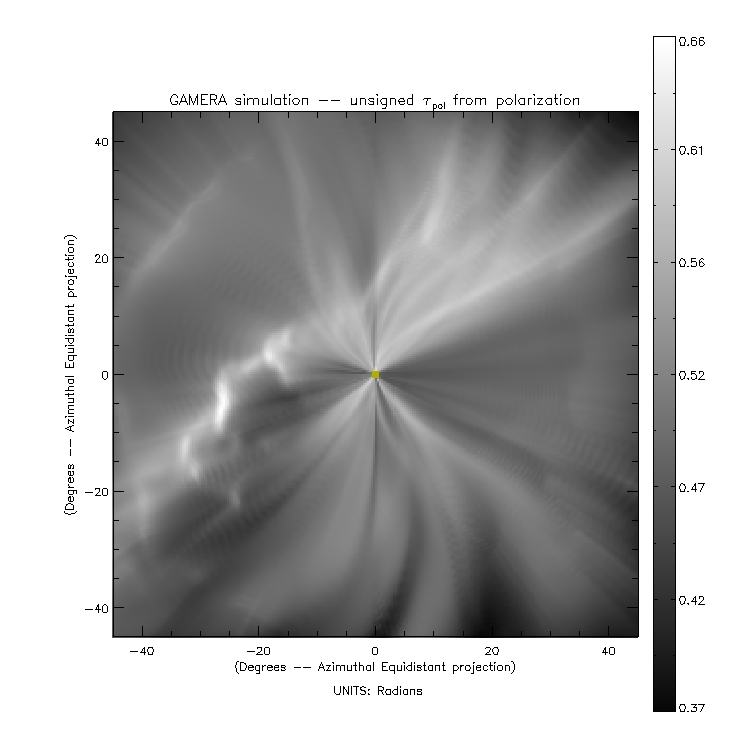}
\caption{Forward-modeled $\tau_{pol}$ for a global solar wind simulation obtained by GAMERA-Helio model of the solar wind driven by the Wang-Sheeley-Arge Air Force Data Assimilative Photospheric Flux Transport (WSA/ADAPT) model output for Carrington rotation 2065 \citep{provornikova_24}. The background (non-structured) solar wind has a non-zero value, consistent with a radially-falling off density, as discussed in the text. }\label{Fig-Gamera}
\end{figure}

 Something close to a $r^{-2}$ falloff is what we would expect to see for a mass-conserving constant-speed solar wind, {so that $\tau_{pol_{r^{-2}}} = \operatorname{asin}(\sqrt (1/4))=0.524$ --  as can be seen in Figures~\ref{Fig-PolPRVol2} and \ref{Fig-TauVolAll45}. Indeed, Figure~\ref{Fig-Gamera} shows this is the central value in the color bar ranging from minimum to maximum density for a global simulation of a forward-modeled complex solar wind}. {The associated polarization degree for a $r^{-2}$ falloff is $p=0.6$ (see Equation~\ref{Eq-polspher}), similar to the asymptotic limit of $0.64$ found in the classical model of \citet{Baumbach1938} (see also \citet{Lamy2020}), and consistent with observations from the ASPIICS spacecraft \citep{zhukov_25b}.}

{In summary:
White light polarization observations will provide a local diagnostic of density radial falloff for the background solar wind, independent from and complementary to non-local techniques. In addition, we have shown that for observations along lines of sights with elongation angles in the range $\sim 1.5-15^\circ$ (the lower part of the PUNCH field of view), the effects of asymmetry are predictable and small enough to distinguish between falloff slopes.}

\subsection{Transitioning from compact to distributed}\label{sec:transition}

It is instructive to consider how $\tau_{pol}$ 
varies for a wedge of constant density centered on $\tau_o$ as it increases its LOS linear width $w$. {Note that {\citet{dekoning_this_issue}} uses a similar approach, but considers features with angular half-width $\Delta$ symmetric about $\tau_o$ (rather than linear half-width $\frac{w}{2}$ symmetric about $\tau_o$), and lets density vary as $r^{-2} (c=2)$ in order to probe how polarization diagnostics 
apply to 2D sheet-like structures associated with SIRs. Our purpose in this section is somewhat different. We consider wedges of constant density ($c=0)$ and vary linear width to calculate the polarization ratio at two extremes -- i.e., delta-function  vs.  infinitely-extended wedge ($w \rightarrow \infty$). The latter is obviously unrealistic, but it allows us to show the transition between the polarization diagnostics described in Section~\ref{sec:local} (appropriate to a 0D, or 1D caustic collection of points) and Section~\ref{sec:spherwind} (appropriate for an infinitely-extended 3D distribution of density). It will also allow us to consider the conditions under which these limits apply for a structure centered on $\tau_o$ (Section~\ref{sec:define_local}).}
As above, we pay special attention to the significance
of asymmetries introduced by the observer's position truncating the line of sight.

\subsubsection{Center of Mass}

The COM transition from compact to infinitely extended masses is trivial for the symmetric case (ignoring the observer), as might be expected from our discussions in Sections~\ref{sec:com_local} and \ref{sec:com_global}.
Figure~\ref{fig-varywidth} (left) illustrates  the constant density case $(c=0)$, which for the symmetric integration (blue line) results in $\tau_{COM}=\tau_o$ for all LOS linear half-widths. For the asymmetric solution (taking the observer's position into account; red line), the solution approaches $\tau_{COM}=-\frac{\pi}{2} (=90^\circ)$ as in Equation~\ref{Eq-tcomasymm}. The departure from the symmetric solution occurs as soon as the wedge LOS linear half-width $\frac{w}{2}$ reaches the distance from the TS to the observer,
\begin{equation}
s_{+_{limit}} 
 = r_{obs} (\cos(\varepsilon) - \sin(\varepsilon) \tan (\tau_o)) \label{Eq-asymmwidth}
 \end{equation}
This position is indicated by the vertical yellow line.

\begin{figure}[ht!]
\centering
\includegraphics[width=0.47\textwidth]{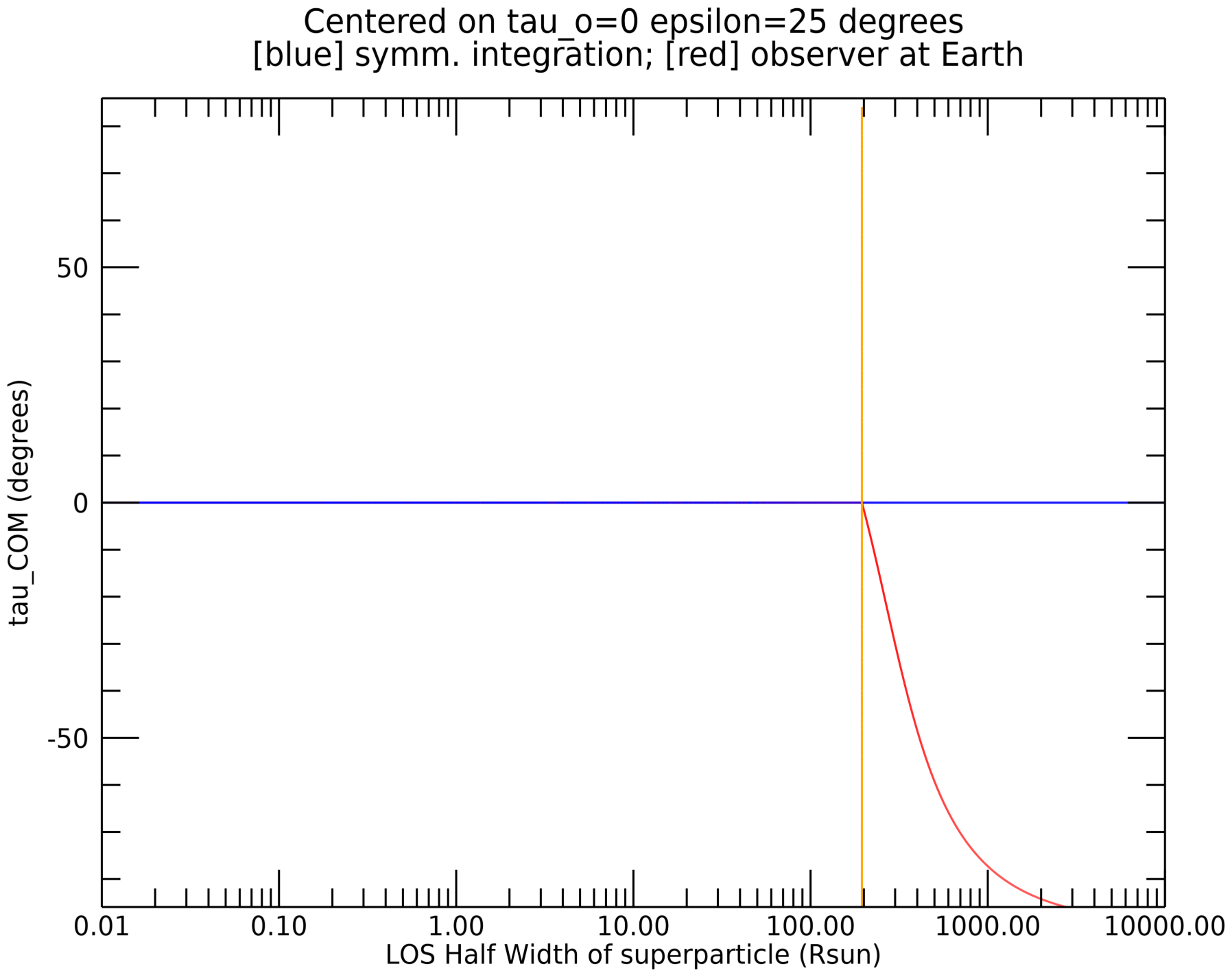}
\includegraphics[width=0.47\textwidth]{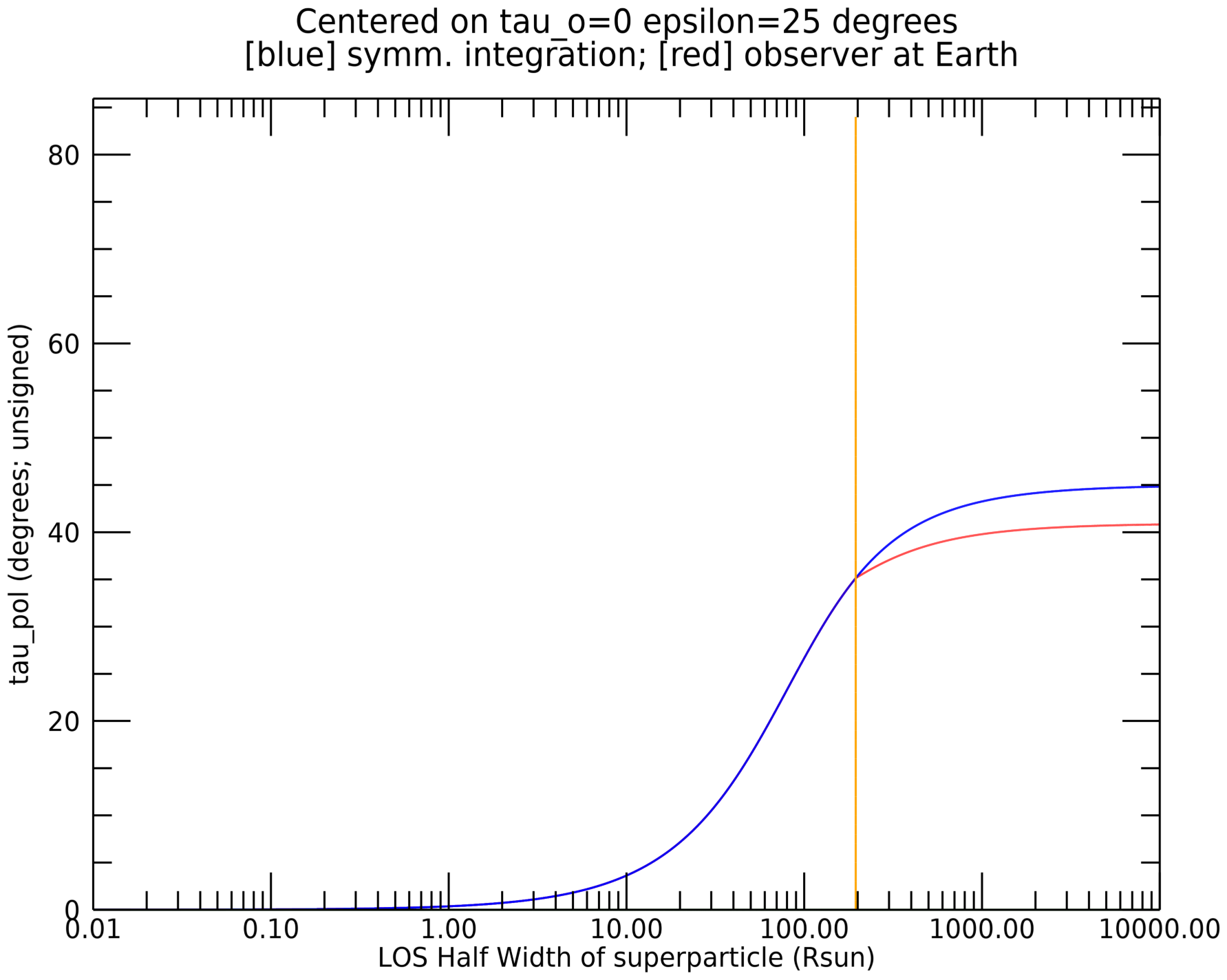}
\caption{(Left) Center of mass angular position $\tau_{COM}$ for a constant-density wedge of varying LOS linear half-width $\frac{w}{2}$, centered on the Thomson Sphere ($\tau_o = 0$), at LOS elongation angle $\varepsilon = 25^\circ$. (Right) Unsigned $\tau_{pol}$ determined from the polarization ratio (PR) as a function of wedge half-width, {which reproduces $\tau_{COM}=\tau_o=0^\circ$ for small LOS widths and asymptotes to $45^\circ$ for the infinite wedge as described in the text.}  The blue lines are the symmetric solution integrated from $-\frac{w}{2}$ to $\frac{w}{2}$, while the red lines are the asymmetric solutions integrated from $-\frac{w}{2}$ to $\frac{w}{2} < s_{+_{limit}}$, defined in Equation~\ref{Eq-asymmwidth}. {The vertical orange lines are placed at $s_{+_{limit}}$
(see text).}}\label{fig-varywidth}
\end{figure}

\subsubsection{Polarization Ratio}

The question of how polarization diagnostics vary with increasing LOS linear half-width $\frac{w}{2}$ 
has clear observational significance.
The constant density $c=0, q=2$ case of Equation~\ref
{Eq-poldensspher}
gives us
\begin{eqnarray} 
p &=&
 \frac{\intom \cos^{2}(\tau)\dtau}{\intom 2 - \cos^{2}(\tau)\dtau} 
 \end{eqnarray}
 
and
\begin{eqnarray} 
PR &=& 
 \frac{\intom \sin^{2}(\tau)\dtau}{\intom \dtau} 
\nonumber \\
&=& \frac{1}{2} \cdot [1 - \frac{\sin(2 \tau_+) - \sin(2 \tau_-)}{2\twidth}]
\end{eqnarray}
leading us to the general equation,
\begin{eqnarray}
\tau_{pol} &=& \operatorname{asin}(\sqrt(\frac{1}{2} \cdot (1 - \frac{(\sin(2 \tau_+) - \sin(2 \tau_-)}{2\twidth})))
\label{Eq-tpolsoln}
\end{eqnarray}
where for the symmetric case,
\begin{eqnarray}
\tau_{\pm} &=& \operatorname{atan}(\frac{s_{\pm}}{d}) = \operatorname{atan}(\tan(\tau_{o}) \pm \frac{w}{2d} )
\label{Eq-tauplusminus}
\end{eqnarray}

{Recalling that $d = r_{obs}\sin(\varepsilon)$, and that $r_{obs} = 215 R_{\odot}$ and $1.5^\circ < \varepsilon < 45^\circ$ in the PUNCH field of view, it is immediately clear that 
if $\frac{w}{2d} >> tan(\tau_o)$,
Equation~\ref{Eq-tauplusminus}  loses sensitivity to $\tau_o$, and $\tau_{\pm}$ go to $\pm \frac{\pi}{2}$. $\tau_{pol}$ thus achieves its asymptotic value of $\frac{\pi}{4} = 45^\circ$, as expected from Equation~\ref{Eq-tpol_spher} for an infinitely extended, uniform density distribution, $c=0$. This is seen in the right panel of Figure~\ref{fig-varywidth}, which shows the case of $\tau_o=0$ (wedge centered on the TS) and $\varepsilon=25^\circ$. It is generally true for all choices of  $\tau_o$ and $\varepsilon$ in the large LOS width (symmetric) limit, as shown in Figure~\ref{fig-tpolvary}.}

\begin{figure}[ht!]
\centering{\includegraphics[width=0.43\textwidth, trim={3 0 3 0},clip]{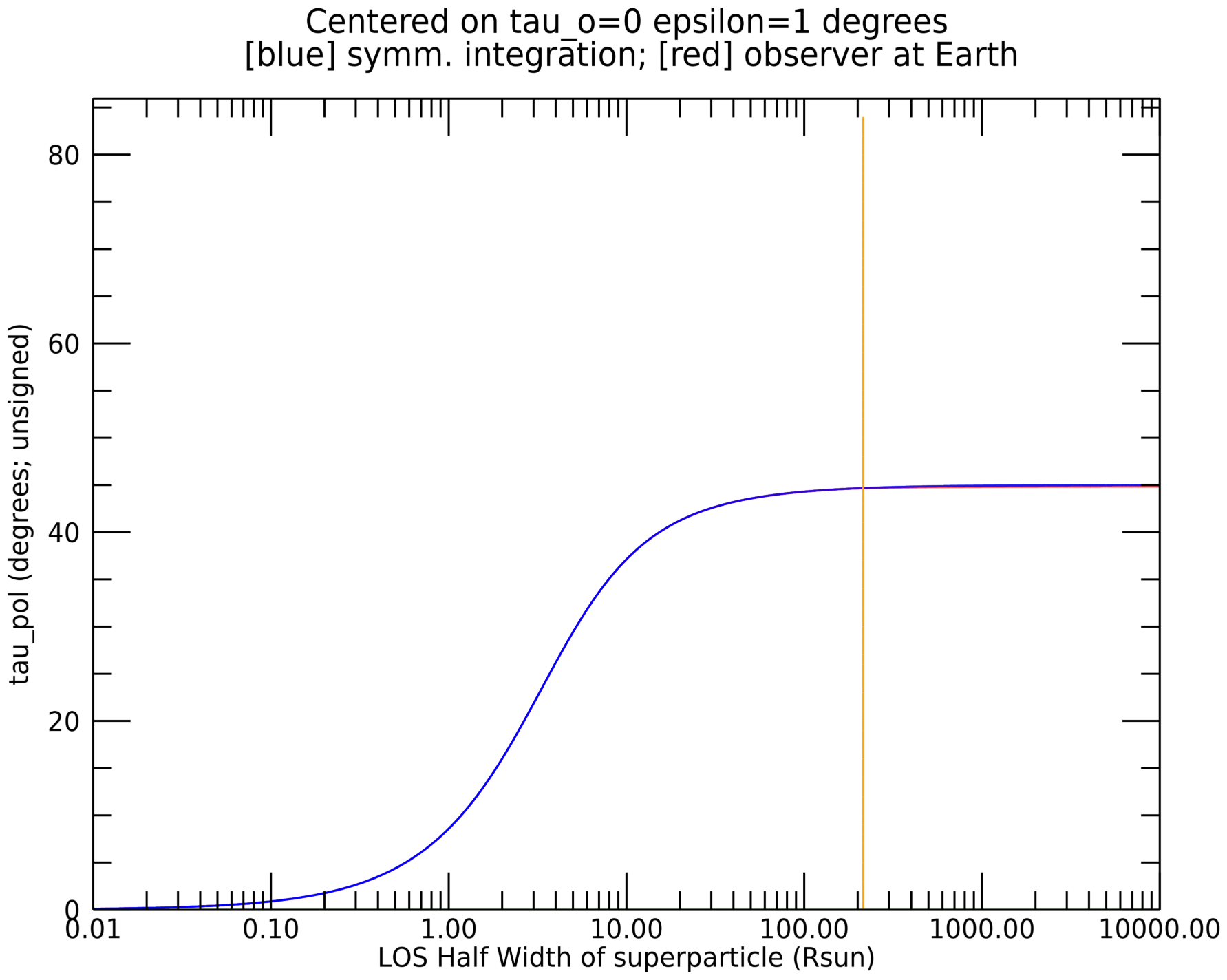}
\includegraphics[width=0.43\textwidth, trim={3 0 3 0},clip]
{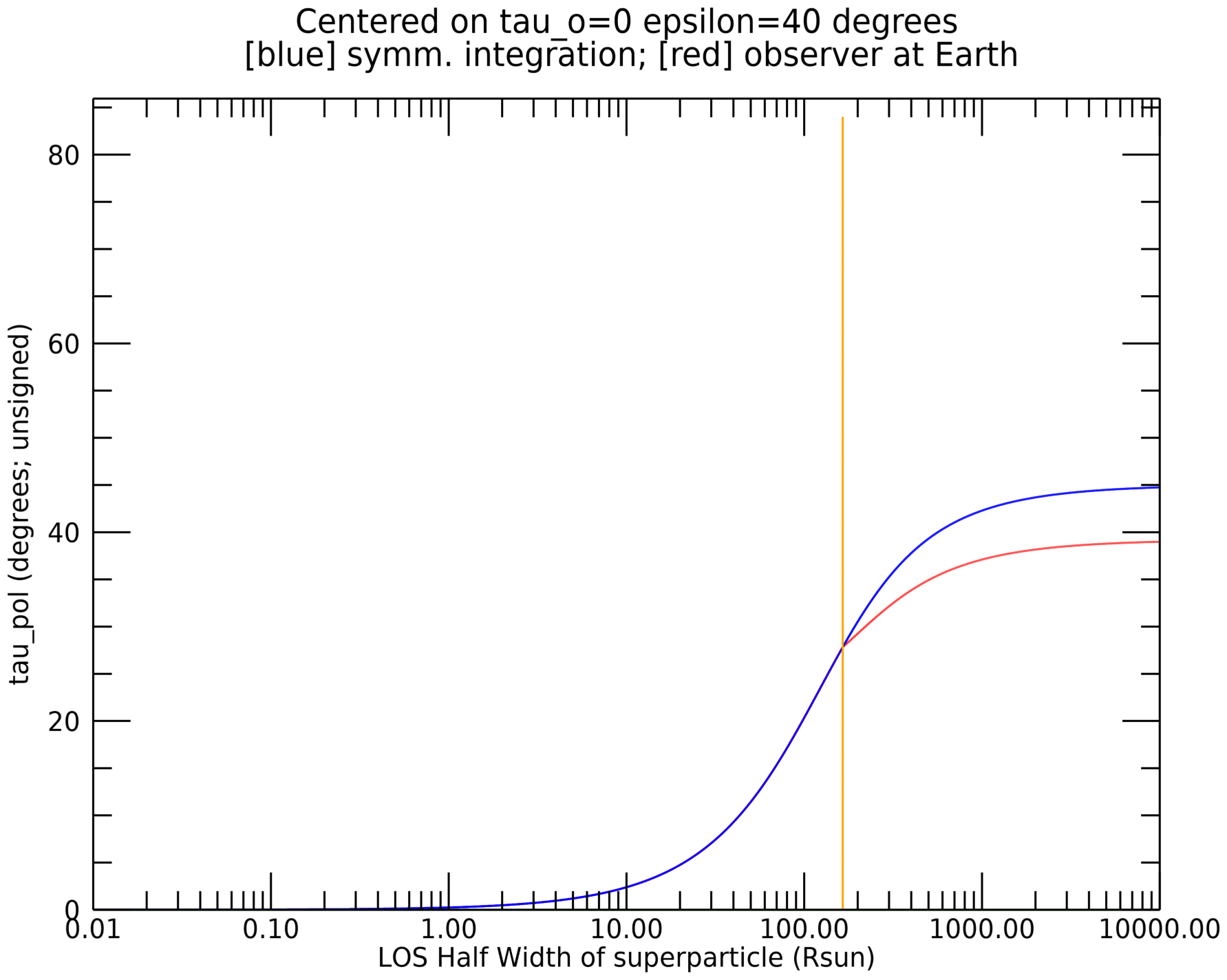}
\includegraphics[width=0.43\textwidth, trim={3 0 3 0},clip]
{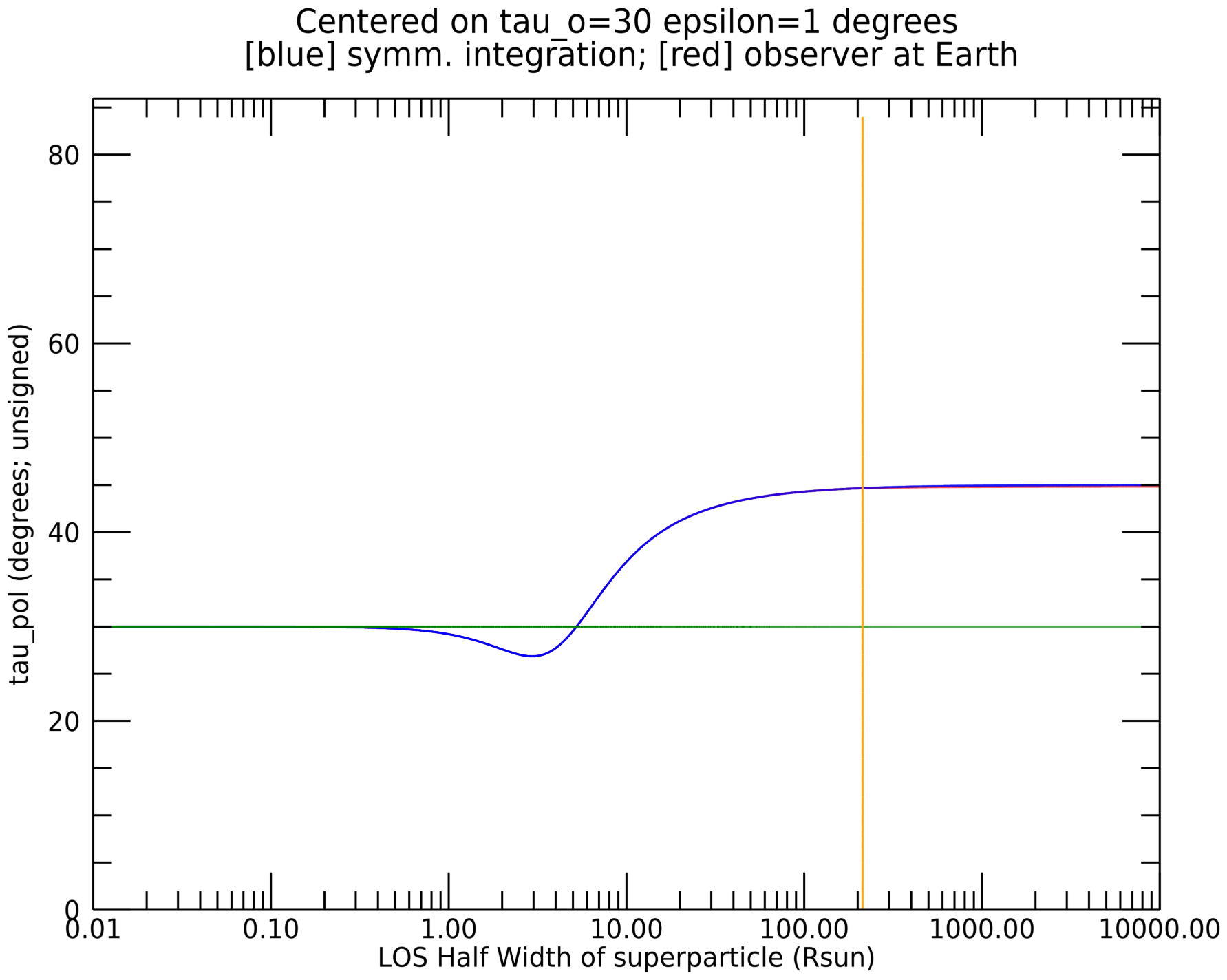}
\includegraphics[width=0.43\textwidth, trim={3 0 3 0},clip]
{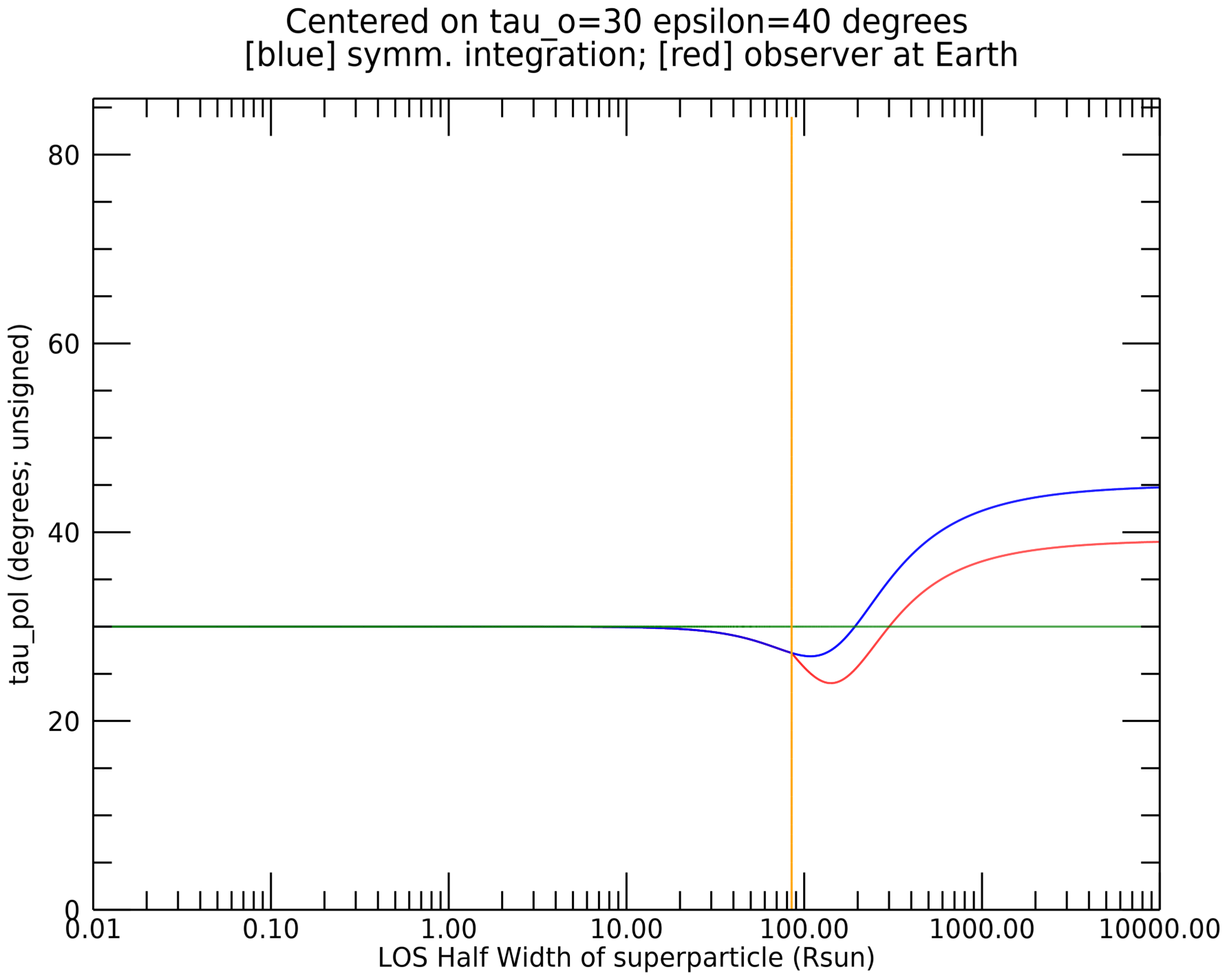}
\includegraphics[width=0.43\textwidth, trim={3 0 3 0},clip]
{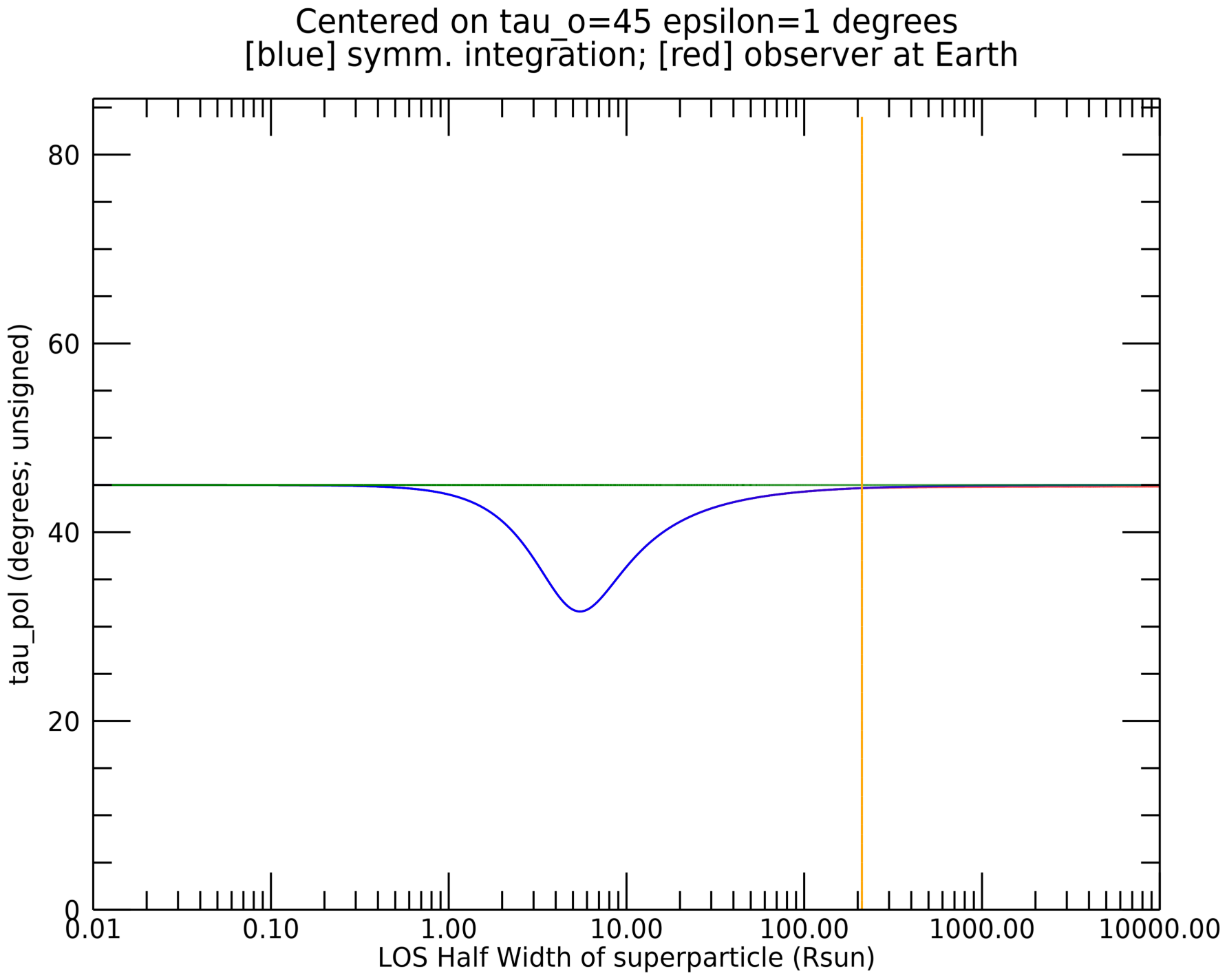}
\includegraphics[width=0.43\textwidth, trim={3 0 3 0},clip]
{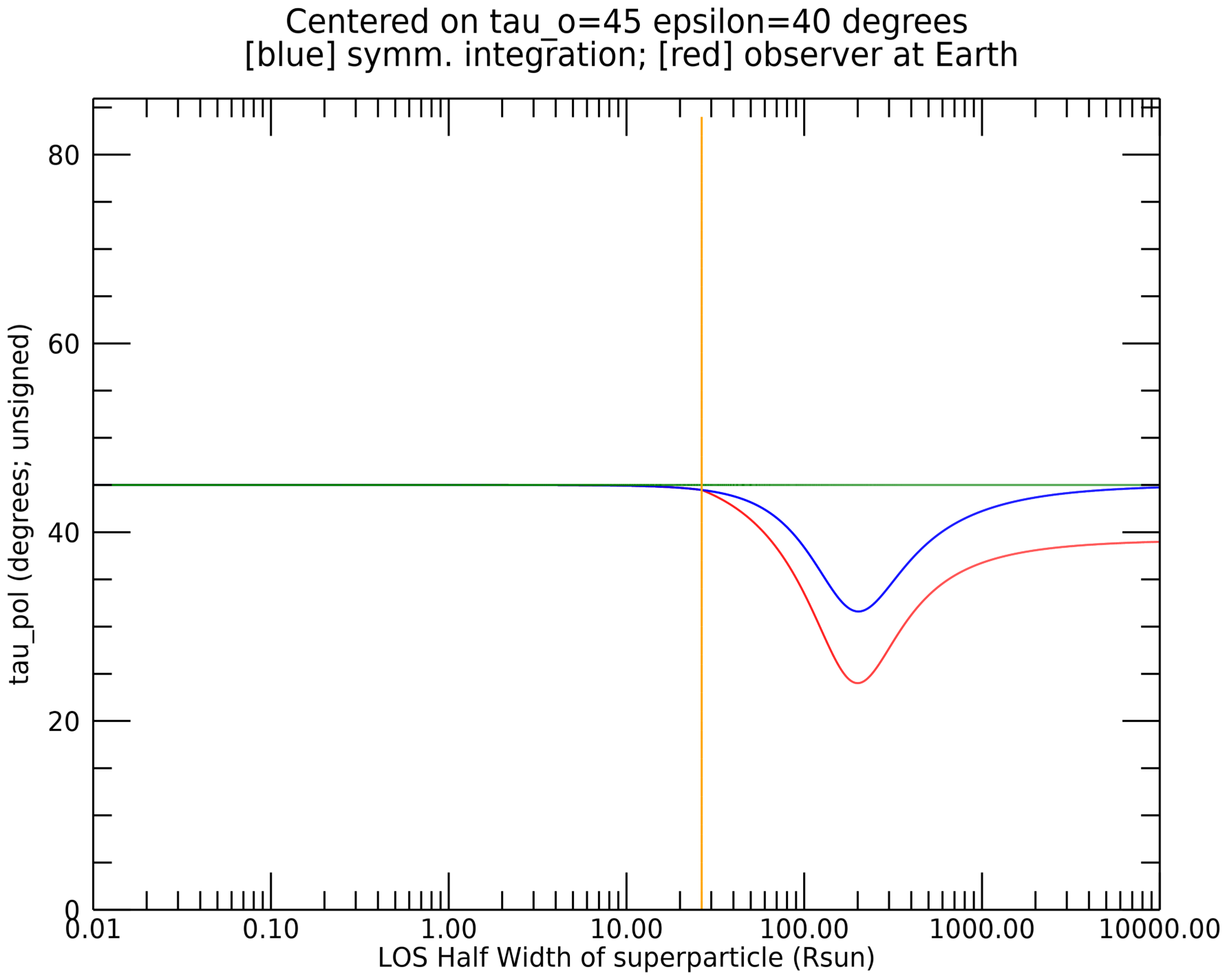}
\includegraphics[width=0.43\textwidth, trim={3 0 3 0},clip]
{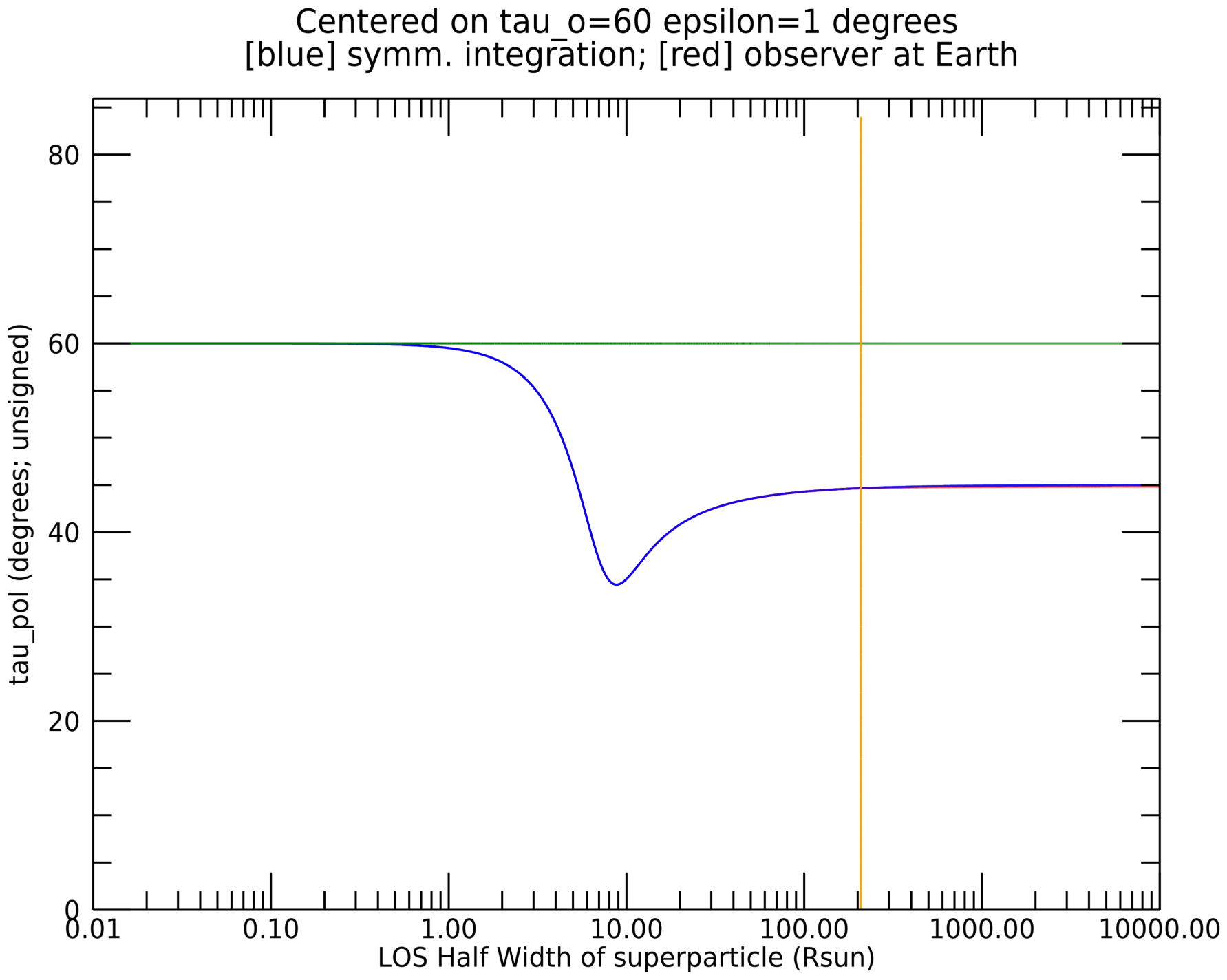}
\includegraphics[width=0.43\textwidth, trim={3 0 3 0},clip]
{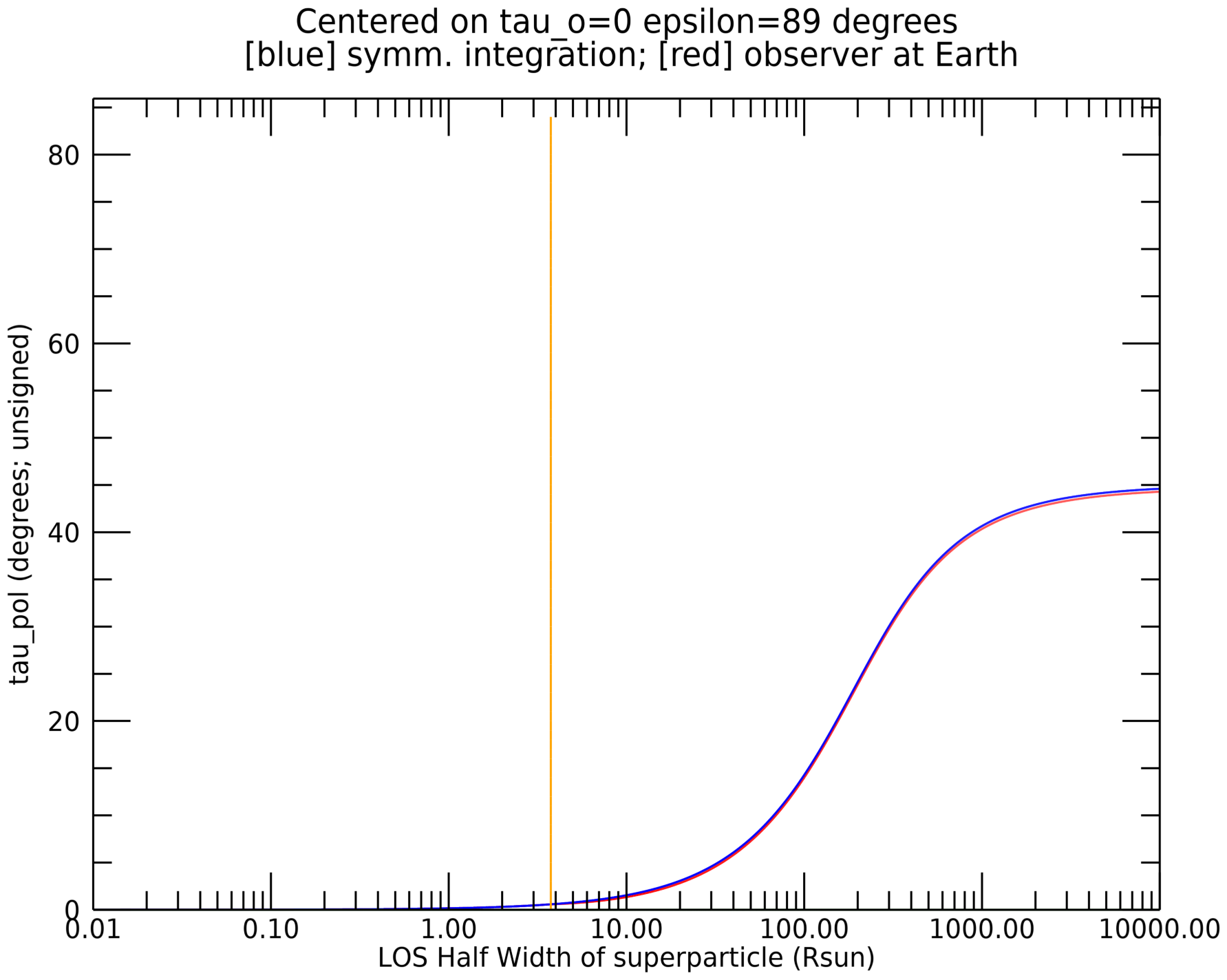}}
\caption{{Polarization ratio angular position $\tau_{pol}$  as in Figure~\ref{fig-varywidth} (right) for (left column) $\varepsilon = 1^\circ$ ($\tau_o=0,30,45,60^\circ$), (right column top three panels) $\varepsilon = 40^\circ$ ($\tau_o=0,30,45^\circ$),
and (bottom right) $\varepsilon= 89^\circ$ ($\tau_o=0^\circ$). Tote that we avoid combinations of $\tau_o$ and $\varepsilon$ that would center the wedge behind the observer due to $\tau_o + \varepsilon > 90^\circ$. Vertical lines are as described in Figure~\ref{fig-varywidth}, and horizontal line is placed at wedge center $\tau_o$.}}
\label{fig-tpolvary}
\end{figure}

The asymmetric case (red lines in Figures~\ref{fig-varywidth} and~\ref{fig-tpolvary}) truncates the integral at the observer, 
\begin{eqnarray}
\tau_{+_{limit}}&=&\operatorname{atan}(\frac{s_{+_{limit}}}{r_{obs}\sin(\varepsilon)}) \nonumber\\
    &=& \operatorname{atan}(\frac{r_{obs}[\cos(\varepsilon)-\sin(\varepsilon)\tan(\tau_o)] }{r_{obs}\sin(\varepsilon)} \nonumber\\
    &=& \operatorname{atan}(\frac{1}{\tan}\varepsilon - \tan(\tau_o)) \nonumber\\
    &=& \frac{\pi}{2} - \varepsilon 
  \label{Eq-taupluslimit}
\end{eqnarray}

which for the $\varepsilon = 25^\circ$ case shown in Figure~\ref{fig-varywidth} results in $\tau_{+_{limit}}=65^\circ$ and an asymptotic approach to
\begin{eqnarray}
    \tau_{pol} &=& \pm \operatorname{asin}[\sqrt(\frac{1}{2}\cdot(1 - \frac{(\sin(\frac{13\pi}{18})+\sin(\pi)}{(\frac{13\pi}{18} + {\pi})})] \nonumber \\
    &=& 0.714 = 40.93^\circ
\end{eqnarray}

{It is interesting to compare the top left ($\varepsilon=1^\circ, \tau_o=0^\circ$) and bottom right ($\varepsilon=89^\circ,\tau_o=0^\circ$) panels of Figure~\ref{fig-tpolvary}, which both show minimal difference between the symmetric (blue) and asymmetric (red) solution -- but for different reasons. In the case of $\varepsilon=1^\circ$, 
lines of sight are essentially parallel so that the solution is intrinsically symmetric, while for the case of $\varepsilon=89^\circ$, the observer is at $\tau_{+_{limit}}=1^\circ$ and the line of sight lies almost entirely outside the Thomson Sphere, $\tau > 0$, such that the solution is intrinsically asymmetric.} 

{To recover the superparticle approximation, we apply a first order Taylor expansion about $\tan(\tau_o)$ for small $\frac{w}{2\rca}$ to Equation~\ref{Eq-tauplusminus}.
\begin{eqnarray}
\tau_{\pm} &\rightarrow&\tau_o \pm \frac{w}{2\rca}\cos^2{\tau_o} \nonumber\\
\sin(2\tau_+) - \sin(2\tau_-) &\rightarrow& 2\cos(2\tau_o)\sin(\frac{w}{d}\cos^2\tau_o)\nonumber\\
2\tau_+ - 2\tau_- &\rightarrow& 2\frac{w}{d}\cos^2\tau_o\nonumber
\end{eqnarray}
Then Equation~\ref{Eq-tpolsoln} reduces to a small-width form,
\begin{eqnarray}
\tau_{pol} &\rightarrow&  \operatorname{asin}(\sqrt(\frac{1}{2} \cdot (1 - \cos(2\tau_o)\frac{sin(\frac{w}{d}\cos^2\tau_o)}{\frac{w}{d}\cos^2\tau_o} )))
\label{Eq-smallwidth}
\end{eqnarray}}

{In the limit $\frac{w}{d}\cos^2\tau_o \rightarrow 0$,
\begin{eqnarray}
\tau_{pol} &=& \operatorname{asin}(\sqrt(\frac{1}{2} \cdot (1 - \cos(2\tau_o))))\nonumber \\
&=&  \operatorname{asin}(\sqrt(\sin^2(\tau_o))) \nonumber \\
&=& |\tau_o|
\end{eqnarray}
This is a general result, and when achieved by LOS width $\rightarrow 0$, it is independent of $\tau_o$ and $\varepsilon$ (see Figure~\ref{fig-tpolvary}). We note in passing $\tau_{pol}= |\tau_o|$ is achieved exactly when $|\tau_o| = 90^\circ$, which is of academic interest only as it involves a wedge placed at infinity.  It is however significant that there is a dependency on $\tau_o$ -- and through $d$, on $\varepsilon$ -- that needs to be considered when determining how small the LOS linear widths must be to be considered ``localized'', as we will discuss in the next section.}

{{In summary: by considering varying LOS linear-width wedges of constant density centered at angular distance $\tau_o$ from the TS, we have demonstrated the transition from a superparticle solution to the spherically-symmetric density solution extended infinitely along the line of sight. In addition, we have quantified how the polarization ratio and $\tau_{pol}$ are modified by asymmetric lines of sight that truncate at the observer's position.}}

\subsection{Defining ``localization'': Quantifying errors and understanding limitations on superparticle analysis}\label{sec:define_local}

{We have demonstrated that the polarization analysis returns ``truth'' for the wedge center position, i.e., $\tau_{pol} = |\tau_o|$, for wedge LOS linear widths $\frac{w}{d} \cos^2(\tau_o) \ll 1$. In this small width limit, the angular half-width 
considered in the treatment of {\citet{dekoning_this_issue}} can be expressed as $\Delta \approx \frac{w}{2d}\cos^2\tau_o$ and will be symmetric about $\tau_o$ (not generally true, as we define $\tau_o$ as the center of the linear width $w$ as opposed to the center of angular width $2\Delta$). Indeed the small-width limit found in \citet{dekoning_this_issue} for a constant density model, $2\Delta \ll 1$, is equivalent to our requirement $\frac{w}{d}\cos^2(\tau_o) \ll 1$.
Structures meeting this condition can thus be said to be localized -- subject to an error that is maximum (in the small-width limit) when the wedge straddles the TS, which we now quantify.}

{For the case shown in Figure~\ref{fig-varywidth}, $\tau_o=0$ and
$\tau_{\pm} = \pm \operatorname{atan}(\frac{w}{2d}).$
Since
\begin{eqnarray}
\sin(2\operatorname{atan}(\frac{w}{2d})) &=&
2\sin(\operatorname{atan}(\frac{w}{2d}))\cos(\operatorname{atan}(\frac{w}{2d}))
\nonumber\\
&=&\frac{\frac{w}{d}}{{1 + (\frac{w}{2d}^2})}\nonumber
\end{eqnarray}
Equation~\ref{Eq-tpolsoln} at the TS then is,
\begin{eqnarray}
\tau_{pol_{TS}}
&=&
\operatorname{asin}(\sqrt(\frac{1}{2} \cdot [1 -\frac{\frac{w}{2d}}{{\operatorname{atan}(\frac{w}{2d})[1+(\frac{w}{2d})^2]}}]))
\label{Eq-tpol_error_TS}
\end{eqnarray}}

\begin{figure}[ht!]
\centering{\includegraphics[width=.95\textwidth]{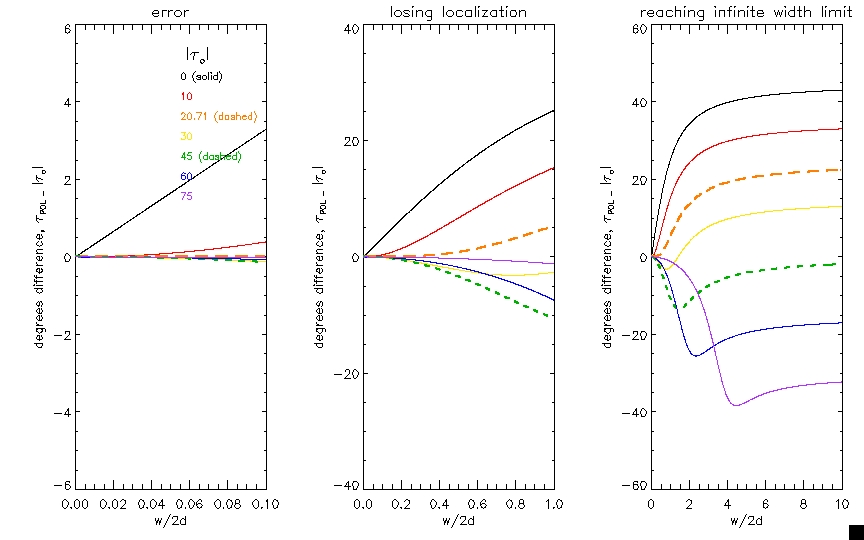}}
\centering{\includegraphics[width=.95\textwidth]{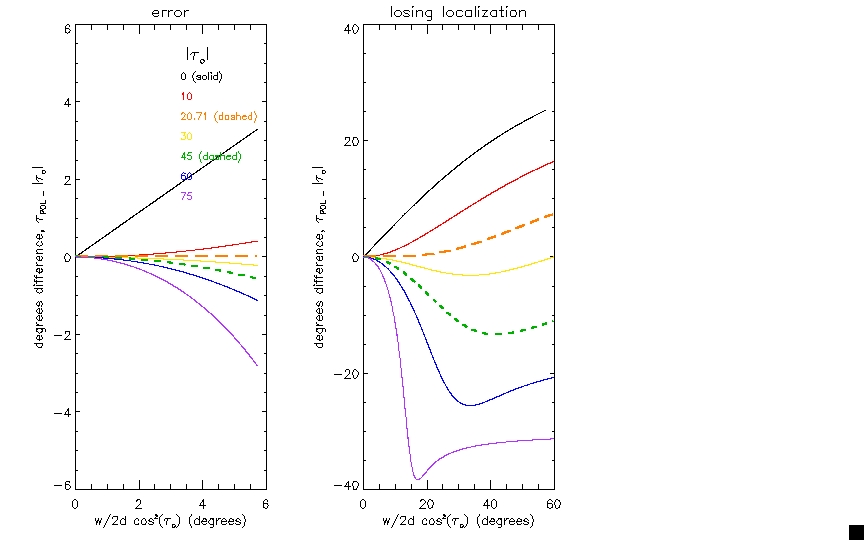}}
\caption{{$\tau_{diff}=\tau_{pol}-|\tau_o|$ plotted vs. (top) linear half-width $\frac{w}{2d}$ and (bottom) $\frac{w}{2d} cos^2(\tau_o)$ (which is the angular half-width $\Delta$ in the small-width limit). The bottom panels have the same horizontal axis range as the top panels, expressed in degrees rather than radians. The solid black line is for a wedge centered on the TS, $\tau_o=0^\circ$, and the colored lines represent different choices of $|\tau_o|$ (shown in degrees in the left panel text inset). 
Panels show results for the small-width regime (left), the middle-width regime (middle), and the large-width (wedge size $\rightarrow \infty$) regime. All of the initial choices of wedge center $\tau_o$ result in $\tau_{diff}$ (the value being plotted) asymptoting to $45^\circ - \tau_o$ in this large-width limit.}}
\label{fig-error}
\end{figure}

\begin{figure}[hb!]
\centering{
\includegraphics[width=0.47\textwidth, trim={10 14.6 10 0},clip]
{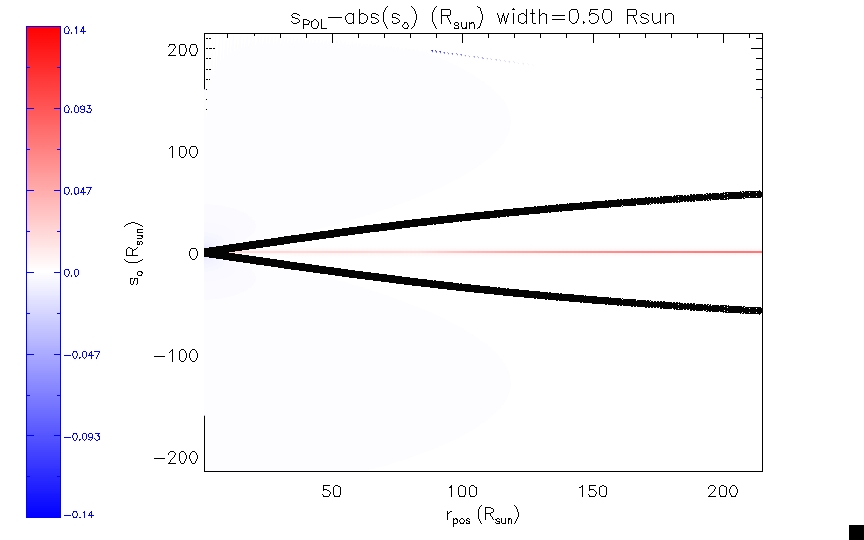}
\includegraphics[width=0.47\textwidth, trim={10 14.6 10 0},clip]
{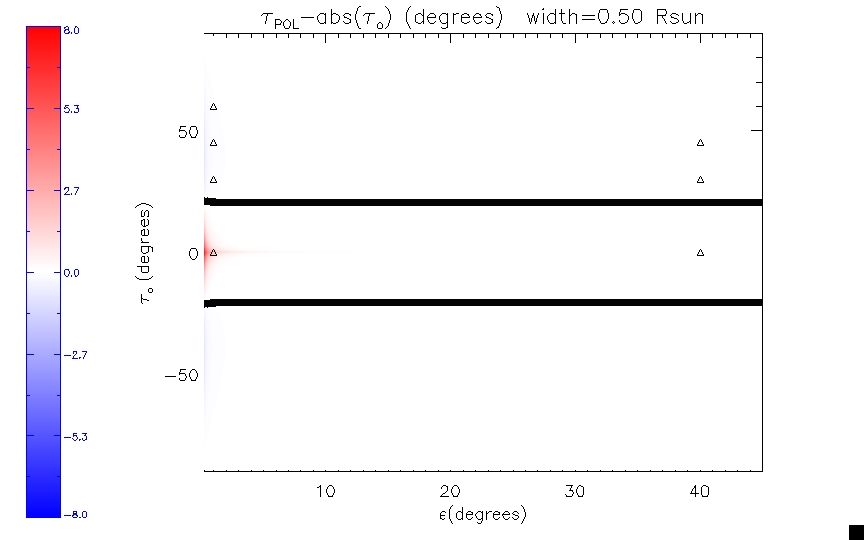}
}
\centering{
\includegraphics[width=0.47\textwidth, trim={10 14.6 10 0},clip]
{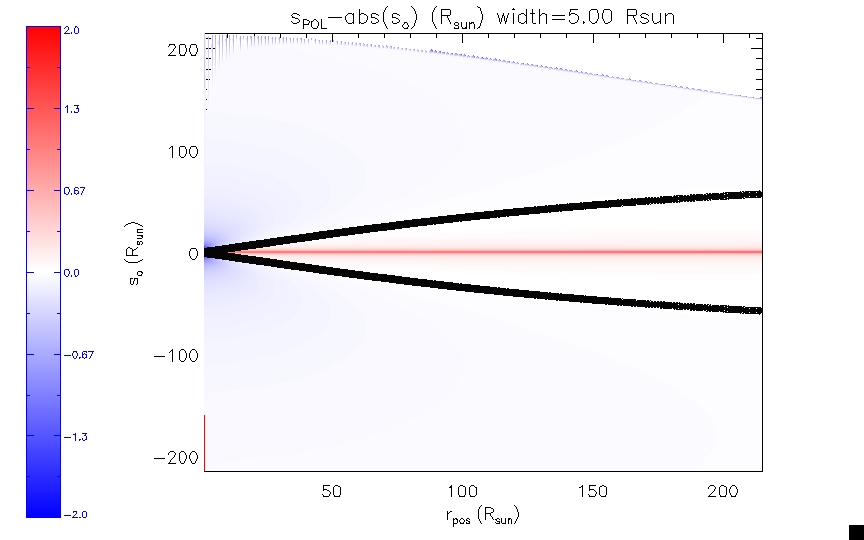}
\includegraphics[width=0.47\textwidth, trim={10 14.6 10 0},clip]
{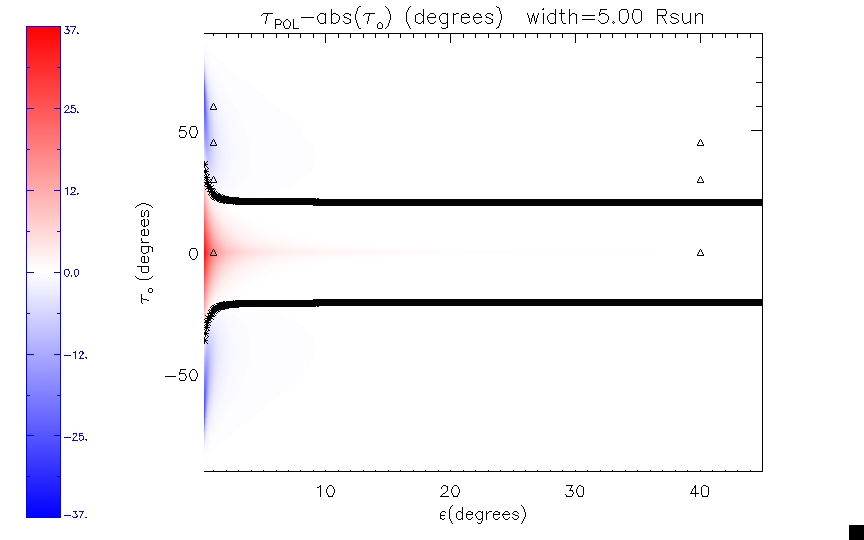}
}
\centering{
\includegraphics[width=0.47\textwidth, trim={10 14.6 10 0},clip]
{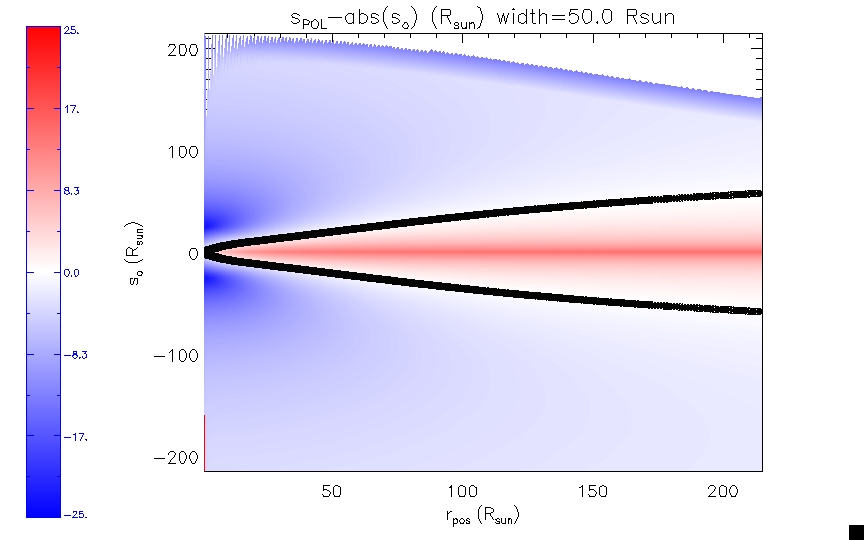}
\includegraphics[width=0.47\textwidth, trim={10 14.6 10 0},clip]
{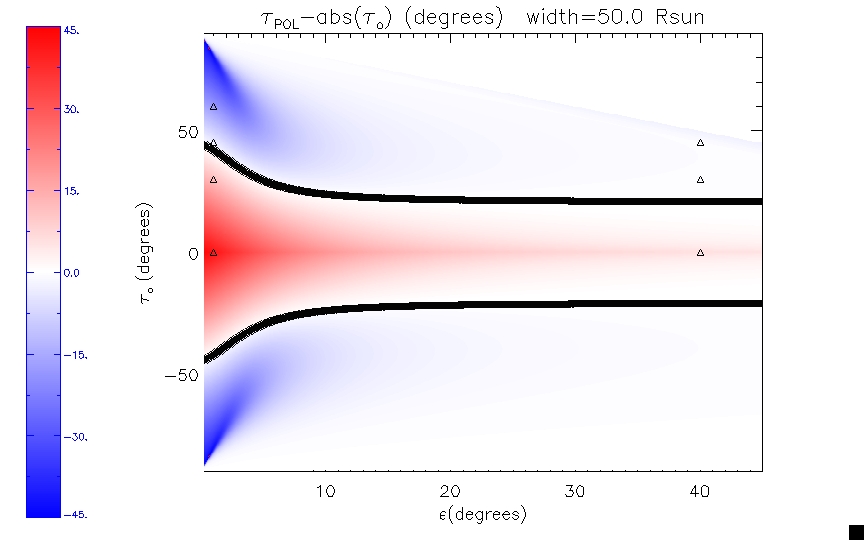}
}
\caption{{Color contours in the left column show
the difference between 
$s_{pol}$ and wedge center linear position $s_o$ in $r_\odot$ as a function of $s_o$ (vertical) and $r_{pos}$ (horizontal)
for LOS linear widths of $w=0.05,0.5, 5.0 r_\odot$ (top to bottom). 
Right column shows the same, but with color contours showing $\tau_{diff}$ in degrees as a function of $\tau_o$ (vertical) and $\varepsilon$ (horizontal). The asymmetry for positive vs. negative $s_o, \tau_o$ (seen in the bottom panels) is due to the requirement that the wedge lie between the TS and the Observer, $\tau_o \le 90^\circ-\varepsilon$. Triangles on the right-hand plots correspond to the choices of $\tau_{o},\varepsilon$ shown in Figure~\ref{fig-tpolvary}. The solid black lines are  solutions to $\tau_{diff}=\tau_{pol}-|\tau_{o}|=0$ as discussed in the text.}}
\label{fig-taudiffvarywidth}
\end{figure}

Equation~\ref{Eq-tpol_error_TS} quantifies the error on the solution of $\tau_{pol}=\tau_o=0$ at the TS, an error that approaches zero as $\frac{w}{2d} \rightarrow 0$ and $45^\circ$ as $\frac{w}{2d} \rightarrow \infty$. 
The general form of this error, $\tau_{diff} = \tau_{pol} - |\tau_o|$, is calculated numerically from Equations~\ref{Eq-tpolsoln} and \ref{Eq-tauplusminus}
{and plotted for different choices of $\tau_o$ in Figure~\ref{fig-error}  
as a function of linear half-width $\frac{w}{2d}$ (top panels) and small-width-limit angular half-width $\frac{w}{2d}\cos^2\tau_o$ (left and middle bottom panels only where $\frac{w}{2d}\cos^2\tau_o \le 1 (radians)$). 
By utilizing the dimensionless parameters $\frac{w}{2d}$ and $\frac{w}{2d}\cos^2\tau_o$, we remove dependence on $\varepsilon$.}

{The top left panel shows the small-width regime, and we see that the $\tau_o=0^\circ$ case (black line) has a significantly larger error than wedges of the same linear width placed at different $\tau_o$ (colored lines),
reaching $>3^\circ$ at $\frac{w}{2d}=0.1$. The bottom left panel makes the comparison for wedges with the same angular width as the TS-centered wedge, still in the small-width regime.
The $|\tau_o|=20.71^\circ$ line (dashed orange) represents a small-width minimum solution of $\tau_{diff}$ and lies tangent to the zero error line until $\frac{w}{2d} \approx 0.4$ or $\frac{w}{2d}\cos^2\tau_o \approx 23^\circ$ (as seen in middle panels). For $|\tau_o| > 20.71^\circ$, $|\tau_{diff}|$ is negative.
The middle panels show the small-width localization being lost and the top right panel shows the asymptotic infinite-width limit being reached for all choices of $|\tau_o|$.
}

{Although there are zero crossings for~$20.71^\circ<|\tau_o|<45^\circ$ (e.g., the $|\tau_o|=30^\circ$ yellow line), they {\it should not be taken as localization}. Rather, they are showing the transition toward the infinite-width solution $\tau_{pol}=45^\circ$, which is the asymptotic behavior for all choices of $\tau_o$ 
as seen in the top right-hand panel of Figure~\ref{fig-error} ($45^\circ$ needs to be subtracted from the absolute value of $\tau_o$ of each colored line to see this). The green dashed line is the case $|\tau_o|=45^\circ$, which in the small-width limit is localized, but in the large-width limit achieves $\tau_{diff}=0$ essentially by the coincidence of being centered on the value associated with an infinite-width wedge. This nonlinearity means that the $\tau_{diff}$ ``error'' on $\tau_{POL}$ is only generally meaningful in the small-width regime.}

{Figure~\ref{fig-taudiffvarywidth} shows how $\tau_{diff}$ and the related linear position difference $s_{diff}=s_{pol}-|s_o|$ vary across the two dimensional ($\tau_o,\epsilon$) space, or equivalently ($s_o,\rpos$) space, 
for a choice of $w=0.5, 5, 50 r_\odot$ (top to bottom). For this we have drawn inspiration from \citet{bemporad_2015}, who used similar plots to discuss polarization uncertainties close to the Sun. }

{For the smallest LOS linear width of $w=0.5 r_\odot$ (top panels), the difference is generally very small (appearing white).
An exception is near $s_o=0$ (left panel) or equivalently $\tau_o=0$ and small $\varepsilon$ (right panel), where the $\tau_{{pol}_{TS}}$ error of Equation~\ref{Eq-tpol_error_TS} manifests. (We note that \citet{bemporad_2015} referred to this error as a ``folding'' effect of the unsigned contributions to the polarization ratio across the TS -- mathematically contained in the $\cos^2\tau_o$ term of Equation~\ref{Eq-smallwidth}.)
For example, for the minimum $\varepsilon=0.2665^\circ $ shown in the plot $(r_{pos} = d = 1 R\odot)$, $\frac{w}{2\rca}=0.25$ and 
$\tau_{diff} = \tau_{{pol}_{TS}}=8.1^\circ$ at the TS. This corresponds to $s_{diff}=d \tan(\tau_{diff})=0.14 R_{\odot}$, a little more than a quarter of the assumed wedge LOS linear width of $0.5 r_\odot$. For larger values of $\varepsilon (r_{pos})$, $\frac{w}{2d}$ and so the angular-width error $\tau_{diff}$ decreases, although the linear-width error $s_{diff}$ at the TS is maintained farther due to its dependency on $d$ and so $\sin(\varepsilon)$. $\tau_{diff}=0$ is indicated by the thick black lines, which intersect the left-hand vertical axis in the top right-hand panel at $21.22^\circ$ (just above the small $\frac{w}{2d}$ limit of $\tau_o=\pm 20.705^\circ$, see Figure~\ref{fig-error}), corresponding to $s_o=\pm0.39 r_\odot$ in the top left-hand panel.}

{The middle panel value of $w = 5 r_\odot$ is similar to that assumed by \citet{bemporad_2015}, and indeed we get essentially identical results to them in the near-Sun, small-$\varepsilon$ regime. At $\tau_o=0$ and the minimum possible $\varepsilon=0.2665^\circ (\rpos = d=1 r\odot)$, $\frac{w}{2d}=2.5$, $\tau_{diff} = \tau_{{pol}_{TS}}=36.6^\circ$, and $s_{diff}=d\cdot\tan(\tau_{diff})=0.74 R_{\odot}$. The thick black line indicating $\tau_{diff}=0$ reaches a value of $\tau_o=\pm 36.1^\circ$ or $s_o=\pm 0.73 r_\odot$ at the (left-hand) vertical axis.}

{As the wedge LOS linear width gets larger still (bottom panels), negative and positive discrepancies grow, consistent with the transition to the infinite-wedge solution. In their near-Sun, small-$\varepsilon$ regime (left-hand sides), the bottom plots have reached this limit, with maximum/minimum values $\tau_{diff}=\pm45^\circ$ and the black $\tau_{diff}=0$ line occurring as expected at $\tau_o=45^\circ$ and $s_o=1 r_\odot$ as $\varepsilon \rightarrow 0$. As discussed above, the black lines $\tau_{diff}=0$ should not be taken as indicators of localization, i.e., ``success'' of the polarization diagnostic in reproducing the location of the wedge center. They occur where $\tau_{o}$ happens to coincide with $\tau_{pol}$  for a given $\frac{w}{2d}$ (see Figure~\ref{fig-error}). True success occurs in the parts of the plots that fade to white for all values of $\tau_o, s_o$ (other than the TS), i.e., where large $\varepsilon$ drives $\frac{w}{2d} << 1$ as on the right-hand sides of Figure~\ref{fig-taudiffvarywidth} panels.}

\begin{figure}[ht!]
\centering{
\includegraphics[width=0.55\textwidth, trim={10 14 10 0},clip]
{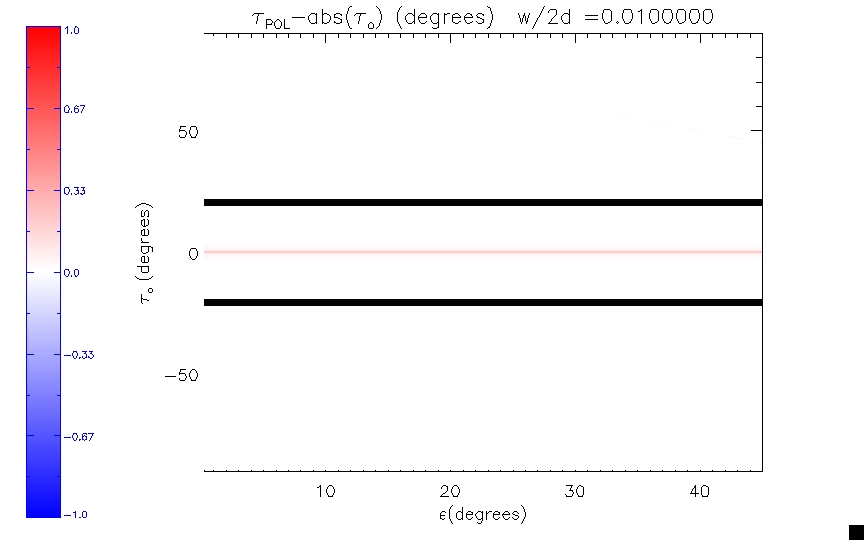}
\includegraphics[width=0.55\textwidth, trim={10 14 10 0},clip]
{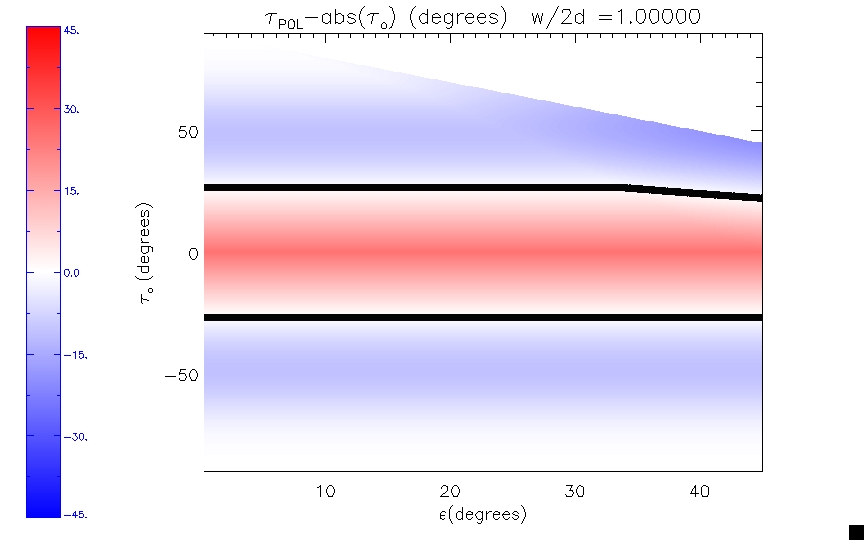}
\includegraphics[width=0.55\textwidth, trim={10 14 10 0},clip]
{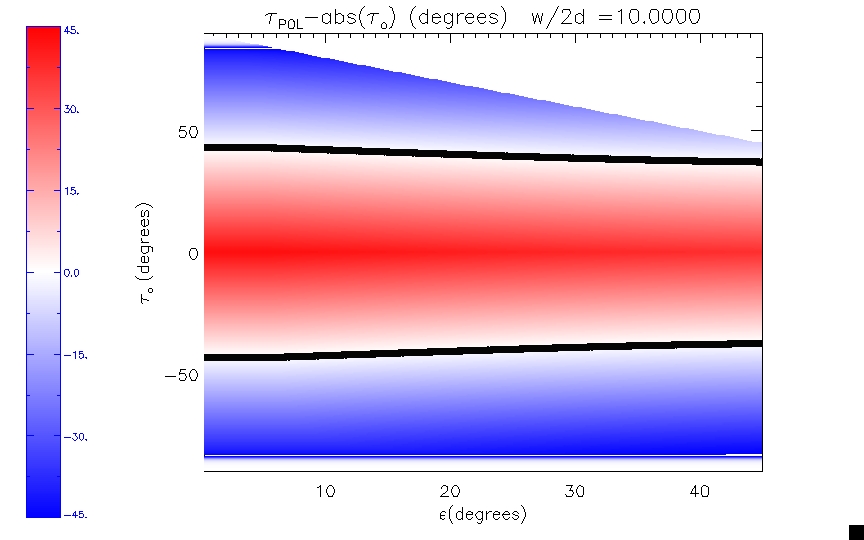}
}
\caption{Color contours show $\tau_{diff}=\tau_{pol}-|\tau_o|$ as in Figure~\ref{fig-taudiffvarywidth} (right column) for choices of $\frac{w}{2d}=0.01$ (top), $\frac{w}{2d}=1.0$ (middle), and $\frac{w}{2d}=10.0$ (bottom). Black lines are solutions to $\tau_{diff}=0$, appearing at $|\tau_o|=20.705^\circ$ in the top panel, and approaching maximum $|\tau_o|= 45^\circ$ by the bottom panels. As in Figure~\ref{fig-taudiffvarywidth}, asymmetries between positive and negative $\tau_o$ occur because the wedge must be centered in front of the observer. In addition, wedge LOS linear widths considered in the bottom panel are enough to shift the black line at large $\varepsilon$ from its symmetric asymptotic value of $45^\circ$ even for negative $\tau_o$ solutions, since the wedges are ``cropped'' as they extend past the observer at $+215 r_\odot$.}
\label{fig-taudiffcritwidth}
\end{figure}

{Figure~\ref{fig-taudiffcritwidth} removes the dependence on $\varepsilon$, as was done in Figure~\ref{fig-error}, by allowing LOS linear widths to vary as a fixed multiple of the LOS linear half-width parameter 
$\frac{w}{2d}$. 
This simplifies the problem greatly, and shows (top panel) how wedges with 
$\frac{w}{2d} << 1$ 
will have essentially zero error $\tau_{diff}$ (not considering other sources of error, e.g. measurement error \citep{deforest_13b}), except at the TS and then to a degree quantifiable by Equation~\ref{Eq-tpol_error_TS}.  Larger values of $\frac{w}{2d}$ ultimately result in the asymptotic solution $\tau_{pol}=45^\circ$, as seen in the position of the black lines on the left hand side of the bottom panel of Figure~\ref{fig-taudiffcritwidth}. 
}

\begin{figure}[ht!]
\centering{
\includegraphics[width=0.55\textwidth, trim={10 14 10 0},clip]
{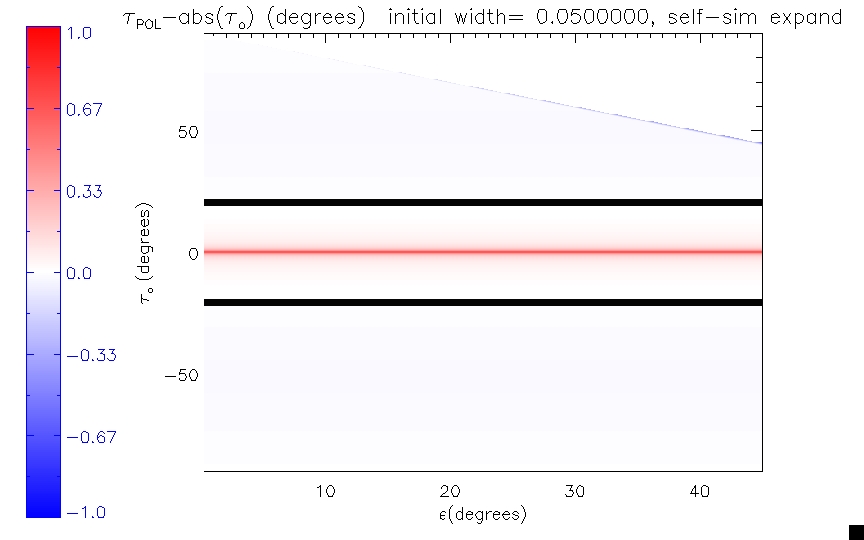}
\includegraphics[width=0.55\textwidth, trim={10 14 10 0},clip]
{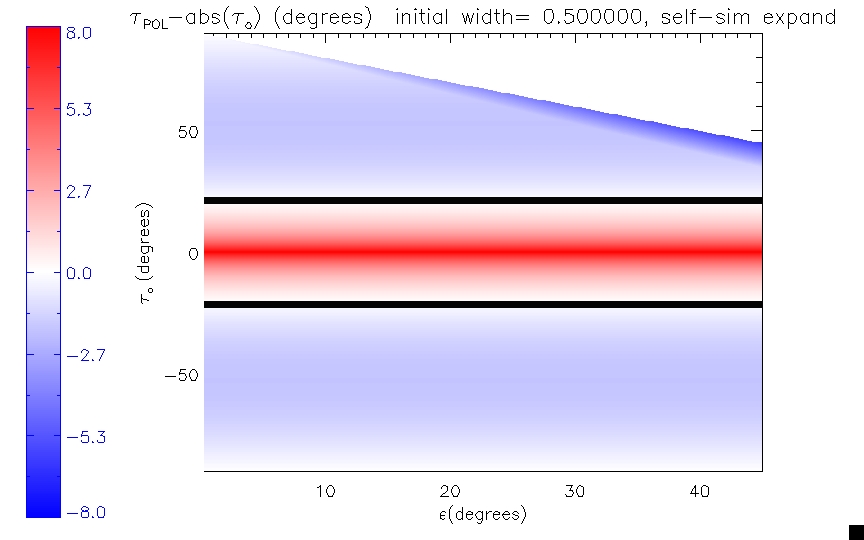}
\includegraphics[width=0.55\textwidth, trim={10 14 10 0},clip]
{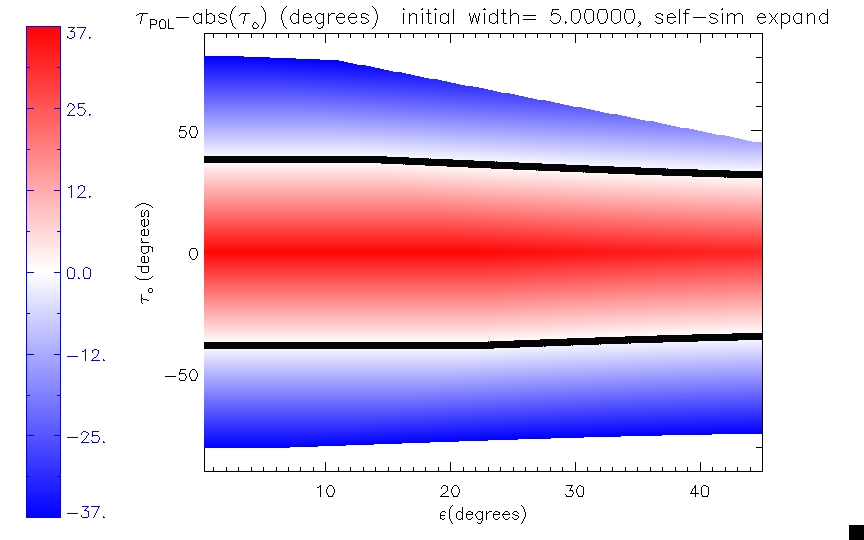}
}
\caption{{As in Figure~\ref{fig-taudiffvarywidth} but with the LOS linear width changing as expected for a self-similarly expanding wedge with initial widths $w_{ss}=0.05, 0.5, 5.0 r_\odot$ at initial height $r_{ss}=1 r_\odot$. }}
\label{fig-taudiffvarywidth-selfsim}
\end{figure}

Observations indicate that structures expand as they move outward, leading to increasing LOS linear width with increasing $\varepsilon$. In Figure~\ref{fig-taudiffvarywidth-selfsim}, we assume that our wedge is an intersection with a sphere of diameter w, which expands self-similarly\footnote{{We recognize that a self-similar expansion is an oversimplification for CMEs, with in-situ statistical evidence for overall expansion at a rate less than self-similar \citep{BothmerSchwenn_1998,Gulisano_2010,zhuang_2023} and variation observed with heliospheric distance \citep{lugaz_2020,zhuang_2023}, which would result in increased localization with heliocentric distance. This could {be} particularly significant for embedded structures such as a prominence, which has been observed to expand much more slowly than the rest of the CME \citep{zhuang_2025}.}} as it moves outward, i.e.,
\begin{equation}
    w(r)  = \frac{w_{ss}}{r_{ss}} \cdot r
\end{equation}
where $w_{ss}$ and $r_{ss}$ are constants of the self-similar expansion. 
Then, since
\begin{equation}
r =\frac{d}{cos(\tau_o)}
\end{equation}
we see that
\begin{equation}
   \frac{w}{2d} = \frac{w_{ss} \cdot cos(\tau_o)}{2r_{ss}}
\end{equation}
Thus the self-similar expanding solution behaves almost like the $\frac{w}{2d}$ solutions of Figure~\ref{fig-taudiffcritwidth}, subject to a $\frac{\cos(\tau_o)}{2}$ scaling, and
a self-similarly expanding structure of initial linear width $w=0.05 r_\odot$, $d=1.0r_\odot$ will remain localized throughout its self-similar expansion. {As in Figure~\ref{fig-taudiffcritwidth}, the middle panel represents a case where the wedge has become too large for localization, and the bottom panel a case that is nearing the infinite LOS width limit where $\tau_{pol} \rightarrow 45^\circ$, regardless of wedge center position $\tau_o$. In this case, the wedge has initial LOS linear width $w=5 r_\odot$ at $d=1.d0r\odot$ (such as that studied in \cite{bemporad_2015}) and so white light polarization data is not a good diagnostic of center position along lines of sight that intersect it over such an extended distance.} 

{In summary: 
the degree of localization of a structure depends not just on the LOS linear width $w$ of the structure being observed but also on the elongation angle $\varepsilon$, as represented by the dimensionless parameter $\frac{w}{2d}$. A self-similarly expanding structure will essentially maintain this critical parameter along its radial trajectory, so in such cases, knowledge of the structure's width at its origin can establish the degree to which white-light polarization data can diagnose the structure's 3D location throughout its trajectory. If it is in the localized, e.g, $\frac{w}{d}\cos^2(\tau_o) << 1$ regime, the error (which is maximum at the TS) can be quantified by Equation~\ref{Eq-smallwidth}.}

\section{Polarization Diagnostics of CMEs}\label{sec:3dpolmodels}

We now use a 3D model of a CME (Section~\ref{sec:forward}) to show how polarization images and movies can remove the front/back ambiguity of CMEs and address ambiguities in trajectory/size (Section~\ref{sec:front-back}). We also use the model to demonstrate how locating the 3D position of CME subtructures might be used to establish chirality, or direction of twist, within a CME (Section~\ref{sec:forward_chi}).

\begin{figure}[ht!]%
\centering
\includegraphics[width=0.9\textwidth, trim={3 3 0 3},clip]
{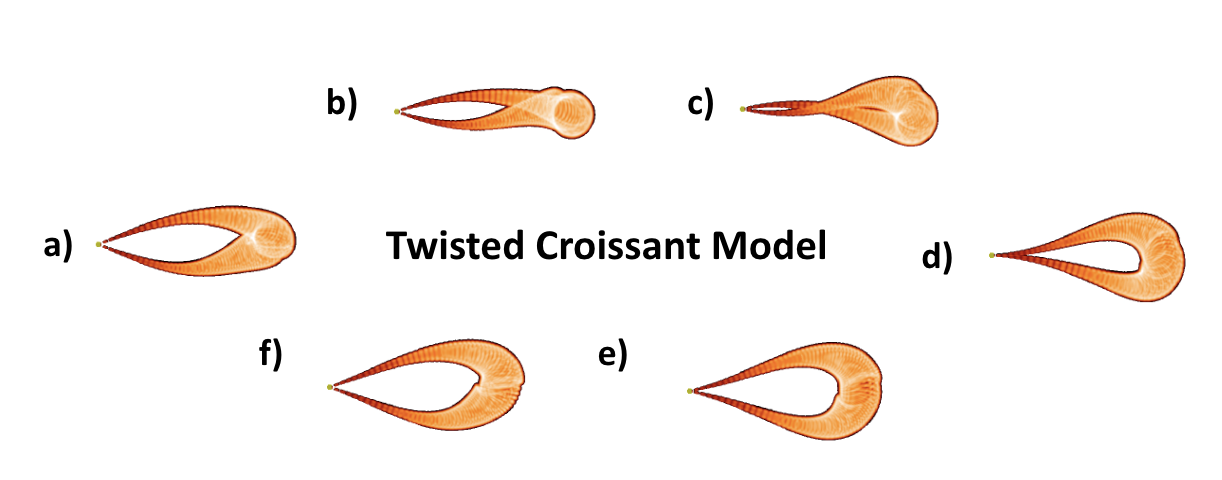}
\caption{pB synthesized twisted croissant model, {shown here from an observing viewpoint that rotates about the direction of CME trajectory starting from (a) an observer at the heliographic equator with CME on the West limb to (b) $30^\circ$, (c) $60^\circ$, (d) $90^\circ$ (North pole), (e) $120^\circ$ (viewing the CME from the solar farside), to (f) $150^\circ$.} \textcolor{red}{MOVIE}}.
\label{fig_croissant_rotate}
\end{figure}

\subsection{Twisted Croissant Model}\label{sec:forward}

Analytic models are useful for building insight and testing our diagnostics. We employ a croissant type CME model — essentially a hollow shell of density with a twist (Figure~\ref{fig_croissant_rotate}). 
The geometry of our model is  similar to the graduated cylindrical shell (GCS) model \citep{thernisien_06,thernisien_09}, but is created with a different approach which offers the potential to easily create more complex shapes. For example, an overall twist can be applied to the croissant model. The model is initialized by defining the location of the inner tube axis, and parameters such as the twist, orientation, or leg separation are applied to this inner axis. The location of the tube sheath is then calculated in relation to the inner axis by defining the radial distance (or flux tube width), which expands from the CME footpoints to the apex. {See Appendix~\ref{secCCroissant} for further details.}

This model has been incorporated into the FORWARD SolarSoft IDL modeling framework \citep{gibson_16}, allowing forward modeling along lines of sight to form 2D projected images of, e.g., $pB$, $tB$, $PR$, $\tau_{pol}$, $x_{pol}$, and the ground truth $\tau_{COM}$ and $x_{COM}$.

\begin{figure}[ht!]%
\centering
\includegraphics[width=0.9\textwidth, trim={3 3 0 3},clip]
{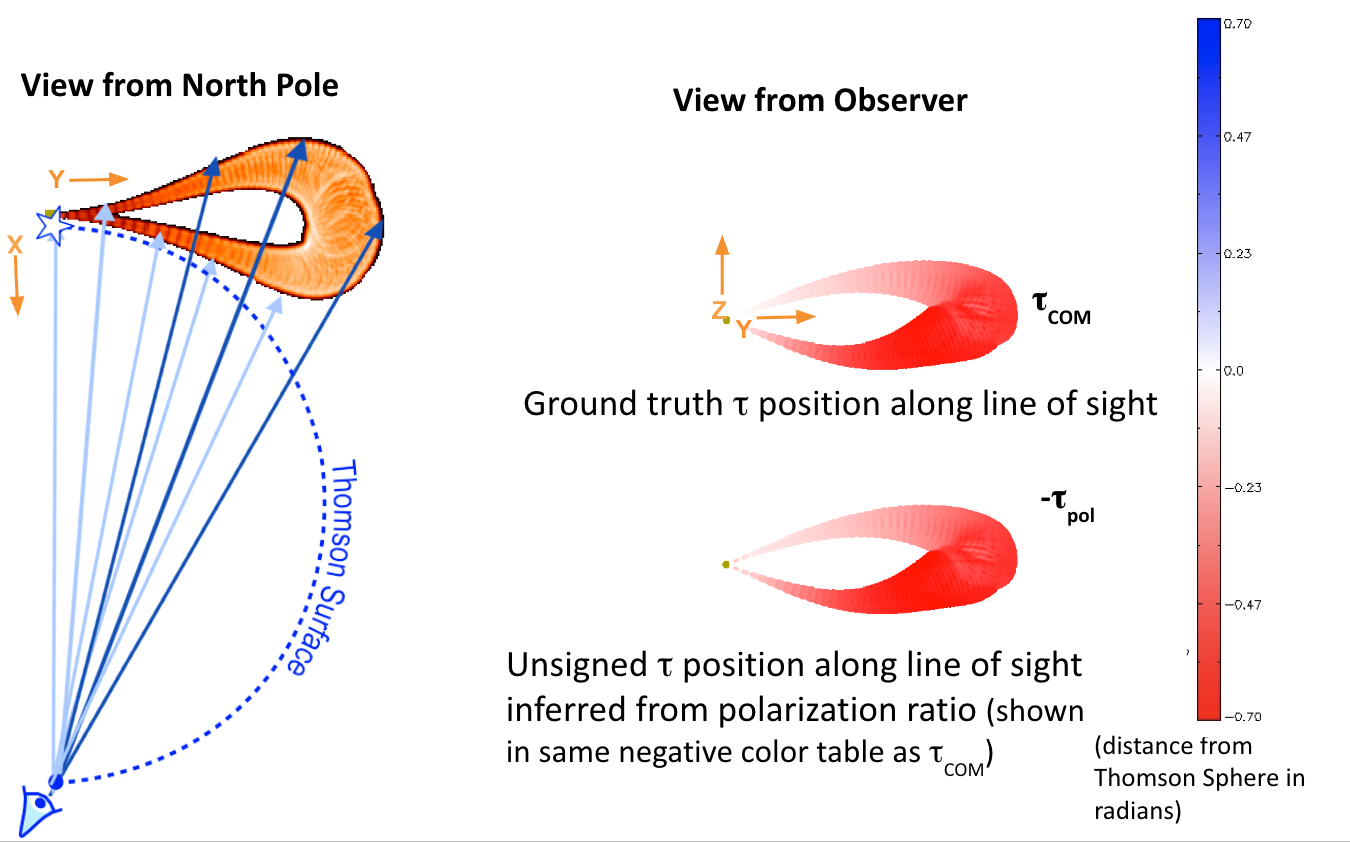}
\caption{Polarization reveals distance from Thomson Sphere (TS). (left) The CME (twisted croissant model) erupting from the West limb, as viewed from the North pole {(the same view shown in Figure~\ref{fig_croissant_rotate} d).} The lines of sight are shaded in progressively darker shades of blue to illustrate distance from the TS of the points of intersection with the croissant. (right top) $\tau_{COM}$ ground-truth calculated center of mass  along the line of sight {as seen by the observer at the equator (the same view shown in Figure~\ref{fig_croissant_rotate} a)}, and (right bottom) $\tau_{pol}$ calculated from forward-modeled synthetic $pB$ and $tB$. The latter is unsigned, and the negative choice here is appropriate for all points along the edge of the CME as it lies completely behind the TS.}.
\label{fig_croissant_tau}
\end{figure}

Figure~\ref{fig_croissant_rotate} shows synthesized $pB$ for a model CME erupting at the solar West limb {equator}. The images $a - f$ show how the LOS-integrated $pB$ changes if the {observer's viewing angle rotates from equator to pole to equator, or equivalently, if the model twisted croissant rotates about the $Y-$axis.} Figure~\ref{fig_croissant_tau} shows that the $\tau_{pol}$ calculated from forward-modeled $pB$ and $tB$ reproduces the (unsigned) distance from the TS, $\tau_{COM}$.  Note that the lines of sight shown in the left side of the figure each only intersect one point at the edge of the twisted croissant, because the edge that is positive $x$ is also positive $z$ (i.e., above the heliographic equator), while the edge that appears behind the $X-$axis from the North pole view of the left-hand side is below the heliographic equator (negative $z$) as viewed from the Earth and as shown in the images on the right-hand side. {Such tangent lines of sight are intrinsically localized as discussed in Section~\ref{sec:define_local}. {Other lines of sight through the middle of the structure do intersect more than one point of the croissant, and this is explicitly accounted for in the LOS integrals of $pB$ and $tB$. The fact that forward-synthesized $\tau_{pol}$ still reproduces the ground truth $\tau_{com}$, as shown in the right-hand-side of Figure~\ref{fig_croissant_tau}, indicates that the polarization ratio can be used in a meaningful way with such superposed structures. We return to this point in Section~\ref{sec-halo}.}}

\begin{figure}[ht!]%
\centering
\includegraphics[width=0.9\textwidth, trim={3 3 0 3},clip]
{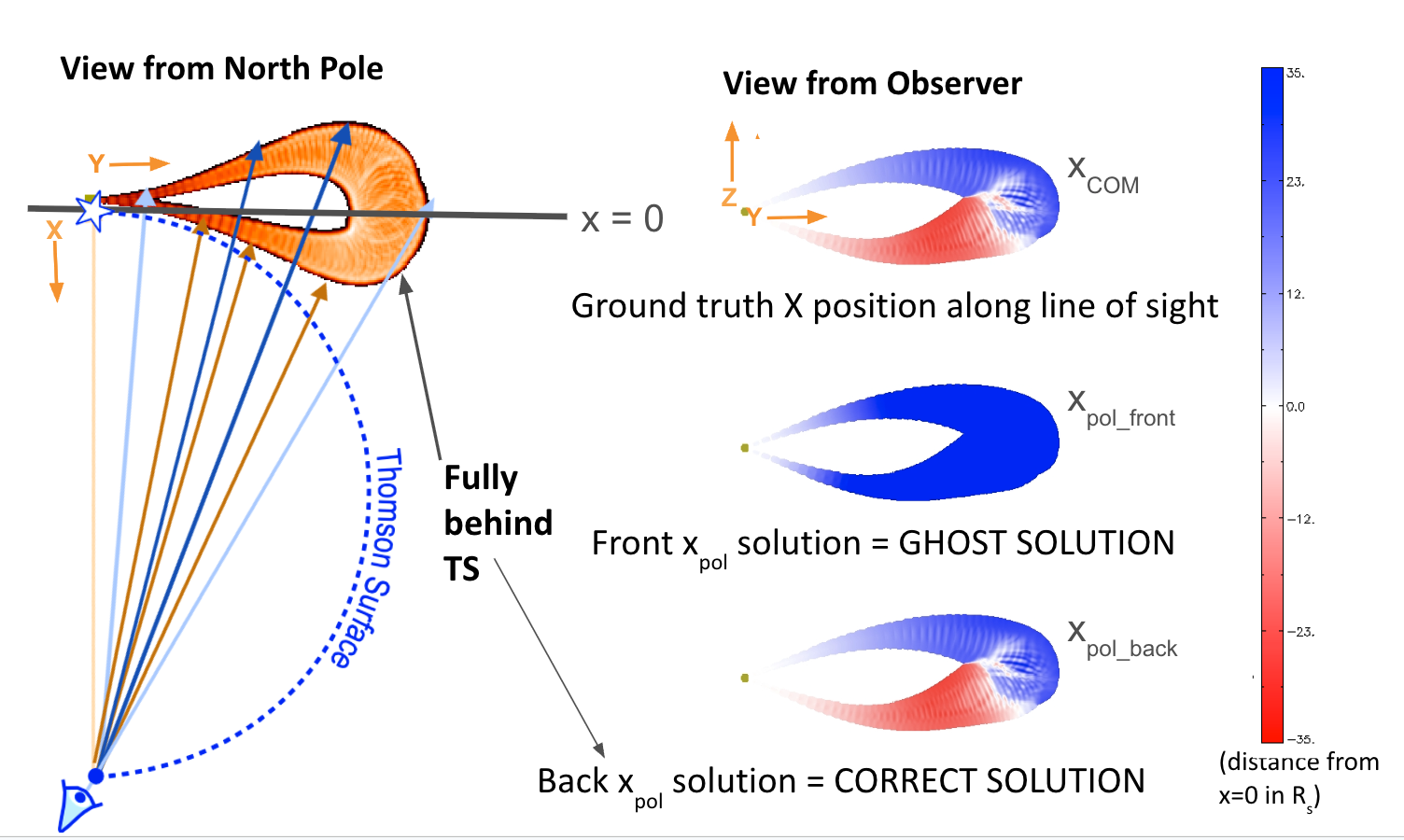}
\caption{Polarization reveals distance from $X=0$ plane ($x_{pol}$). (Left) The twisted croissant CME model as in Figure~\ref{fig_croissant_tau}. The lines of sight here are shaded blue/red to indicate the croissant's edge distance (negative/positive) from $x=0$. (right) $x_{COM}$ from the observer (eye shown in left image) and $x_{pol}$ calculated from forward-modeled synthetic $pB$ and $tB$. The latter has two solutions as discussed in the text (Equation~\ref{Eq-Xpol}) and shown in the middle and bottom images. The choice corresponding to negative $\tau_{pol}$ (bottom) is appropriate for this CME as it lies completely behind the TS.}.
\label{fig_croissant_xpol}
\end{figure}

In order to specify the position of the croissant edge in three dimensions, it is useful to calculate it in terms of distance from the $x=0$ plane, i.e., $x_{pol}$. This then can be combined with a projection of the image onto the Observing Plane of Sky ($YZ$; see Figure~\ref{fig-geom}) along with knowledge of how the observers lines of sight intersect said plane, in order to yield $X-Y-Z$ coordinates of the edge (Equations~\ref{Eq-Xpol}-\ref{Eq-Zpol}).
Equation~\ref{Eq-Xpol} has two solutions for $x_{pol}$ corresponding to the $\tau_{pol}$ in front of/behind the TS — we will refer to these as the ``front" and ``back" solutions. In the case shown in Figure~\ref{fig_croissant_xpol}, the CME lies entirely behind the TS and so the $x_{pol}$ corresponding to the back solution is the correct one, while the $x_{pol}$ corresponding to the front solution is the so-called ``ghost" solution \citep{deforest_13b}, which we now discuss.



\begin{figure}[t!]%
\centering
\includegraphics[width=0.8\textwidth, trim={3 3 0 3},clip]
{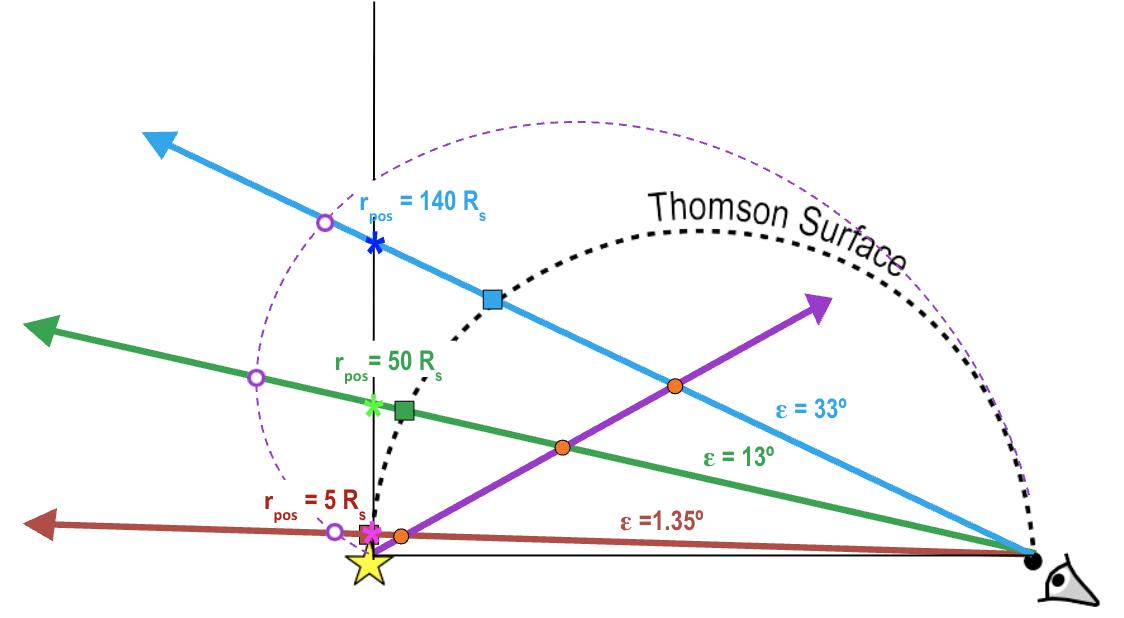}
\caption{Two distinct trajectories arise from time series of polarization data. The solid purple line shows the true trajectory of a CME leading edge as it moves toward the observer, while the dashed purple curve shows a ``ghost'' trajectory as described in the text.  The red, green and blue lines of sight intersect the TS at the positions indicated by colored squares, and the $x=0$ plane as indicated by the colored $*$. The distance from the TS determined from $PR$, i.e., $\tau_{pol}$, is indicated by orange dots in front of the TS (front solution), and purple circles behind the TS (back solution). Figure adapted from \citet{deforest_13b}.}
\label{fig_approach_ghost}
\end{figure}

\subsection{Tracking CME Trajectories}\label{sec:front-back}

\citet{deforest_13b} explored the front/back of TS ambiguity implicit to analyzing white-light polarization data and demonstrated that the ambiguity can be removed for moving structures in wide-field heliographic images. We briefly review this now, and show how both movies and single images of CME $x_{pol}$ might be used to establish whether a CME is approaching or receding from the observer at Earth.

\begin{figure}[b!]%
\centering
\includegraphics[width=0.8\textwidth, trim={3 3 0 3},clip]
{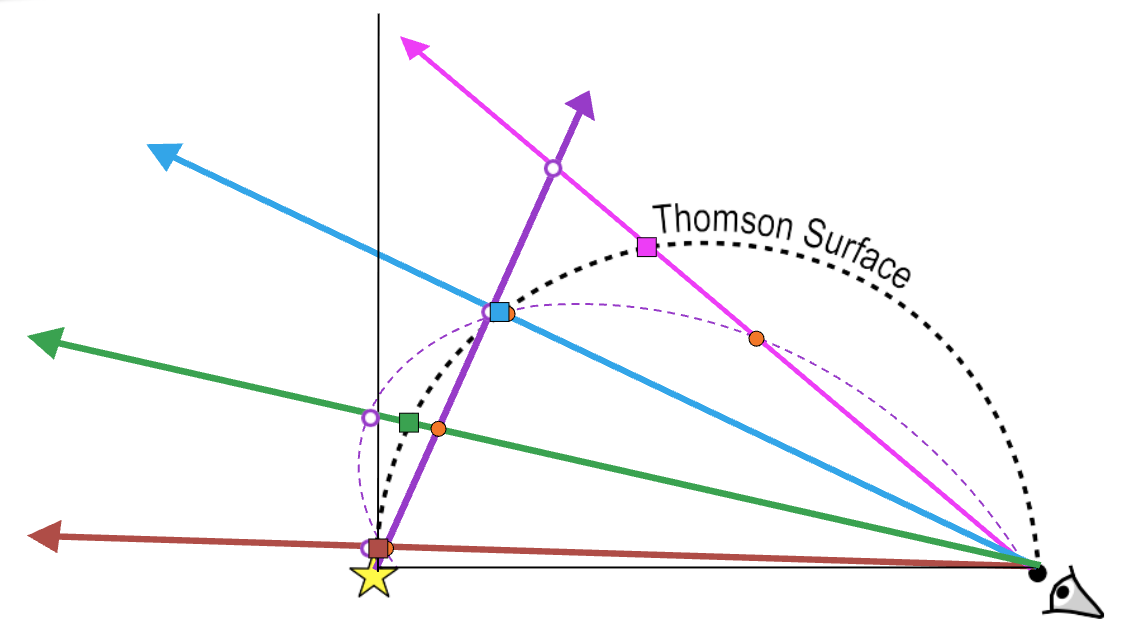}
\caption{Approaching CME front as in Figure~\ref{fig_approach_ghost} but with a higher inclination trajectory that crosses the TS closer to the Sun. Note that the first two points along the true trajectory (solid purple line) are solid orange dots (positive $\tau_{pol}$ or front solutions), while the next two points are purple circles (negative $\tau_{pol}$ or back solutions), and the converse is true for the ghost solution (dashed purple line). Figure adapted from \citet{deforest_13b}.}
\label{fig_approach_ghost_high}
\end{figure}

\subsubsection{Revealing the ghost trajectory}

As we have discussed above, the more compact a structure is, the more straightforward it is to locate its position in 3D.  
An important case is that of the CME leading edge, which by definition has a tangent (compact) intersection with any given line of sight as discussed above \citep{Howard2013}. 
This is generally true for any surface, not just spherical ones, as it is a consequence of the interior-extremum theorem of calculus. {Note that this 2D  tangent to the line of sight (that we refer to as the ``leading edge'') is slightly different than the CME front propagation location  that is radially farthest from the Sun (sometimes referred to as the CME ``nose''; see, e.g., \citet{moestl_2012,moestl_2013}) -- an effect that leads to an uncertainty in CME direction and propagation speed, even when the CME nose location is fully known \citep{malanushenko_this_issue}.
Note also that our croissant model does not attempt to reproduce the outermost sheath of the CME or associated shock, which are distinct from the magnetic ejecta that drives it and the ICME's leading edge \citep{kilpua_2017,temmer_2022}. It would be well worth doing a full analysis of how the inclusion of polarization localization might improve uncertainties in white-light reconstruction techniques for realistic CME models (e.g., \citep{verbeke_2023}), but this lies outside the scope of this paper.}

For coronagraphic images, there is a front-back ambiguity inherent to Equation~\ref{Eq-Xpol} that is unavoidable, because the TS is nearly aligned with the $x=0$ plane, and thus the two solutions are essentially symmetric about this observing plane of sky. Because heliospheric imagers have a wide field of view, however, the curvature of the TS breaks the symmetry, so that a time series of $x_{pol}$ for an Earth-approaching CME produces a radial ``true" trajectory, and a curved ``ghost" trajectory as shown in Figure~\ref{fig_approach_ghost}. From this it can be seen that the front solution (orange dots) follow the correct trajectory with monotonically increasing radial distance and always positive $x_{pol}$, while the back solution (purple circles) trace out the ghost trajectory that first recedes and then advances toward the observer.  In this case, all points of the true trajectory lie in front of the TS (positive $\tau_{pos}$ —  orange dots), while all the points shown on the ghost trajectory are behind the TS (negative $\tau_{pos}$ —  purple circles). 

Figure~\ref{fig_approach_ghost_high} shows a steeper trajectory that crosses the TS closer to the Sun. In this case, the true trajectory includes some points from the front solution and some from the back, as does the ghost trajectory. Nevertheless, 
the true solution can be found by making the choice for each time step of the trajectory closest to radial {See {\citet{malanushenko_this_issue}} for a demonstration using simulated PUNCH data synthesized from a global MHD model}. 

The solutions for an Earth-receding CME are qualitatively different than the Earth-advancing CMEs shown in Figures~\ref{fig_approach_ghost} and \ref{fig_approach_ghost_high}. Figure~\ref{fig_retreat_ghost} shows that the true trajectory of a receding CME leading edge is radial and always negative $x_{pol}$, while the ghost trajectory is nonradial and always positive $x_{pol}$. The former is made up of $\tau_{pol}$ solutions behind the TS, and the latter of solutions in front of the TS.

\begin{figure}[ht!]%
\centering
\includegraphics[width=0.9\textwidth, trim={3 3 0 3},clip]
{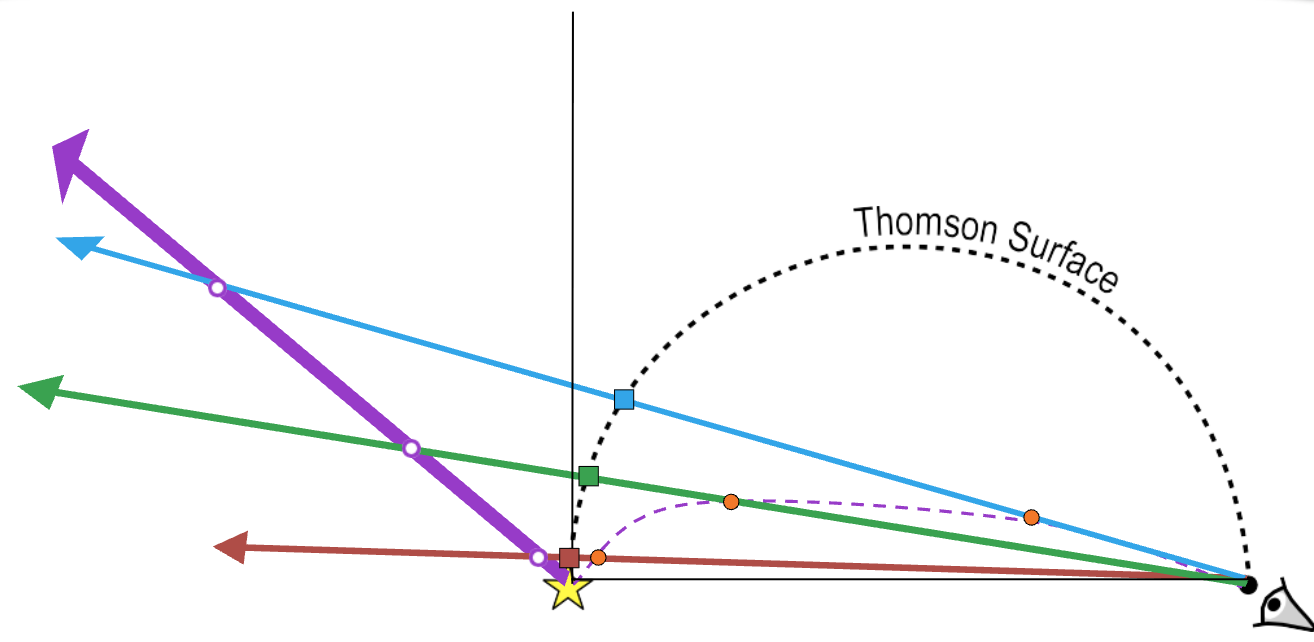}
\caption{CME moving away from observer. The solid purple line shows the true trajectory of the CME leading edge, and the dashed purple line shows the ghost trajectory.}
\label{fig_retreat_ghost}
\end{figure}

In summary: if the two (front/back of TS) solutions for position $x_{pol}$ for the CME leading edge are both positive at any point in the CME's passage, the CME must be Earth-approaching. If one is positive and one negative, it is generally necessary to follow the time series of the radial position of the front and back solutions (calculated via Equations~\ref{Eq-Xpol}-\ref{Eq-Zpol}) in order 
to establish a radial trajectory made up of some combination of front and back solutions; the sign of $x$ in that trajectory then determines whether the CME is Earth-approaching or Earth-receding.

\subsubsection{Imaging halo CMEs}\label{sec-halo}

As described above, and discussed in \citet{deforest_13b}, disambiguation for a CME leading edge is possible with a time series by plotting the radial position vs. time of the front and back solutions for the leading edge \citep{laurent_25}. But CMEs are more than a single leading edge, and information about their location is encoded in the polarization data from their full extent. {This has been demonstrated by \citet{jarolim_25}, who used a physics-informed-neural-network approach to track CME trajectory from synthetic images of eruptions.} 
In this section, we consider in general how images and movies of CME $x_{pol}$ might be used to establish whether a CME is advancing toward or receding from the observer.

\begin{figure}[ht!]%
\centering
\includegraphics[width=0.9\textwidth, trim={3 3 0 3},clip]
{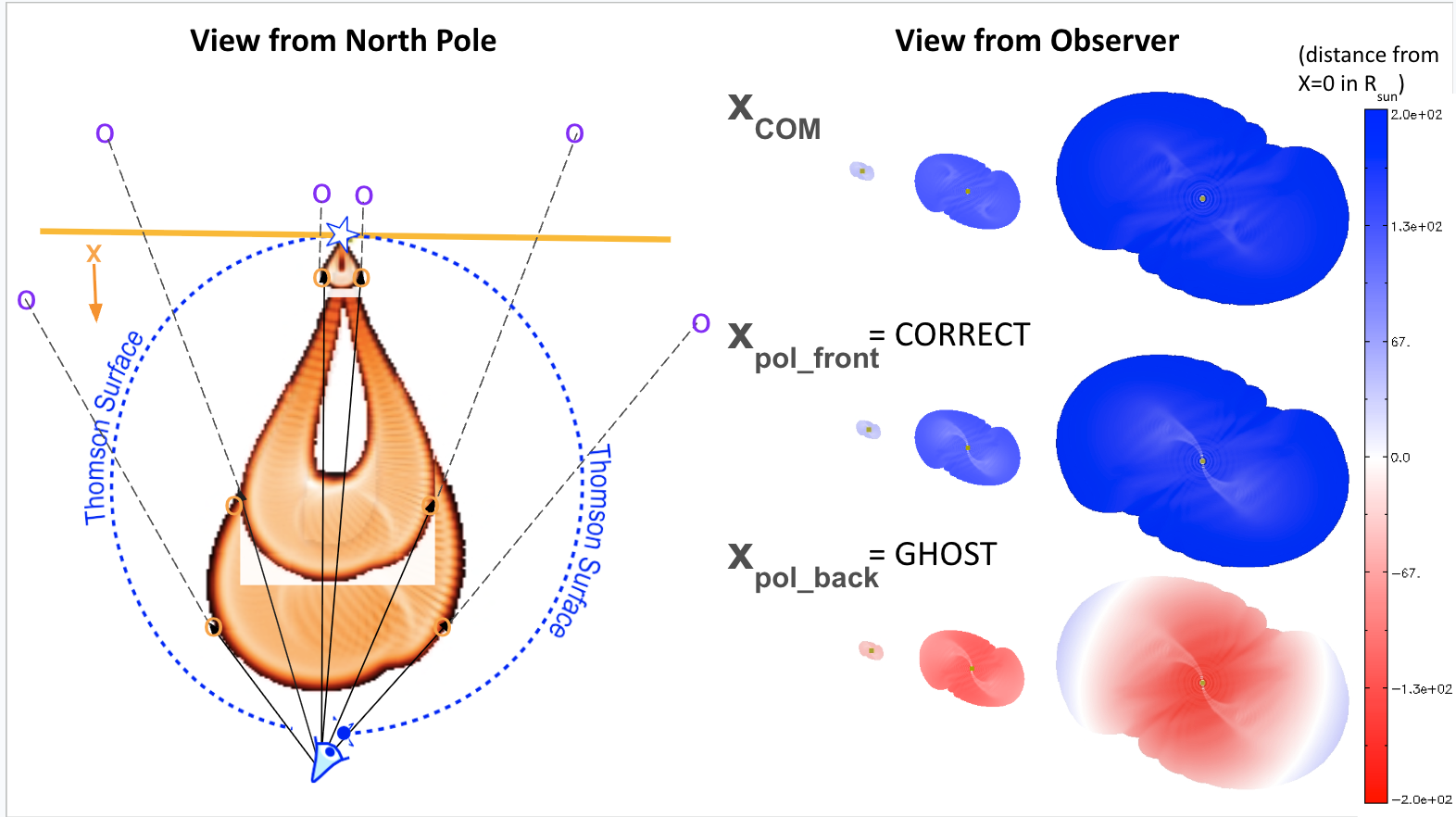}
\caption{Halo CME advancing toward the observer. Left: view from the North pole of the CME in $pB$ for three time steps. Orange circles show where lines of sight  intersect the flux rope edge in the $x_{pol_{front}}$ solution, purple circles show the ghost  $x_{pol_{back}}$ solution. Right: View from the observer for the three time steps for ground truth $x_{COM}$ (top), the correct, positive-$x$ solution (middle), and the ghost primarily negative-$x$ solution (bottom). The fact that parts of the most expanded CME shows both signs in $x_{pol_{back}}$ is a smoking-gun signature of an Earth-approaching halo CME.}
\label{fig_halo_approach_xpol}
\end{figure}

\begin{figure}[ht!]%
\centering
\includegraphics[width=0.9\textwidth, trim={3 3 0 3},clip]
{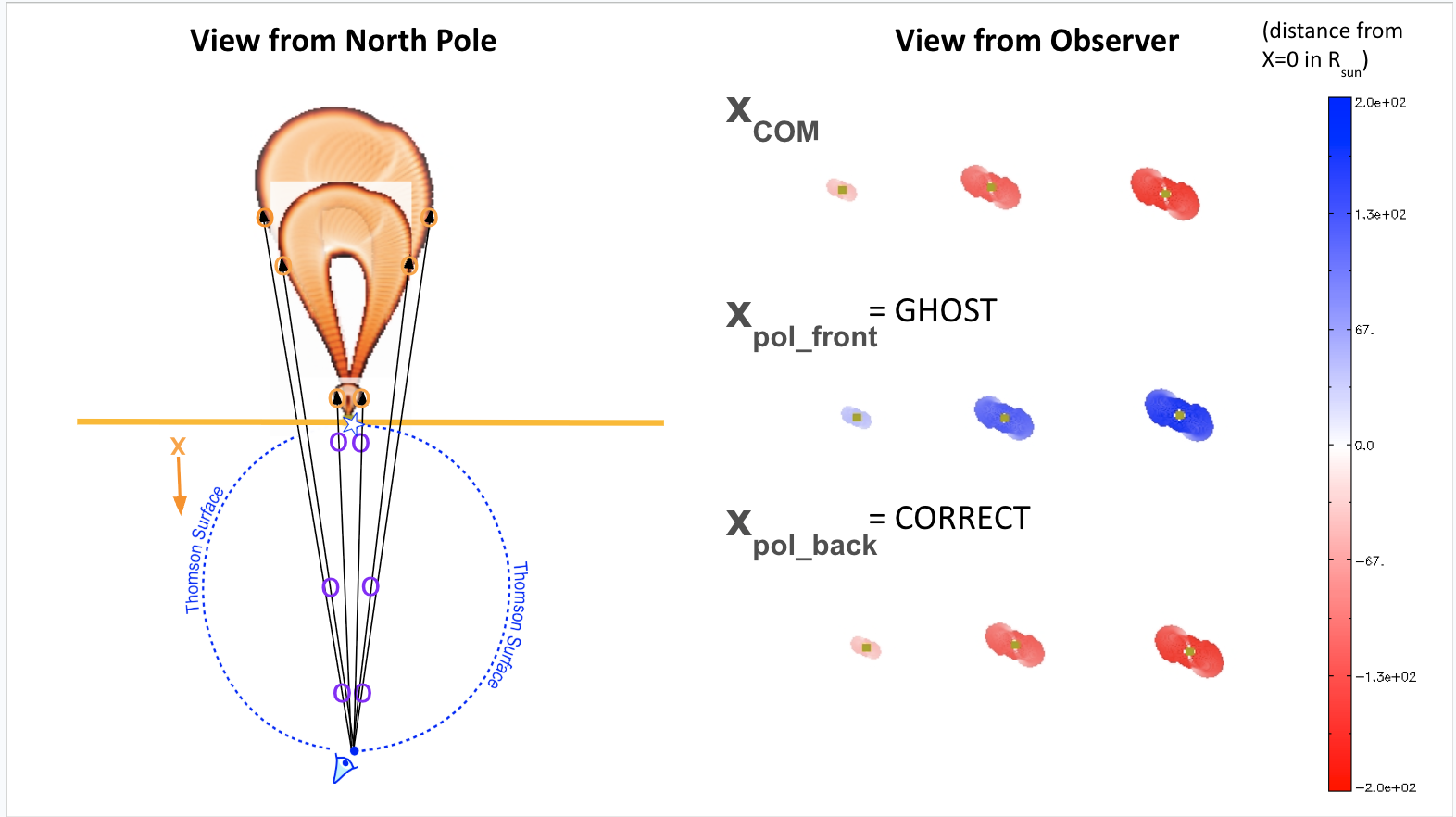}
\caption{Halo CME receding from the observer. As in Figure~\ref{fig_halo_approach_xpol}, but with $x_{pol_{back}}$ being the correct solution and $x_{pol_{front}}$ the ghost solution. Note that the solutions asymptote to a maximum apparent size as lines of sight become effectively parallel.}
\label{fig_halo_recede_xpol}
\end{figure}

\begin{figure}[ht!]%
\centering
\includegraphics[width=0.85\textwidth]
{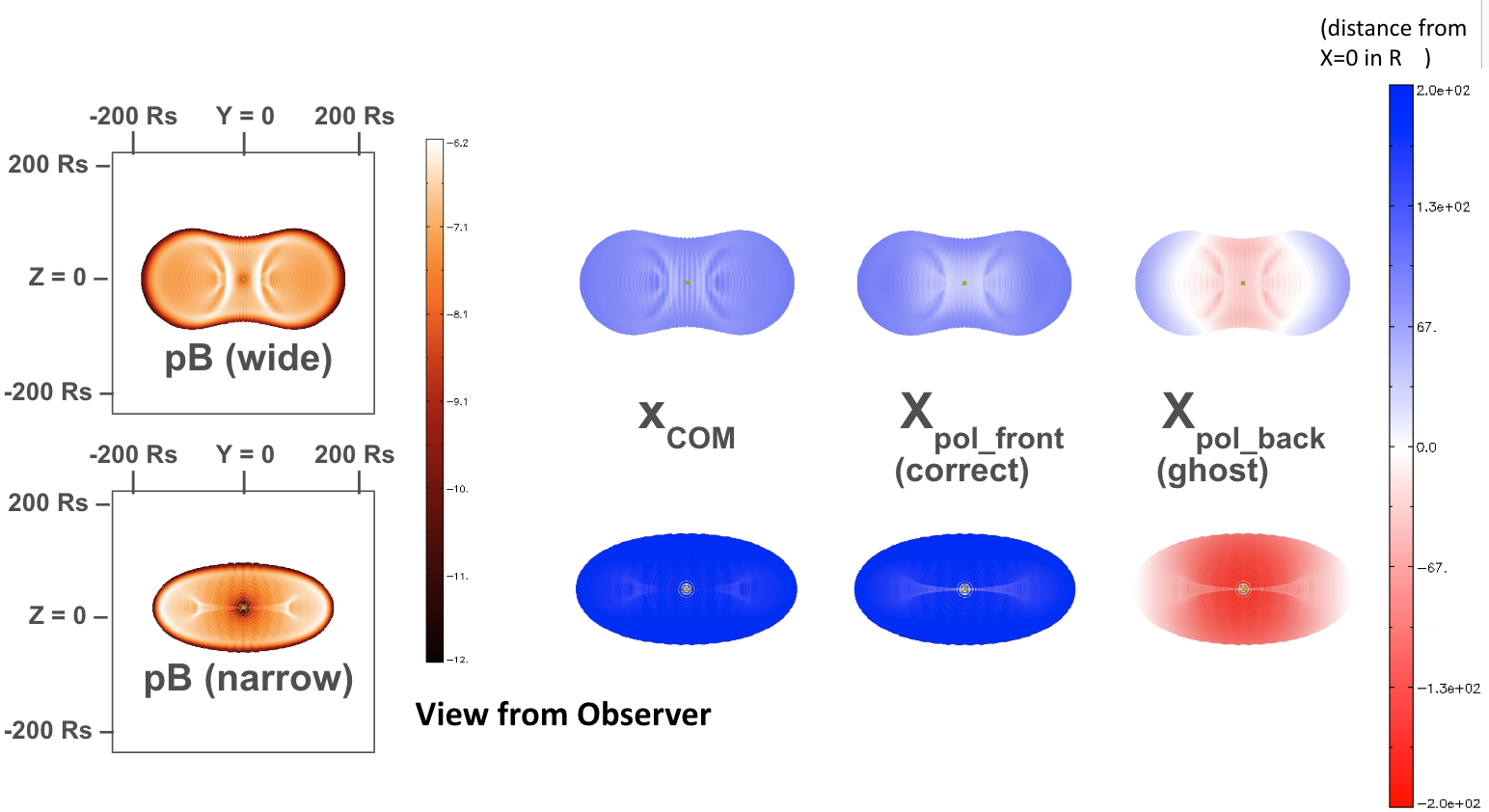}
\caption{Wide vs. narrow halo CME approaching the observer. Left column shows wide (top) and narrow (bottom) halo CMEs (see Figure~\ref{fig_widenarrow_north}) from the observer's point of view. Right three columns show that despite the fact they have the same projected apparent size and aspect ratio, the polarization ratio yields very different results and immediately shows the wide CME (top row) is farther from the observer, as evidenced by the lower distance from $x=0$ (paler red/blue in images).}
\label{fig_widenarrow}
\end{figure}

\begin{figure}[hb!]%
\centering
\includegraphics[width=0.6\textwidth]
{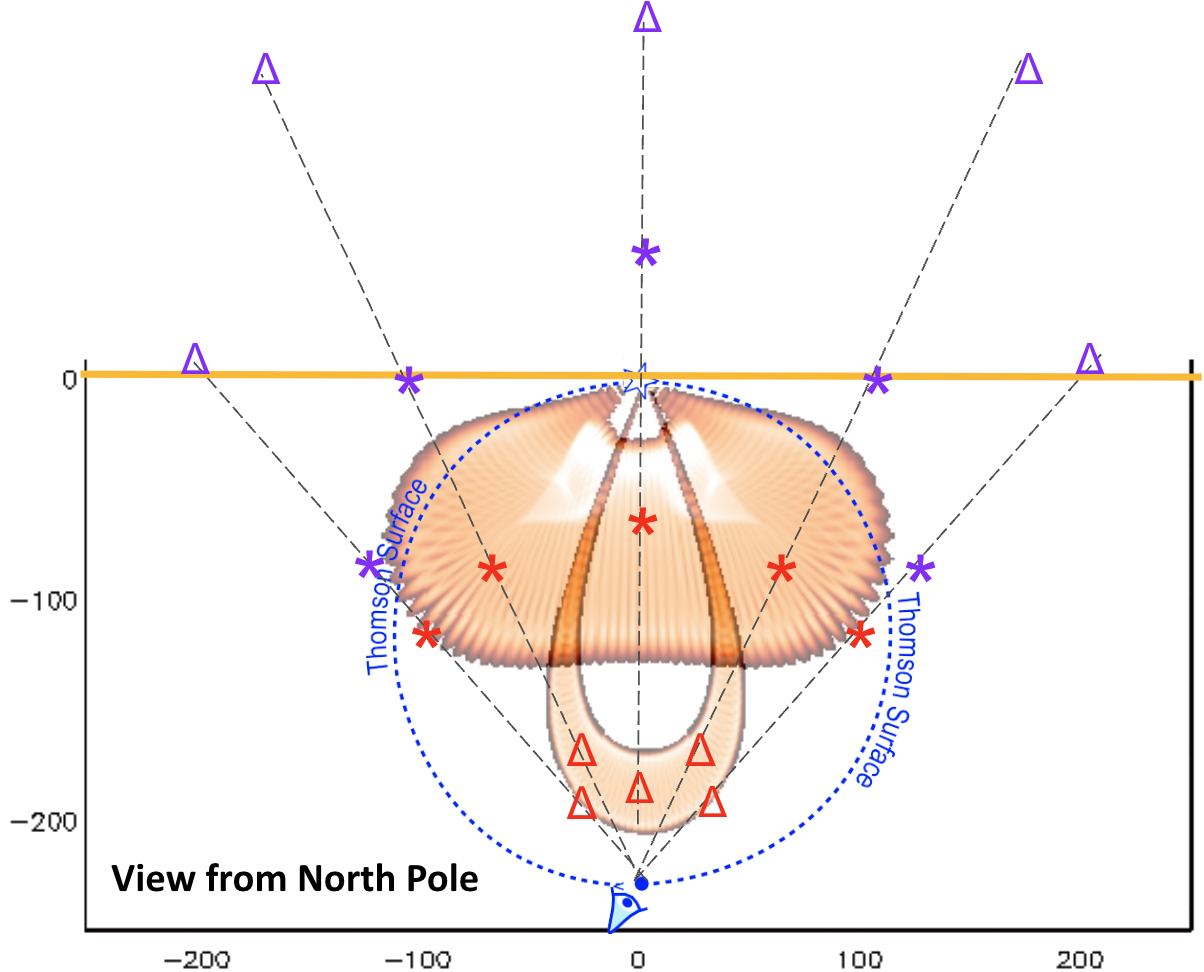}
\caption{Wide vs. narrow halo CMEs can be disambiguated by polarization. Red triangles show the correct $x_{pol_{front}}$ solution for the various lines of sight through the narrow CME, while purple triangles show the ghost $x_{pol_{back}}$ solutions on the opposite side of the TS. Similarly, red asterisks show the correct $x_{pol_{front}}$ solutions for the wide CME, while purple asterisks show the ghost $x_{pol_{back}}$ solutions. In this case, only the leading edges for each CME can be thought of as superparticles and located by $x_{pol_{front}}$, while the other lines of sight intersect multiple structures and result in a representation of center of mass equivalent to $x_{COM}$. }
\label{fig_widenarrow_north}
\end{figure}

Figure~\ref{fig_halo_approach_xpol} shows images for a Earth-approaching case similar to Figure~\ref{fig_approach_ghost}. On the left, the view from the North pole shows images of $pB$ for the CME for three time steps as the flux rope expands self-similarly. The arrows connect the observer to points tangent to the {flux rope} (orange circles) and extend toward the plane of the sky with dashed lines, ending in the purple circles  that indicate the ghost position on the other side of the TS. The view from the observer is shown on the right side, with the ground truth $X_{COM}$ showing the true location of the flux-rope edge in front of the $x=0$ plane (red), captured also in $x_{pol_{front}}$. 

The ghost solution $x_{pol_{back}}$ is by definition always behind the $TS$, but the most expanded stage has leading-edge ghost positions in front of the $x=0$ plane, yielding a multicolored image (bottom right of Figure~\ref{fig_halo_approach_xpol}).  This can only happen for an Earth-approaching halo CME. As Figure~\ref{fig_halo_recede_xpol} shows, a halo CME receding from the observer will maintain opposite colors for $x_{pol_{front}}$ and $x_{pol_{back}}$, and will asymptotically approach an unchanging apparent size.

Figure~\ref{fig_widenarrow} introduces another use for polarization, the disambiguation of situations  where non-parallel lines of sight lead to a similar appearance for narrow, fast CMEs vs. wide, slow CMEs. The two cases shown correspond to wide and narrow halo CMEs (also shown from a North Pole view in Figure~\ref{fig_widenarrow_north}), and illustrate how polarization data yield very different results despite similarity in $pB$ projected size and shape. The time series of these polarization images can be used to establish the leading edge trajectory as described above, establishing CME speed and predicting time of arrival. But even a single image is enough to show that the wide CME is much closer to $x=0$ than the narrow CME, and thus moving more slowly.

We note that Figures~\ref{fig_halo_approach_xpol}-\ref{fig_widenarrow_north} represent cases where lines of sight intersect multiple structures, in particular, as Figure~\ref{fig_widenarrow_north} demonstrates explicitly, all but the leading edge (tangent) lines of sight intersect both sides of the croissant. Nevertheless, Figure~\ref{fig_widenarrow} shows that $x_{pol_{front}}$ reproduces $x_{COM}$ for most points in both projected CMEs.
{See }{\citet{dekoning_this_issue}} {for further demonstration of the resiliency of the superparticle approximation to the superposition of multiple structures along the line of sight.)}


In summary: even a single snapshot of a halo CME can yield information about whether it is Earth-approaching or Earth-receding. It can also provide immediate information about whether it is wide and slow vs. narrow and fast.

\subsection{Polarization Diagnostics of CME Substructures}\label{sec:forward_chi}

CMEs are generally expected to incorporate coherently-twisted magnetic fields, i.e., magnetic flux ropes.  The chirality, or handedness of magnetic twist of the flux rope is significant, because they provide information about the direction of the magnetic fields that may interact with the Earth's magnetic field  \citep{cane_2000,yurchyshyn_2003,Deforest_17,pal_2022}. {Figure~\ref{fig_chiral} illustrates this for two examples of right- and left-handed flux ropes expanding out from the Sun.}

 \begin{figure}[ht!]%
\centering
\includegraphics[angle=270,width=0.9\textwidth]
{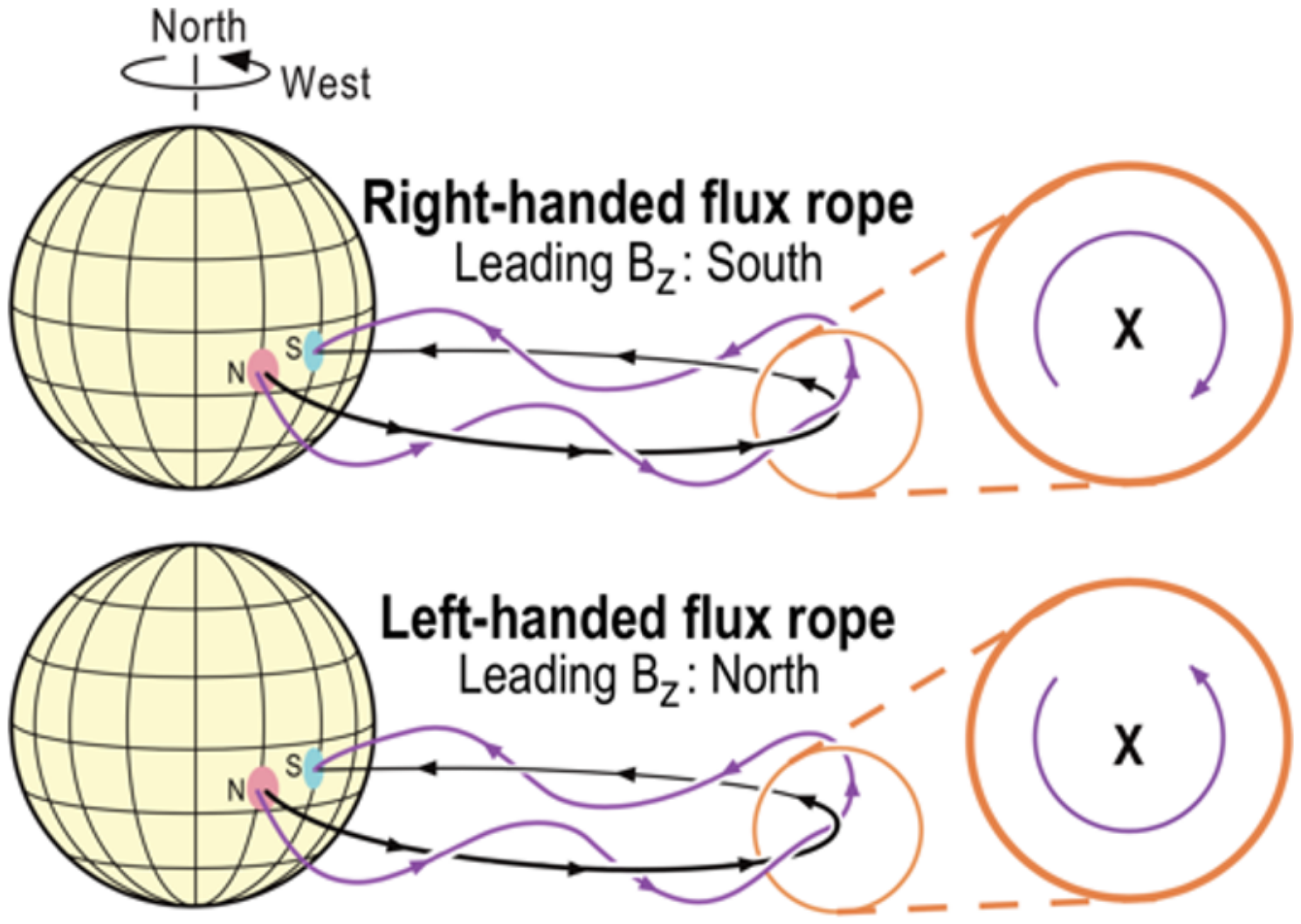}
\caption{{Representation of two flux ropes, right-handed and left-handed, with the same polarity footpoints at the Sun. As they expand into interplanetary space, we see the right-handed flux rope has South Bz at its front, while the left-handed flux rope has North Bz at its front.}}
\label{fig_chiral}
\end{figure}

 \begin{figure}[hb!]%
\centering
\includegraphics[width=0.9\textwidth]
{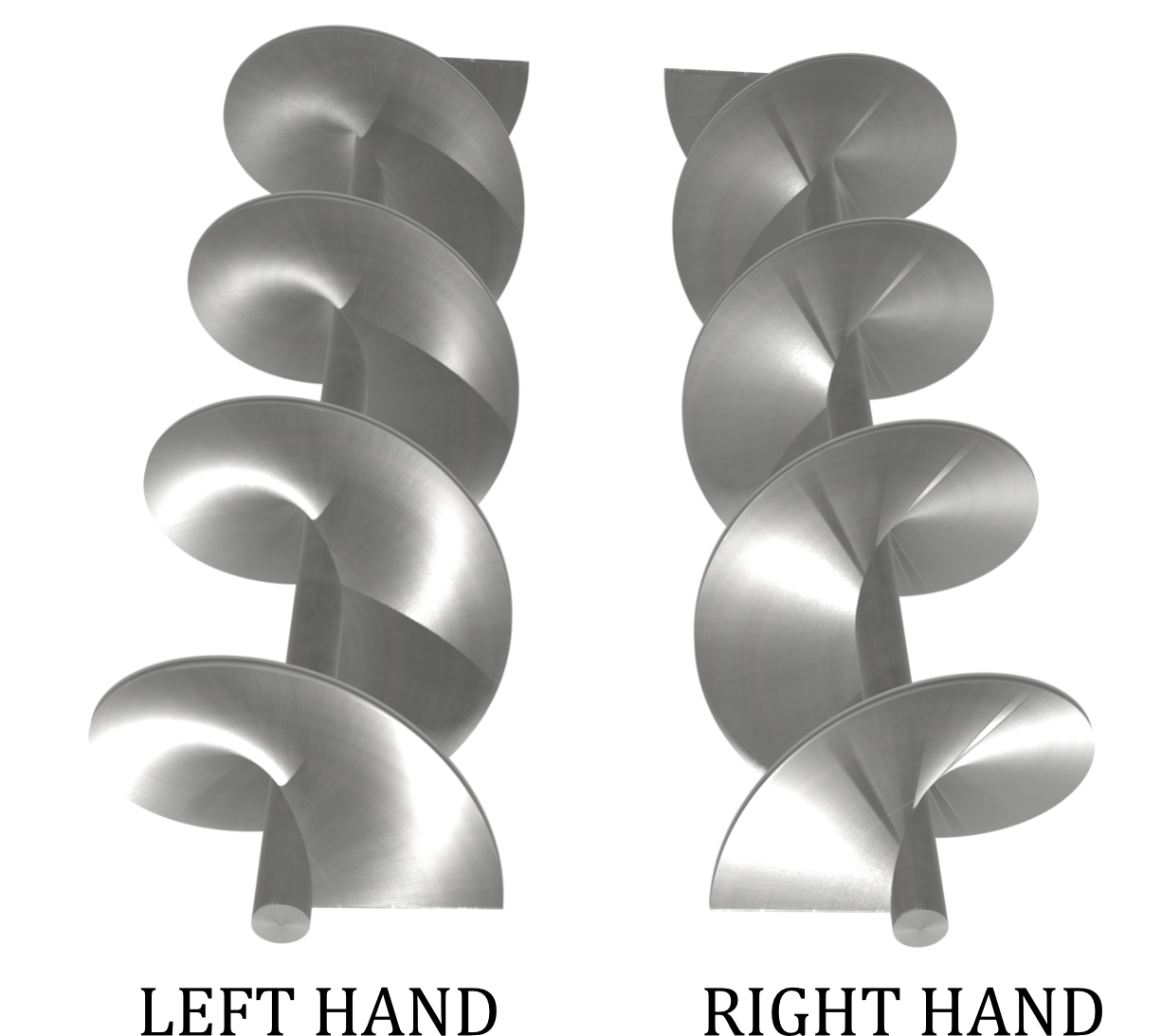}
\caption{Chirality can be determined by considering the direction a structure curves when moving from front to back, clockwise (right-handed) or counterclockwise (left-handed).}
\label{fig_screw}
\end{figure}

\subsubsection{Diagnosing chirality}

By establishing the location of helical CME substructures using polarization, it is possible to establish chirality from the direction the structure curves when traced from front to back. If clockwise, it is right-handed, while if counterclockwise, it is left-handed (see Figure~\ref{fig_screw}).
The ability to do this was demonstrated using coronagraph observations in \citet{Deforest_17}, where white light polarization data was used to establish the distance from the $x=0$ plane for a substructure within the erupting CME. Since the CME was observed from two vantage points (SOHO and STEREO), the front vs. back solution was easily disambiguated. The $x$ position of the substructure was then traced from front to back along a C-shaped curve, establishing a clockwise direction, and thus right-hand twist. This particular CME was later observed as a magnetic cloud, which also showed right-handed twist, consistent with the determination made from the coronal polarization data.

\begin{figure}[h!]
\centering
\includegraphics[width=0.9\textwidth]{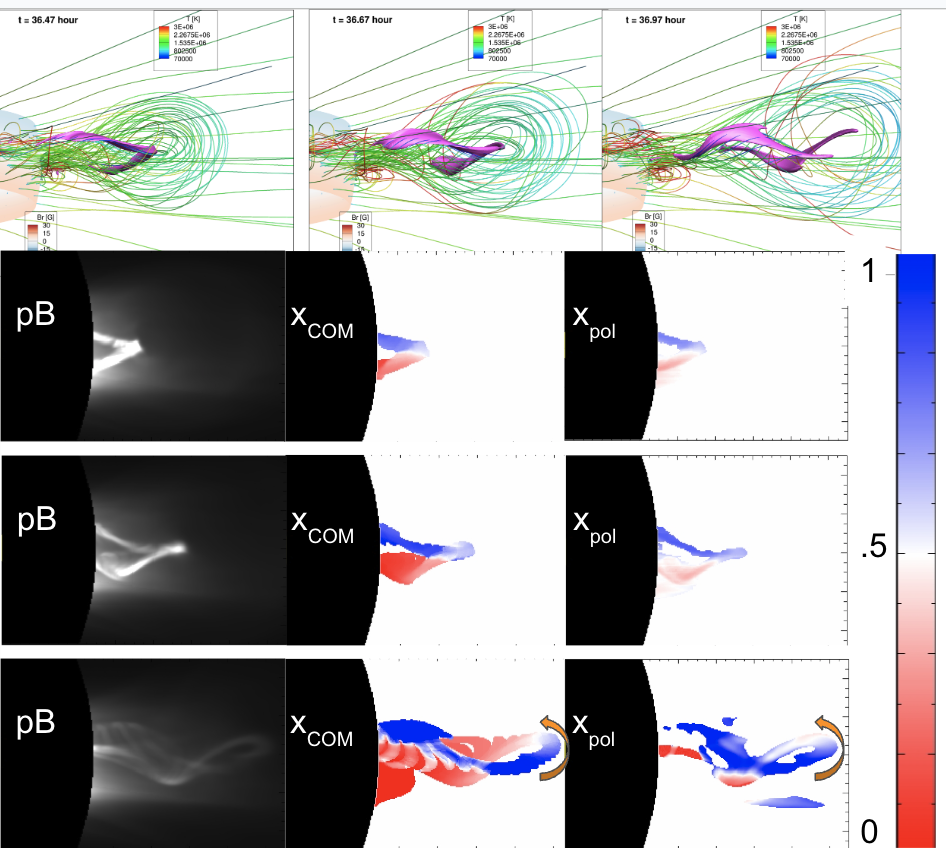}
\caption{Locating the structure of an erupting prominence/filament based on forward modeling of the \citet{fan_18} model of an erupting prominence-carrying flux rope. 
Top row: magnetic field lines of the erupting flux rope and the associated prominence
material outlined by the purple isosurface of temperature (at T = 80,000 K)
at three successive time instances indicated in the images.
Next three rows:
Left column shows the corresponding forward modeled white light pB, where the dense
prominence material carried by the erupting flux rope is clearly seen as the bright
core of the CME.
Middle column shows the position along the line of sight of the density center of mass ($x_{COM}$),
thresholded $n_e > 10^8$, $10^8$, $5 \times 10^7 {\rm cm}^{-3}$
for the three time instances, and the right column shows the position
extracted from background-subtracted synthetic polarization data as described in the text,
with thresholded $pB > 12$, $9$, $0.6 \times 10^{-8} B_{\odot}$ for the three time instances.
Distance from the $x = 0$ plane of sky is indicated by color, in solar radii, offset so that the change from red
to blue occurs at the center of the flux rope, x = 0.5. The arrow in the bottom row traces the
direction the filament curves at the apex: counterclockwise from front to back,
and thus left-handed chirality, which is consistent with the writhing of the (purple) filament
shown in the top row. Occulting disk is set at $1.5 r_{\odot}.$
}\label{fig_yuhong}
\end{figure}

To demonstrate this, Figure~\ref{fig_yuhong} forward models a simulation of an erupting magnetic flux rope {close to the Sun}  \citep{fan_18}. This MHD model incorporates simple empirical coronal heating, optically thin radiative cooling, and field-aligned thermal conduction, and thus allows the formation of prominence condensations, which we will locate using polarization data.   For three sample time steps, {we first calculated the density center of mass ($x_{COM}$; middle column) and then forward modeled synthetic $pB$ and $tB$
 and calculated $x_{pol}$ (right column). Note that thresholding has been employed to isolate the substructure from the background. In particular, only densities greater than the threshold values indicated in the figure caption are used in the center-of-mass calculation. For the polarization calculation, we have subtracted  an earlier timestep (at $t = 36.17$ hour) and then have run the calculation using only pixels with $pB$ greater than a thresholded value (indicated in the figure caption).}

\begin{figure}[ht!]%
\centering
\includegraphics[width=0.9\textwidth]{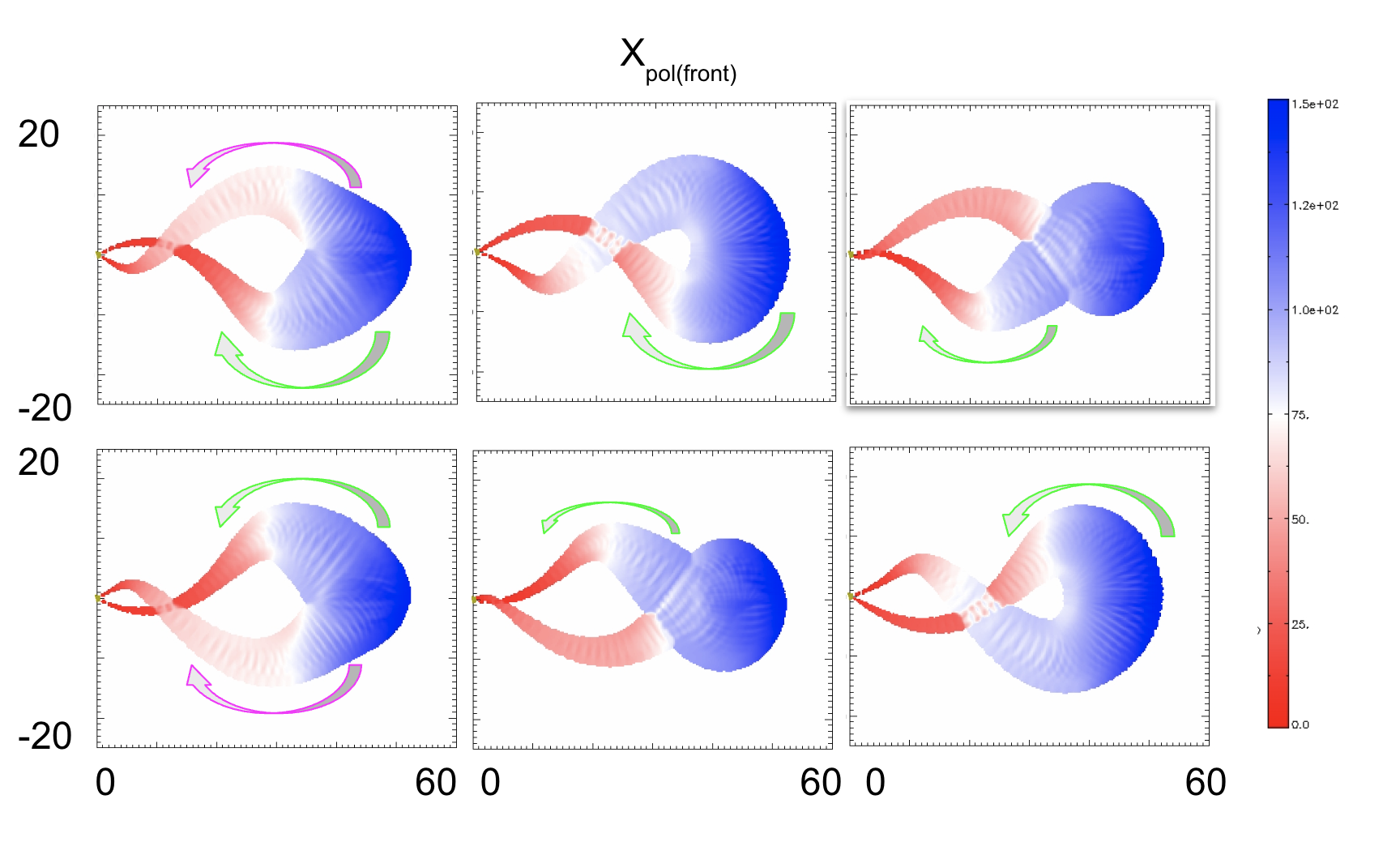}
\caption{Twisted croissant CME model for right-handed (top) and left-handed (bottom) flux ropes, for three orientations of the ropes (rotating about the $Y-$axis. Axes are plotted in units of $R_{\circ}$, and color bar shows $x_{pol}$ in the same units. The green curved arrows show circulation about an axis that establishes flux rope chirality, while the pink arrow tracks a misleading apparent circulation of the opposite sign (see text). \textcolor{red}{MOVIE}.}\label{fig_twistcroissant}
\end{figure}

{The $x_{pol}$ we have used is the front solution, as we } have centered the flux rope 20 degrees in front of the $x=0$ plane so that it lies entirely in front of the TS, and assume that this has been established by, { e.g., EUV observations projected against the solar disk. }
In addition, the rope is rotated 10 degrees clockwise around the Y axis. The position along the $X$-axis is indicated by the color bar: note that unlike previous figures, the color bar is not centered on $x=0$ but rather the center of the rope, in order to make it easier to see the direction of twist.

 The first thing to note is the general agreement of $x_{COM}$ and $x_{pol}$. {This substructure is localized, 
 but some small difference is expected due to the error associated with proximity to the TS, and indeed we see as expected $x_{pol} > x_{COM}$ near $x=0$, manifesting as red $\rightarrow$ pink. } Circulation about the flux rope axis is harder to determine, especially for the first two time steps. However, by the third time step, the rope has expanded out far enough to indicate a counterclockwise rotation when measured front to back (blue to red). {An analysis of the rotation of the simulated flux rope (counter-clockwise when viewed from above) provides further, complementary evidence for left-handed twist \citep{green_07,demoulin_07,gibfan_08}.}

Polarization analysis of erupting structures such as the prominence modeled here may be done in the low-to-middle corona using the $pB$ and $tB$ data from ESA's Proba-3 ASPIICS mission \citep{zhukov_25,zhukov_25b}. {In this region, the TS lies along the $x=0$ plane and does not have the symmetry-breaking curvature found in a wider field of view. The front-back ambiguity is thus harder to disentangle (see, e.g. \citet{Dai_2014}). One way around this is to use multiple viewpoints, e.g., polarization data from the STEREO spacecraft \citep{Dai_2020}. Another way is to monitor an eruption starting in the ASPIICS field of view as it continues into the PUNCH field of view, and use the polarization data from PUNCH to help disambiguate the polarization information in the lower, ASPIICS field of view.}

To simulate twist in the outer, PUNCH field of view, we turn to our twisted croissant model. 
Figure~\ref{fig_twistcroissant} plots a somewhat more twisted model than Figure~\ref{fig_croissant_rotate}. Again, we have positioned this structure so that it lies entirely $x > 0$, and make the assumption that this has been established by following the CME trajectory as described above. Also again, we have centered the color table on the center of the rope, rather than on $x=0$. 

\begin{figure}[ht!]%
\centering
\includegraphics[width=0.9\textwidth]{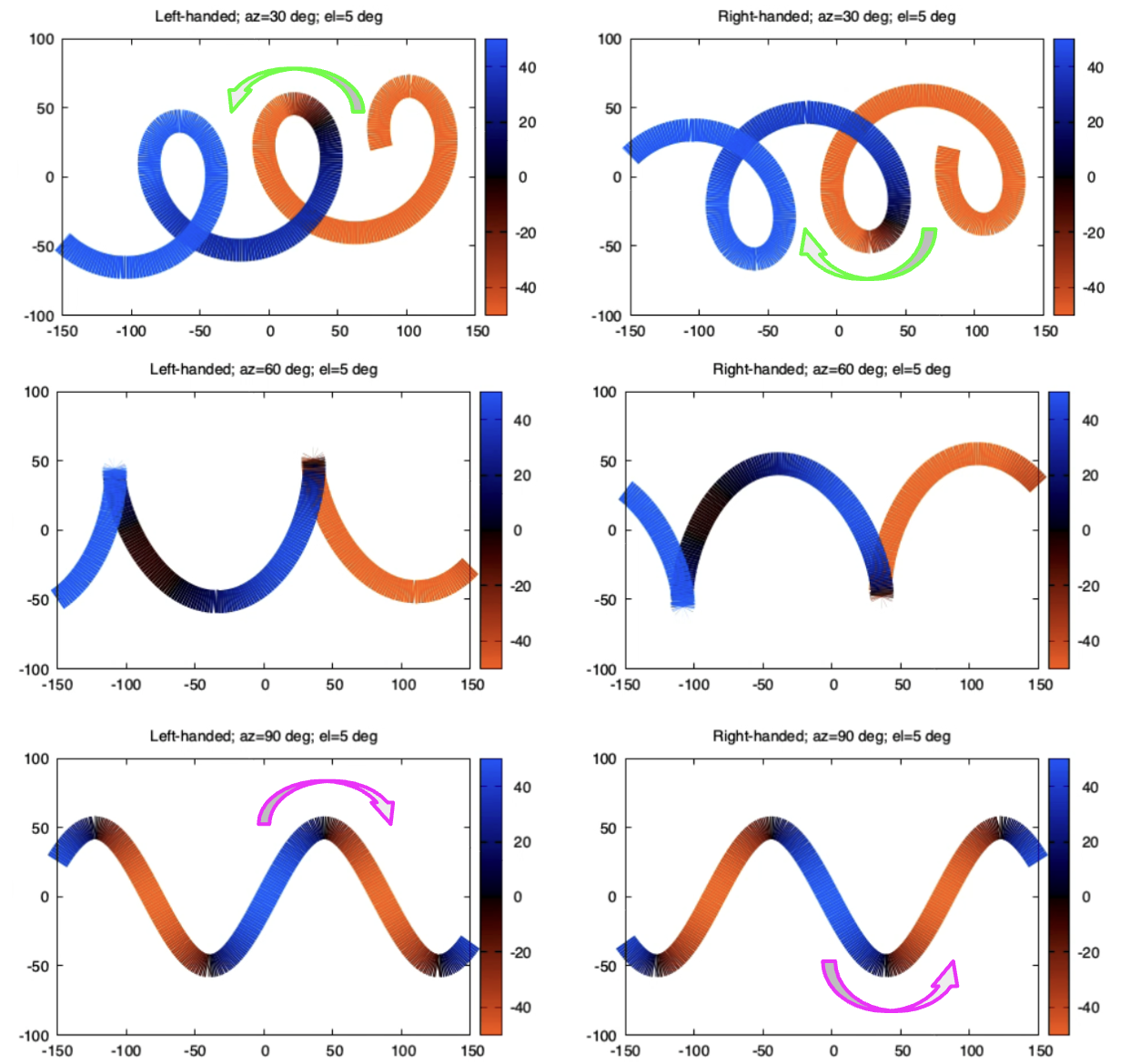}
\caption{Simple helical model with blue-red color table indicating distance in the foreground (blue) and background (red). Left (right) column is left(right)-handed rope. Orientation of the rope is rotated clockwise about the vertical axis. The green curved arrows show circulation about an axis that establishes flux rope chirality, while the pink arrow tracks a misleading apparent circulation of the opposite sign (see text). \textcolor{red}{MOVIE}}\label{fig_toy}
\end{figure}

The $x$ position of the center of mass is well captured from synthetic polarization data as described above, and the positive chirality can generally be determined by following curves from front to back (blue to red), as shown by the green curved arrows. However, the pink curved arrow shows a potential ambiguity for the first orientation of the rope, which appears to rotate counterclockwise front-to-back in its northern portion as opposed to the clockwise rotation seen in its southern portion throughout (and vice verse for the left-handed rope in the bottom row). 

To shed light on why this occurs, we present an even more idealized model in Figure~\ref{fig_toy}. As in Figure~\ref{fig_twistcroissant}, for some orientations the axis of rotation is clear, and the circulation about it correctly indicates the sign of chirality. However, the bottom panel shows a view perpendicular to the axis where following a curve from front to back yields the wrong sign. 

In order to remove this discrepancy, it is necessary to not rely on any one projection, but rather to use the full 3D location of the substructure (Equations~\ref{Eq-Xpol}-\ref{Eq-Zpol}) to establish circulation about an axis.

In summary: To locate the position of a CME substructure in 3D, it is first necessary
to eliminate the ghost solution. After this, its 4D space-time trajectory may be obtained and potentially used to diagnose chirality.

\section{Discussion and Conclusions}\label{sec:conclusions}

In this paper, we have  demonstrated how the polarization ratio $PR$ contains useful information about the two limits discussed in this paper, i.e., that of a superparticle, or compact mass along the line of sight, and that of spherically symmetric density distribution.  The former is dependent on the degree to which the mass is localized, which we have shown has a non-trivial dependence on LOS linear half-width and LOS viewing angle {as characterized by the critical width parameter, $\frac{w}{2d}$}.  The latter represents a local diagnostic of the global density radial falloff that might be used to study variations in the background solar wind as a function of time -- for example global responses to large CMEs, or solar cycle timescale variation.

We have used forward models of CMEs to illustrate how the 3d location of the edges of CMEs can be determined and have demonstrated that even with multiple compact masses along a line of sight, the COM density position can be reproduced. We have shown how individual polarization images can establish whether a CME is approaching or receding, and whether it is wide and slow vs. narrow and fast. We have demonstrated that while flux rope chirality (or handedness) can be deduced by tracing the curvature of CME substructures, it requires consideration of all projections of the 3D surface mapped out by polarization.

The models used in this paper are idealized, as our goal has been to establish  proof-of-concept for the methods described and to deepen our understanding of the sources of uncertainty for polarization diagnostics.  We refer the reader to \citet{malanushenko_this_issue}{ for an application of these diagnostics to white-light CME data synthesized from more sophisticated global MHD simulations.} Of course, real data introduces a range of additional practical difficulties. In particular, imaging data from PUNCH and elsewhere are subject to noise. As described in Section 3.2 of \citep{deforest_this_issue}, PUNCH has been designed with sensitivity requirements intended to allow 3D tracking of features of expected brightness. There is also the problem of disentangling background structures from CMEs, which time-differencing as employed in Figure~\ref{fig_yuhong} cannot fully address. However, we note sophisticated means exist for isolating CME signal from the background without amplifying noise \citep{morgan_2012}, and these have been applied successfully to real time operations (see, e.g., \url{https://solarphysics.aber.ac.uk/Archives/cme_realtime/realtime_catalogue.html}). {Finally, there are the uncertainties intrinsic to 3D reconstruction techniques as referenced above.}

{Setting aside noise {and uncertainty}, we have demonstrated some intrinsic error sources associated with polarization feature localization, which can be quantified if we are in the small-width limit. Just as importantly, we have explored and demonstrated the limits of the superparticle approximation for polarized 3D analysis, including where and how the small-width limit breaks down so that there potential for the information from the polarization to be misinterpreted. As the critical width parameter $\frac{w}{2d}$ transitions toward the infinite width regime, $\tau_{pol}$ loses information about the wedge center position $\tau_o$, and the apparent over- and under-estimate in $\tau_{diff}=\tau_{pol}-|\tau_o|$ no longer represents an error on position localization, but simply the difference between any given $\tau_o$ and the value of $\tau_{pol}$ that happens to characterize that $\frac{w}{2d}$. For this reason, the exact solutions, $\tau_{diff}=0$ (black lines in Figures~\ref{fig-taudiffvarywidth}-\ref{fig-taudiffvarywidth-selfsim}), are essentially meaningless and potentially misleading: A broken clock is accurate twice a day. }

{It is important to emphasize that even such extended features, if visible in an image plane, can be localized along the lines of sight near their image-plane edges. As we have discussed, we can safely assume that CME ``leading edges'' are localized at a point of tangency between the line of sight and the CME envelope, and the superparticle approximation therefore applies. For CME substructure or mesoscale ``blobs'' in the solar wind, one should establish $\frac{w}{2d}$, or, analogously, the near-Sun width for a self-similarly expanding feature, in order to establish how localized a feature is and to what degree the superparticle approximation may be used to characterize its structure overall. We note that features that expand at a less than a self-similar rate can actually become more localized as they travel outward, e.g., the compact prominence tracked by \citet{zhuang_2025}.}


{SIRs are an intermediate case in that they have global scale but form comparatively thin manifolds and hence may be localized along some lines of sight where they are visible, {for example when the lines of sight are tangent to the stream interface}. In general, our assumption of a uniform density wedge is not generally appropriate for SIRs. {\citet{dekoning_this_issue}} instead consider features of a given angular width $\Delta$ -- equivalent to our $\frac{w}{2d} \cos^2(\tau_o)$ in the small-width limit 
and possessing the same definition of compactness under the assumption of uniform density 
(i.e., $2\Delta << 1$, equivalent to $\frac{w}{d} \cos^2(\tau_o)<< 1$), but find a somewhat more stringent localization requirement when they assume a density profile of $r^{-2}$ within their feature, i.e., $4\Delta << 1$. In between, they examine how $PR$ might be combined with knowledge of either feature angular width or feature central angle to establish the missing variable. }

{There is thus clear benefit in supplementing polarization diagnostics with information about the size and structure of a feature being studied. Some promising approaches include geometric analyses of feature dimensions \citep{Moraes_2024ApJ,Lopez_2025}, multiple viewpoints from different spacecraft and/or tomographic reconstructions \citep{morgancook_2020,Dai_2020, jackson_2023,jarolim_25}, and global MHD simulations \citep{sachdeva_2021,lionello_2023,odstrcil_2023,pogorolev_2024,linan_2025,provornikova_this_issue}.}

In final summary, white light polarization entrains important information about the structure and distribution of plasma in the young solar wind. By exploring the analytic limits of polarization analysis, we have determined practical limits of when and how polarization data may be used in the context of solar wind imaging -- including with PUNCH. 
 
\section{Acknowledgments}

{PUNCH is a heliophysics mission to study the corona, solar wind, and space
weather as an integrated system, and is part of NASA’s Explorers program
(Contract 80GSFC14C0014). S.E.G., Y. F., and A.M. acknowledge support of the National Center for
Atmospheric Research, a major facility sponsored by the National Science Foundation under Cooperative Agreement No. 1852977. C. A. dK. acknowledges support from NASA grant N99055DS and AFOSR YIP program FA9550-21-1-0276, and  C. A. dK., C. E. D., A. M., and S. E. G. acknowledge funding from AFOSR grant FA9550-21-1-0457.
S. E. G. and C. E. D. acknowledge support from NASA grant 80NSSC21K1860 (US-coIs of the ASPIICS mission). E.P is supported by the AFOSR grant FA9550-21-1-0457 and AFOSR YIP program FA9550-21-1-0276.}

\newpage
\begin{appendices}

\section{Definitions}\label{secAdef}

See Figure~\ref{fig-geom} and \cite{deforest_this_issue} for illustration and further discussion of the quantities listed below. 
Bold-faced  font indicates {quantities known or directly measurable by PUNCH;} regular font indicates derived quantities.

\begin{itemize}
     \item  Radial-Tangential Normal (RTN) coordinate system    \citep{burlaga_84,franz_02}.
     \begin{itemize}
         \item  {\bf X axis} = Sun-observer line (positive/negative in front/behind Sun)
     \item {\bf Z axis} = projected solar rotation axis
     \item {\bf Y axis} = perpendicular to both X and Z in a right-handed system
     \end{itemize}
 \end{itemize}
 \begin{itemize}
     \item {\boldmath r$_{obs}$} = distance between observer and Sun
     \item {\boldmath $\varepsilon$} = elongation angle of line of sight measured from X axis
     \item {\boldmath $\alpha$} = polar angle of intersection of line of sight with Y-Z plane (Observing Plane of Sky), measured counterclockwise from Z axis about X axis.
    {\item {\boldmath $p$} = degree of polarization, $\frac{pB}{tB}$.}
\end{itemize}

\begin{itemize}
     \item $r_{pos}$ = distance from Sun center to intersection of line of sight with Y-Z plane 
     \begin{itemize} 
        \item $r_{pos} = r_{obs}~\tan \varepsilon$
        \item With $\alpha$, this specifies the location of the LOS-integrated observable on the Observing Plane of Sky
     \end{itemize}
       \item $d$ = distance of closest approach to Sun center of any line of sight (impact factor; occurs at intersection of line of sight and TS)
     \begin{itemize} 
        \item $d = r_{obs}~\sin(\varepsilon)$
     \end{itemize}
   \item {$PR$ = polarization ratio}
    \begin{itemize}
        \item $PR = \frac{1 - p}{1 + p}$
    \end{itemize}
    \item $\chi$ = scattering angle
    \begin{itemize}
        \item {$\chi_{pol}=\pm \operatorname{acos}(\sqrt{PR})$}
    \end{itemize}
    \item $\tau$ = angle along line of sight; $\tau=0$ is intersection of line of sight with TS, positive toward observer, negative away.
     \begin{itemize}
       \item {$\tau_{pol}=\pm \operatorname{asin}(\sqrt{PR})$}
       \item {$\tau = \frac{\pi}{2}- \chi$}
    \end{itemize}
    \item $\xi$ = out-of-sky-plane angle
    \begin{itemize}
       \item $\xi(\tau) = \varepsilon + \tau$
    \end{itemize}
    \item $\psi$ = complement to $\xi$
    \begin{itemize}
        \item $\psi(\tau) = 90 - \xi$
    \end{itemize}
        \item $r$ = distance to scattering point from center of Sun
    \begin{itemize}
        \item $r(\tau) = d / \cos(\tau)$
    \end{itemize}
    \item $s$ = distance from TS to scattering point for a given line of sight (positive toward observer, negative away)
    \begin{itemize}
        \item $s(\tau) = d~\tan(\tau)$
    \end{itemize}
    \item $\ell$ = distance from scattering point to observer (always positive)
    \begin{itemize}
        \item $\ell(\tau) = r_{obs} \cos (\varepsilon) - s$
    \end{itemize}
\end{itemize}

\section{{Croissant Model}}\label{secCCroissant}

\begin{figure}[hb!]
\centering
\includegraphics[width=1.0\textwidth]{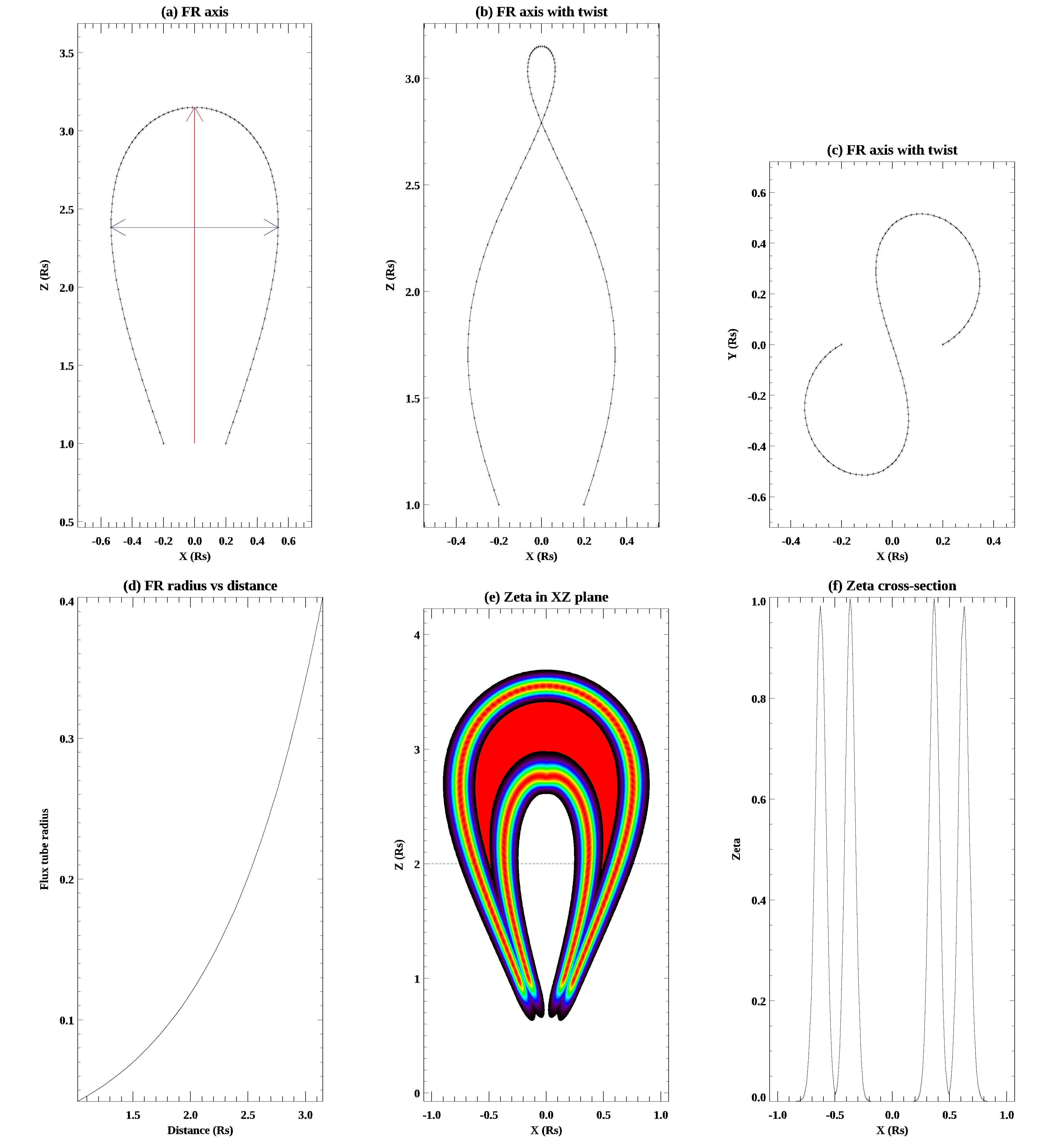}
\caption{(a) The initial geometry of the CME flux rope central axis, as described by equation \ref{ceq1}. The red arrow shows $r_{mx}$, or the maximum distance at the axis apex. The blue arrows illustrate $\Delta X$, set in this example to $\Delta X = 1$. (b) Example of applying a twist to the central axis, as viewed along the $y$ axis and (c) along the $z$ axis. (d) The flux rope radius $w$ as a function of distance as given by equation \ref{ceq2}. (e) $\zeta$, or the normalized spatial weighting of density given by equation \ref{ceq3}, defining the sheath around the central axis in the $xz$ plane (for a non-twisted central axis). (f) $\zeta$ plotted along the dashed horizontal line in (e).}
\label{croissant}
\end{figure}


{The model is initialized by defining the coordinates of a set of 100 points which give the location of an axis which lies along the center of the flux tube. These are defined as
\begin{eqnarray}
\label{ceq1}
x=(r_{mx}-1) sin(t) \Delta X + t l/\pi \\
y=0 \\
z=1+(r_{mx}-1) cos(t/2)
\end{eqnarray}
where $t$ is a parameter with 100 points increasing linearly from $-\pi$ to $\pi$, $r_{mx}$ is the maximum heliocentric distance (or apex) of the flux tube axis, $\Delta X$ is a parameter controlling the overall width of the flux tube (i.e. angular width separating the legs), and $l$ is a parameter that controls how rapidly the legs expand with height to the defined overall width. In this initial configuration, the flux tube axis is embedded in the $xz$ plane and is symmetric around $x=0$ (or aligned with the north pole), as shown in Figure \ref{croissant}a. If required, a twist can easily be applied by a rotation around the $z$ axis, with the rotation angle increasing from zero at $z=1$ (minimum distance) to the maximum defined twist angle at $z=r_{mx}$, as shown in Figures \ref{croissant}b and c. An overall orientation can also be applied by rigid rotation around the $z$ axis. A 3D rotation of the entire axis to the required central longitude and latitude relative to an observer, from it's initial location centered on the $z$ axis (or north pole), is then made.}

{The radius of the flux tube around the central axis is defined as
\begin{equation}
\label{ceq2}
\rho=\rho_0 \exp{ \left( [(r/r_{mx})-1]/h_\rho \right) } 
\end{equation}
where $\rho_0$ is a parameter controlling the maximum radius of the flux tube (at the maximum or apex distance with default value 0.4 for $r_{mx}=3.15$), $r$ is the heliocentric distance of each point along the axis, and $h_{\rho}$ is a parameter controlling how rapidly the flux tube expands with distance from the footpoints, with default value $h_{\rho} = 0.3$. $\rho$ is shown as a function of distance in Figure \ref{croissant}d.}

{Once the geometry of the axis and the flux tube radii are defined the density distribution of the flux tube sheath is calculated. There are very efficient ways to do this computationally, but for compatibility with FORWARD calculations this is made on a full 3D grid of voxels defining the corona. The orientation of the local normal to the axis relative to solar $x,y,z$ is calculated at each axis point. Looping through each point, indexed $i$, the coordinates of the voxels are translated to a new origin at that axis point, and the voxel coordinates rotated so that the $yz$ plane is tangent to the curve of the axis at that point, giving a new set of coordinates for each voxel: $x_n, y_n, z_n$. A weighting at each voxel is defined as
\begin{equation}
\label{ceq3}
\zeta = \exp{ \left( -[d_i-\rho_i]^2 / \sigma_i   - x_n^2 / 2\Delta s \right) },
\end{equation}
with $d_i = \sqrt{y_n^2+z_n^2}$ and $\Delta s$ the distance between each axis point. $\sigma_i$ is a parameter controlling the thickness of the sheath, and is set by default to $0.0016 r_i $ (so that the sheath thickness increases with distance). Looping through all hundred flux axis points, the maximum recorded $\zeta$ is retained at each voxel.}

{$\zeta$ is a spatial weighting ranging from zero to one: a two-dimensional cross-section is shown in Figure \ref{croissant}e, and a one-dimensional cross-section is plotted in Figure \ref{croissant}f. We set the total mass of the CME, and distribute that mass appropriately by $\zeta$. Furthermore, we apply a smooth function of height that distributes more of the mass at the CME apex compared to the footpoints. Without this function the CME legs dominate the brightness when FORWARD applies LOS integrations. In principle, more realistic density distributions may be sought, although CME masses and density distributions are currently poorly constrained by observations.}

\end{appendices}
\bibliographystyle{spr-mp-sola}
\bibliography{mybibliography}%
\end{document}